\documentclass[letterpaper,11pt,usenames,preprintnumbers,dvipsnames,cmyk]{article}
\pdfoutput=1

\usepackage[totalwidth=480pt, totalheight=680pt]{geometry}
\usepackage{amsmath, array}
\usepackage{amsfonts}
\usepackage[colorlinks=true, linktoc=page, citecolor=blue, urlcolor=blue]{hyperref}
\usepackage[retainorgcmds]{IEEEtrantools}
\usepackage{scalerel}
\usepackage{verbatim}
\usepackage[bulletsep]{collref}
\usepackage{float}

\newcommand{\aaa}{\mathfrak{a}}
\newcommand{\bbb}{\mathfrak{b}}
\newcommand{\ccc}{\mathfrak{c}}

\hypersetup{pdfstartview={XYZ null null 1}}

\begin{document}
\begin{flushright}
APCTP Pre2024-021
\end{flushright}
\title{{\color{Blue}\textbf{Turbulent aspects of BMN membrane dynamics}}\\[12pt]}\date{}
\author{\textbf{Minos Axenides},$^1$ \textbf{Emmanuel Floratos},$^{1,2,3}$ \textbf{and Georgios Linardopoulos}$^{4}$\footnote{E-mails: \href{mailto:axenides@inp.demokritos.gr}{axenides@inp.demokritos.gr}, \href{mailto:mflorato@phys.uoa.gr}{mflorato@phys.uoa.gr}, \href{mailto:georgios.linardopoulos@apctp.org}{georgios.linardopoulos@apctp.org}.}\\[12pt]
$^1$ Institute of Nuclear and Particle Physics \\ National Centre for Scientific Research, "Demokritos" \\ 153 10, Agia Paraskevi, Greece. \\[6pt]
$^2$ Department of Nuclear and Particle Physics \\ National and Kapodistrian University of Athens \\ 157 84, Athens, Greece. \\[6pt]
$^3$ Research Office of Mathematical Physics and Quantum Information \\ Academy of Athens, Division of Natural Sciences, Athens 10679, Greece. \\[6pt]
$^4$ Asia Pacific Center for Theoretical Physics (APCTP) \\ Hogil Kim Memorial Building, \#501 POSTECH \\ 77 Cheongam-Ro Nam-gu, Pohang Gyeongsangbuk-do 37673, Korea. \\[12pt]}
{\let\newpage\relax\maketitle}
\begin{abstract}
\normalsize{\noindent We investigate the large-$N$ limit of the BMN matrix model with classical bosonic membranes which have spherical topologies and spin inside the 11-dimensional maximally supersymmetric plane-wave background. First we classify all possible M2-brane configurations based on the distribution of their components inside the $SO(3)\times SO(6)$ symmetric plane-wave spacetime. We then formulate a number of simple but very representative ansätze of dielectric tops that rotate in this space. We examine the leading-order radial and angular/multipole stability for a wide range of these configurations. By analyzing perturbations at the next-to-leading order, we find that they exhibit the phenomenon of turbulent cascading of instabilities. Thereby, long-wavelength perturbations generate higher-order multipole instabilities through their nonlinear couplings. \\}
\end{abstract}
\newpage\tableofcontents\normalsize
\section[Introduction and motivation]{Introduction and motivation}
\noindent Plane-fronted (gravitational) waves with parallel rays (pp-waves for short) were introduced in 1925 by Brinkmann \cite{Brinkmann25} as solutions of the 4-dimensional (vacuum) Einstein equations. The metric
\begin{IEEEeqnarray}{c}
ds^2 = 2 du dv + H(u,x,y) du^2 + dx^2 + dy^2, \qquad \nabla^2 H(u,x,y) = 0, \label{MetricBrinkmann1}
\end{IEEEeqnarray}
expresses pp-waves in the so-called Brinkmann coordinate system. Ehlers and Kundt \cite{EhlersKundt62} equivalently defined pp-waves as spacetimes which afford a covariantly constant null Killing vector $k_n$:
\begin{IEEEeqnarray}{c}
\nabla_m k_n = 0, \qquad k_n k^n = 0.\footnote{$m, n \in \left\{u,v,x,y\right\}$. Setting $k_{n} \equiv g_{n v}$, the nullity $k_{n}k^{n} = g_{n v}g^{n v} = g_{vv} = 0$ and covariant constancy $\nabla_{m}k_{n} = \nabla_{m}g_{n v} = 0$ of the vector $k_{n}$ essentially follow from the independence of the pp-wave metric \eqref{MetricBrinkmann1} from the light-cone coordinate $v$. Evidently, $k_{n}$ also satisfies the Killing equation $\nabla_{m}k_{n} + \nabla_{n}k_{m} = 0$.} \label{NullKillingProperty}
\end{IEEEeqnarray}
The term "plane-fronted" refers to the fact that pp-wave spacetimes can be completely covered by 2-dimensional wave fronts that are orthogonal to the Killing vector $k$. Since $k$ is a constant, the wave fronts are planes that propagate parallel to each other in the direction of $k$ ("parallel rays"). \\[6pt]
\indent The mere existence of a wave vector $k$ already signifies a close relationship between pp-waves and our familiar electromagnetic waves. In fact \eqref{MetricBrinkmann1} also solves Einstein-Maxwell theory upon appropriately choosing the function $H(u,x,y)$. Plane waves are special pp-waves for which
\begin{IEEEeqnarray}{c}
H(u,x,y) = a\left(u\right)\left(x^2 - y^2\right) + 2b\left(u\right)xy + c\left(u\right)\left(x^2 + y^2\right), \label{MetricPlaneWaves1}
\end{IEEEeqnarray}
($c\left(u\right) = 0$ in the vacuum). Plane-waves are the gravitational analogs of plane electromagnetic waves. As such they provide the gravitational field of finite gravity sources very far away from them. \\[6pt]
\indent Pp-waves can be generalized to higher-dimensional spacetimes with or without supersymmetry. The most general metric of a $d + 1$ dimensional spacetime with a covariantly constant null Killing vector is:
\begin{IEEEeqnarray}{c}
ds^2 = -2 dx^+ dx^- - F(x^+,x^i) dx^+ dx^+ + 2 A_j(x^+,x^i) dx^+ dx^j + g_{jk}(x^+,x^i) dx^j dx^k, \label{MetricGeneral}
\end{IEEEeqnarray}
where $i,j = 1, 2, \ldots d - 1$ and
\begin{IEEEeqnarray}{c}
x^{\pm} \equiv \frac{1}{\sqrt{2}}\left(x^0 \pm x^{d}\right).
\end{IEEEeqnarray}
The functions $F(u,x^i)$, $A_j(u,x^i)$, $g_{jk}(u,x^i)$ are determined from the equations of motion of 11-dimensional supergravity which are in turn satisfied by the solution \eqref{MetricGeneral}. \\[6pt]
$\bullet$ For $A_j = 0$, $g_{jk} = \delta_{jk}$, we retrieve the $d+1$ dimensional Brinkmann metric (cf.\ \eqref{MetricBrinkmann1}):
\begin{IEEEeqnarray}{c}
ds^2 = -2 dx^+ dx^- - F(x^+,x^i) dx^+ dx^+ + dx^i dx^i. \label{MetricBrinkmann2}
\end{IEEEeqnarray}
$\bullet$ As before (cf.\ \eqref{MetricPlaneWaves1}) plane-waves are pp-waves with $F(x^+,x^i) = f_{ij}(x^+) x^i x^j$, $A_j = 0$ and $g_{jk} = \delta_{jk}$:
\begin{IEEEeqnarray}{c}
ds^2 = -2 dx^+ dx^- - f_{ij}(x^+) x^i x^j dx^+ dx^+ + dx^i dx^i. \label{MetricPlaneWaves2}
\end{IEEEeqnarray}
$\bullet$ Homogeneous plane-waves have a constant $f_{ij}(x^+) = \mu_{ij}^2$:
\begin{IEEEeqnarray}{c}
ds^2 = -2 dx^+ dx^- - \mu^2_{ij} x^i x^j dx^+ dx^+ + dx^i dx^i. \label{MetricPlaneWavesHomogeneous}
\end{IEEEeqnarray}
$\bullet$ Homogeneous and isotropic plane-waves have $\mu_{ij} = \mu$, on top of the previous attributes:
\begin{IEEEeqnarray}{c}
ds^2 = -2 dx^+ dx^- - \mu^2 x^i x^i dx^+ dx^+ + dx^i dx^i. \label{MetricPlaneWavesHomogeneousIsotropic}
\end{IEEEeqnarray}
\indent Pp and plane-wave spacetimes stand out thanks to a set of remarkable properties. Perhaps the most important one among them is that they (in their Brinkmann form \eqref{MetricBrinkmann2}) can be obtained from any given metric by means of the Penrose limiting procedure \cite{Penrose76}. The Penrose limit consists in blowing up the spacetime around null geodesics (effectively "zooming in" to them). This way, new exact solutions of Einstein's equations can be constructed from known ones. The Penrose limit has been generalized to string theory and supergravity by G\"{u}ven \cite{Gueven00}. \\[6pt]
\indent Further noticing that the pp-wave metric \eqref{MetricBrinkmann2} can be written in the form
\begin{IEEEeqnarray}{c}
g_{mn} = \eta_{mn} + h_{mn}, \qquad h_{mn} \equiv - F(x^+,x^i) k_m k_n, \label{MetricLinearizedGravity}
\end{IEEEeqnarray}
which resembles linearized gravity, it can be shown that \eqref{MetricLinearizedGravity} is an exact solution of Einstein's equations even when $h_{mn}$ is not a perturbation (i.e.\ "small"). As a consequence, many properties of flat Minkowski spaces can often be uplifted to pp-wave backgrounds with only minor modifications. Conversely, it has been shown by Penrose \cite{Penrose65a} that pp-wave spacetimes are not globally hyperbolic (as opposed to flat Minkowski space), since no global Cauchy hypersurface can be defined at any point. In addition, pp and plane wave spacetimes never contain black holes or equivalently event horizons \cite{HubenyRangamani02c}, although the opposite is always true as we saw above, i.e.\ singular spacetimes always have a (singularity-free) plane-wave limit. The upshot is that pp-wave backgrounds are excellent probes to the properties of curved spaces without being much more complicated than flat spaces.\footnote{For example, the quantization of superstrings on AdS$_5\times\text{S}^5$, in the context of the AdS/CFT correspondence \cite{Maldacena97}, is very complicated. Solving and quantizing superstrings on the 10-dimensional maximally supersymmetric plane-wave background, that is the Penrose-G\"{u}ven limit of AdS$_5\times\text{S}^5$, is much simpler \cite{Metsaev01b, MetsaevTseytlin02}. A similar limiting procedure can also be carried out on the CFT side of the correspondence, leading to the so called BMN sector of $\mathcal{N} = 4$ super Yang-Mills theory \cite{BMN02}. The AdS$_5$/CFT$_4$ correspondence has been exhaustively studied in the BMN limit (see e.g.\ \cite{Pankiewicz03, Plefka03, Sadri-SheikhJabbari04, RussoTanzini04} for reviews).} \\[6pt]
\indent Another important property of spacetimes \eqref{MetricGeneral} that admit a covariantly constant null Killing vector, is that all their scalar invariants (constructed from the Riemann tensor and its covariant derivatives) vanish. These are generally known as vanishing scalar invariant (VSI) spacetimes. In the form \eqref{MetricBrinkmann2} pp-waves are also $\alpha'$-exact solutions of supergravity/string theory \cite{AmatiKlimcik88a, HorowitzSteif90}, i.e.\ they correspond to exactly conformal field theories. \\[6pt]
\indent As it turns out, the Penrose-G\"{u}ven limit preserves the supersymmetries of the original space so that the maximally supersymmetric backgrounds AdS$_{4/5/7}\times\text{S}^{7/5/4}$ of type IIB and 11-dimensional supergravity (32 supercharges) give rise to two maximally supersymmetric homogeneous plane-wave solutions of the form \eqref{MetricPlaneWavesHomogeneous} in 10 and 11 dimensions \cite{OFarrillPapadopoulos02b}.\footnote{The latter have been dubbed Hpp-waves, that is Cahen-Wallach (CW) plane-waves with homogeneous fluxes. Along with 10 and 11-dimensional flat space, these are the 7 maximally supersymmetric backgrounds in 10 and 11 dimensions.}$^,$\footnote{See also \cite{FloratosKehagias02} for a study of Penrose limits for various AdS$_a\times\text{S}^b$ orbifolds.} In 11 dimensions the maximally supersymmetric homogeneous plane-wave background is part of the Kowalski-Glikman (KG) solution \cite{KowalskiGlikman84a}:
\begin{IEEEeqnarray}{l}
ds^2 = -2 dx^{+} dx^{-} - \left[\frac{\mu^2}{9}\sum_{i=1}^3 x_i x_i + \frac{\mu^2}{36}\sum_{j=1}^6 y_j y_j\right] dx^+ dx^+ + \sum_{i=1}^3 dx_i dx_i + \sum_{j=1}^6 dy_j dy_j \label{MaximallySupersymmetricBackground1} \\[6pt]
F_{123+} = \mu. \label{MaximallySupersymmetricBackground2}
\end{IEEEeqnarray}
\paragraph{Matrix models} Yet another occurrence of the plane-wave geometry \eqref{MaximallySupersymmetricBackground1}--\eqref{MaximallySupersymmetricBackground2} arises in connection with the matrix model of Berenstein, Maldacena and Nastase (BMN) \cite{BMN02}:
\begin{IEEEeqnarray}{c}
H = H_0 + \frac{R}{2}\cdot\text{Tr}\left[\sum_{i = 1}^3\frac{m^2}{9} \, \textbf{X}_i^2 + \sum_{j = 4}^9\frac{m^2}{36} \, \textbf{X}_j^2 + \sum_{i,j,k = 1}^3 \frac{2m}{3} \, i\epsilon_{ijk}\textbf{X}_i\textbf{X}_j\textbf{X}_k - \frac{m}{2} i \Psi^T\gamma_{123}\Psi\right], \qquad \quad \label{BMN_MatrixModel}
\end{IEEEeqnarray}
which constitutes a deformation of the Banks-Fischler-Shenker-Susskind (BFSS) matrix theory \cite{BFSS97},
\begin{IEEEeqnarray}{c}
H_0 = \frac{R}{2}\cdot\text{Tr}\left[\dot{\textbf{X}}^2 - \frac{1}{2}\left[\textbf{X}_A,\textbf{X}_B\right]^2 - \Psi^T\gamma_A[\textbf{X}_A,\Psi]\right], \quad A, B = 1,\ldots,9, \qquad \label{BFSS_MatrixModel}
\end{IEEEeqnarray}
by mass terms and a Myers term \cite{Myers99b}.\footnote{In \eqref{BMN_MatrixModel}--\eqref{BFSS_MatrixModel}, each component of the 9d vector $\textbf{X}_A$ and the 16d Majorana spinor $\Psi$ is a $N \times N$ Hermitian matrix. $\gamma_A$ are the 9d ($16\times16$) Euclidean Dirac matrices, $R$ is the DLCQ compactification radius and $m \equiv \mu/R$.} The BMN matrix model \eqref{BMN_MatrixModel} describes the discrete light-cone quantization (DLCQ) of M-theory on the 11-dimensional maximally supersymmetric plane-wave background \eqref{MaximallySupersymmetricBackground1}--\eqref{MaximallySupersymmetricBackground2}. As shown in \cite{DasguptaJabbariRaamsdonk02}, the Hamiltonian \eqref{BMN_MatrixModel} can be derived by regularizing the light-cone supermembrane in the same background. Equivalently, the light-cone supermembrane on the maximally supersymmetric plane-wave background can be seen as the continuum ($N\rightarrow\infty$) limit of the BMN matrix model. Similar results have been known for BFSS theory \cite{deWitHoppeNicolai88}. \\[6pt]
\indent The mass terms of the BMN matrix model \eqref{BMN_MatrixModel} lift the flat directions of BFSS matrix theory \eqref{BFSS_MatrixModel} making the supermembrane spectrum discrete. On the other hand, the Myers term allows for static fuzzy sphere solutions:
\begin{IEEEeqnarray}{c}
\textbf{X}_i = r \cdot J_i, \quad i = 1,2,3 \qquad \& \qquad \textbf{X}_j = 0, \quad j =4,\ldots,9, \label{FuzzySphere1}
\end{IEEEeqnarray}
where the matrices $J_i$ furnish a $N$-dimensional representation of $\mathfrak{su}\left(2\right)$. The radii,
\begin{IEEEeqnarray}{c}
r = 0, \qquad r = \frac{\mu}{3}, \qquad r = \frac{\mu}{6}, \label{FuzzySphere2}
\end{IEEEeqnarray}
correspond to the maximally supersymmetric vacuum $\textbf{X}_i = 0$, the $1/2$-BPS solution and an unstable, non-supersymmetric, positive-energy solution respectively. BPS configurations of the BMN matrix model have been studied in \cite{Bak02a, Mikhailov02b, Park02b, BakKimLee05, HoppeLee07}. Matrix and membrane ($N \rightarrow \infty$) solutions in pp-wave backgrounds can be found in \cite{ArnlindHoppe03b, ArnlindHoppeTheisen04, BerensteinDzienkowskiLashof-Regas15, Hoppe15}.
\paragraph{Black holes} The study of chaotic phenomena that take place in the vicinity of black holes (BHs) has become relatively popular in recent years, mainly because these phenomena are related to the paradox of information loss \cite{ShenkerStanford13a, Hawking14, MaldacenaShenkerStanford15}. The observations that are made by infalling observers (fifos), get scrambled by the microscopic degrees of freedom in the near-horizon region \cite{SusskindLindesay05, PapadodimasRaju12} and reach fiducial observers (fidos) in the form of chaotically processed information. Meanwhile the outgoing Hawking radiation (soft+hard) has its own random correlation to the information that is apparently lost \cite{CarneyChauretteNeuenfeldSemenoff17, Strominger17}. \\[6pt]
\indent Sekino and Susskind \cite{SekinoSusskind08} conjectured that black holes are the fastest information scramblers in nature, provided that the dynamics of their microscopic degrees of freedom on their horizons is both manifestly chaotic and nonlocal. The BFSS matrix model \eqref{BFSS_MatrixModel} encapsulates both of these properties, superseding any conceivable local field theory description, as well as the good old phenomenological BH membrane paradigm \cite{Damour78, ThornePriceMacdonald86}. Strong evidence has accumulated in support of the claim that the quantized (BMN) matrix theory \eqref{BMN_MatrixModel}--\eqref{BFSS_MatrixModel} (reducing to the BFSS matrix model in the $m \rightarrow 0$ limit) provides a valid description of the chaotic and non-local dynamics of the microscopic degrees of freedom that are present on the horizons of BHs. This implies that superfast propagation and mixing of in-falling information ("fast scrambling") \cite{HaydenPreskill07} are emergent features of quantum matrix models. The stable fuzzy sphere solutions of the BMN matrix model in \eqref{FuzzySphere1}--\eqref{FuzzySphere2} should then be capable of describing a wide range of turbulent phenomena on the horizons of BHs. As such they should be useful in the study of fluctuations on BH horizons \cite{Gur-AriHanadaShenker15, AsanoKawaiYoshida15}, in relation to the scrambling hypothesis. \\[6pt]
\indent In this paper, we continue a program that was started in \cite{AxenidesFloratos07}\,\footnote{See also the thesis \cite{Linardopoulos15b}.} and consisted in finding explicit non-perturbative soliton solutions of Mp-branes in flat as well as in curved spacetimes. In \cite{AxenidesFloratosLinardopoulos13a} the focus was on integrability and the stringy properties of M2-branes, while in \cite{AxenidesFloratosLinardopoulos15a} a presumably brany limit of spiky strings on S$^2$ was found.\footnote{These systems might have a useful role to play in the quantization of M2-branes in AdS backgrounds. See \cite{Tseytlin23b, GiombiKurlyandTseytlin24}.} Here we go one step further by addressing one of the open problems that were posed in \cite{AxenidesFloratos07}, namely that of writing down analytic classical M2-brane solutions that spin inside the 11-dimensional maximally supersymmetric plane-wave background \eqref{MaximallySupersymmetricBackground1}--\eqref{MaximallySupersymmetricBackground2}. Our primary goal is to provide a complete classification of spinning spherical dielectric membrane configurations in $SO(3) \times SO(6)$ symmetric backgrounds such as \eqref{MaximallySupersymmetricBackground1}--\eqref{MaximallySupersymmetricBackground2} (see also the talk \cite{AxenidesFloratosKatsinisLinardopoulos20}). \\[6pt]
\indent Secondly, we take on the systematic study of the chaotic properties of the BMN matrix model in its large-$N$ limit \cite{AxenidesFloratosLinardopoulos17a, AxenidesFloratosLinardopoulos17b}, which is known to be described by a theory of supermembranes \cite{DasguptaJabbariRaamsdonk02}. This is accomplished by means of a detailed stability analysis for the aforementioned spherical dielectric membrane configurations at the leading and the next-to-leading order of perturbation theory. At leading order, we examine both radial and angular/multipole perturbations in the full $SO(3)\times SO(6)$ geometric background. At the next-to-leading order we analyze the angular/multipole spectrum in the $SO(3)$ counterpart of $SO(3)\times SO(6)$. We uncover the familiar from hydrodynamics phenomenon of turbulent instability cascade by which dipole (spin $j=1$) and quadrupole ($j=2$) instabilities propagate from leading order ($n=1$) towards all higher multipoles ($j = 1,2,\ldots$) of higher-order perturbation theory ($n = 2,3,\ldots$). Long-wavelength (small $j$) perturbations couple nonlinearly to small-wavelength (large $j$) ones, inducing instabilities to all higher perturbative orders. This phenomenon suggests that turbulence might be a dynamical property of the microscopic degrees of freedom on black hole horizons, rendering them the fastest scramblers of information in nature. We should notice however that we do not discuss the role of classical multimembrane configurations that are contained in the matrix model \cite{Maldacena24talk}, as well as their possibly important role in the chaotic dynamics of black hole horizons. Also, we do not discuss quantum effects for the propagation of chaos among the horizon degrees of freedom.\footnote{The authors would like to thank an anonymous referee for his/her valuable comments regarding the significance of multimembrane and quantum instabilities of the BMN matrix model to the dynamics of information processing which takes place on the horizons of BHs.} These important issues will hopefully be dealt with in future work. \\[6pt]
\indent Here is the outline of our paper. In section \ref{Section:DielectricMembranes} below we introduce our ansatz for membranes in the maximally supersymmetric plane wave background \eqref{MaximallySupersymmetricBackground1}--\eqref{MaximallySupersymmetricBackground2}. The ansatz describes (bosonic, spherically symmetric) configurations which fall into three main classes, types I, II and III. Of these, we will be interested in two representative cases of type III solutions, namely the static dielectric membrane in $SO(3)$ and the axially symmetric membrane in $SO(3)\times SO(6)$. In section \ref{Section:LOperturbations} we perturb these two type III solutions to leading order in perturbation theory and analyze the resulting radial and angular/multipole spectra. In section \ref{Section:NLOperturbations} we carry out angular/multipole perturbations for the $SO(3)$ symmetric membrane at the next-to-leading order. Our conclusions can be found in section \ref{Section:ConclusionsDiscussion}.
\section[Dielectric membranes]{Dielectric membranes \label{Section:DielectricMembranes}}
\noindent The starting point of our analysis is the Hamiltonian of a bosonic relativistic membrane in the 11-dimensional maximally supersymmetric plane-wave background \eqref{MaximallySupersymmetricBackground1}--\eqref{MaximallySupersymmetricBackground2} which describes the continuum (or large $N$) limit of the BMN matrix model \eqref{BMN_MatrixModel}--\eqref{BFSS_MatrixModel} \cite{BMN02}. The Hamiltonian takes the following form in the so-called light-cone gauge $x^+ = \tau$ \cite{DasguptaJabbariRaamsdonk02}:\footnote{The indices of the coordinates $x_i$ are implicitly taken to run from 1 to 3, while those of the coordinates $y_j$ run from 1 to 6. Also there's no distinction between upper/lower indices, so these are used interchangeably throughout the text.}
\begin{IEEEeqnarray}{ll}
H = \frac{T}{2}\int d^2\sigma\left[\pi_i^2 + \frac{1}{2}\left\{x_i,x_j\right\}^2 + \frac{1}{2}\left\{y_i,y_j\right\}^2 + \left\{x_i,y_j\right\}^2 + \frac{\mu^2 x^2}{9} + \frac{\mu^2 y^2}{36} - \frac{\mu}{3}\,\epsilon_{ijk}\left\{x_i,x_j\right\}x_k\right], \qquad \label{ppWaveHamiltonian1}
\end{IEEEeqnarray}
which can also be expressed as a sum of squares:
\begin{IEEEeqnarray}{ll}
H = \frac{T}{2}\int d^2\sigma\left[\pi^2 + \Big(\frac{\mu}{3}\,x_i - \frac{1}{2}\epsilon_{ijk}\left\{x_j,x_k\right\}\Big)^2 + \frac{1}{2}\left\{y_i,y_j\right\}^2 + \frac{\mu^2}{36}\,y_j y_j + \left\{x_i,y_j\right\}^2\right]. \label{ppWaveHamiltonian2}
\end{IEEEeqnarray}
In \eqref{ppWaveHamiltonian1}--\eqref{ppWaveHamiltonian2}, $T$ is just the membrane tension, while we have also defined,
\begin{IEEEeqnarray}{ll}
\pi_i^2 \equiv \sum_{i=1}^3 \dot{x}_i \dot{x}_i + \sum_{j=1}^6 \dot{y}_j \dot{y}_j, \qquad x^2 \equiv \sum_{i=1}^3 x_i x_i, \qquad y^2 \equiv \sum_{j=1}^6 y_j y_j.
\end{IEEEeqnarray}
Let us also spell out the definition of the Poisson bracket $\{\;,\;\}$ (see e.g.\ \cite{NicolaiHelling98}),
\begin{equation}
\left\{f\,,\,g\right\} \equiv \frac{\epsilon_{rs}}{\sqrt{w\left(\boldsymbol\sigma\right)}} \, \partial_r f \; \partial_s g = \frac{1}{\sqrt{w\left(\boldsymbol\sigma\right)}} \, \left(\partial_1 f \; \partial_2 g - \partial_2 f \; \partial_1 g\right), \label{PoissonBracket}
\end{equation}
where $d^2\sigma = \sqrt{w\left(\boldsymbol\sigma\right)} \; d\sigma_1 \, d\sigma_2$ corresponds to the spatial worldvolume and $\epsilon_{rs}$ is the Levi-Civita symbol in 2 dimensions. In a flat worldvolume, $w\left(\boldsymbol\sigma\right) = 1$ and the common definition of Poisson brackets is recovered. \\[6pt]
\indent Here are the equations of motion for the target space coordinates $x$ and $y$ which follow from the Hamiltonian \eqref{ppWaveHamiltonian1}--\eqref{ppWaveHamiltonian2}:
\begin{IEEEeqnarray}{l}
\ddot{x}_i = \left\{\left\{x_i,x_j\right\},x_j\right\} + \left\{\left\{x_i,y_j\right\},y_j\right\} - \frac{\mu^2}{9}\,x_i + \frac{\mu}{2}\epsilon_{ijk}\left\{x_j,x_k\right\} \label{xEquation} \\[6pt]
\ddot{y}_i = \left\{\left\{y_i,y_j\right\},y_j\right\} + \left\{\left\{y_i,x_j\right\},x_j\right\} - \frac{\mu^2}{36}\,y_i. \label{yEquation}
\end{IEEEeqnarray}
The spatial $SO(3)\times SO(6)$ coordinates $x$ and $y$ should satisfy the Gauss law constraint:
\begin{IEEEeqnarray}{l}
\sum_{i = 1}^3\left\{\dot{x}_i, x_i\right\} + \sum_{j = 1}^6\left\{\dot{y}_j, y_j\right\} = 0. \label{GaussLaw1}
\end{IEEEeqnarray}
\paragraph{The spherical ansatz} Following the seminal work \cite{CollinsTucker76}, we propose an ansatz for the spatial coordinates of the membrane $x$ and $y$. We dubbed the ansatz "spherical" for obvious reasons:
\begin{IEEEeqnarray}{lll}
\textbf{x}_1: \ x_i \equiv x_{1i} = \tilde{x}_{1i}\left(\tau\right)e_1\left(\sigma\right), \quad & i = 1,\ldots,q_1 \label{SphericalAnsatz1x} \\[6pt]
\textbf{x}_2: \ x_{q_1+j} \equiv x_{2j} = \tilde{x}_{2j}\left(\tau\right)e_2\left(\sigma\right), \quad & j = 1,\ldots,q_2 \qquad \& \quad & q_1 + q_2 + q_3 = 3 \label{SphericalAnsatz2x} \\[6pt]
\textbf{x}_3: \ x_{q_1+q_2+k} \equiv x_{3k} = \tilde{x}_{3k}\left(\tau\right)e_3\left(\sigma\right), \quad & k = 1,\ldots,q_3 \label{SphericalAnsatz3x}
\end{IEEEeqnarray}
and
\begin{IEEEeqnarray}{lll}
\textbf{y}_1: \ y_i \equiv y_{1i} = \tilde{y}_{1i}\left(\tau\right)e_1\left(\sigma\right), \quad & i = 1,\ldots,s_1 \label{SphericalAnsatz1y} \\[6pt]
\textbf{y}_2: \ y_{s_1+j} \equiv y_{2j} = \tilde{y}_{2j}\left(\tau\right)e_2\left(\sigma\right), \quad & j = 1,\ldots,s_2 \qquad \& \quad & s_1 + s_2 + s_3 = 6 \label{SphericalAnsatz2y} \\[6pt]
\textbf{y}_3: \ y_{s_1+s_2+k} \equiv y_{3k} = \tilde{y}_{3k}\left(\tau\right)e_3\left(\sigma\right), \quad & k = 1,\ldots,s_3, \label{SphericalAnsatz3y}
\end{IEEEeqnarray}
where, denoting the spherical coordinates by $\left(\sigma_1, \sigma_2\right) \rightarrow \left(\theta, \phi\right)$, we have defined:
\begin{IEEEeqnarray}{c}
(e_1, e_2, e_3) = (\cos\phi \sin\theta, \sin\phi \sin\theta, \cos\theta), \qquad \{e_i, e_j\} = \epsilon_{ijk} \, e_k, \qquad \int e_i \, e_j\,d^2\sigma = \frac{4\pi}{3} \, \delta_{ij}, \qquad \label{EpsilonDefinition1}
\end{IEEEeqnarray}
for $\phi \in [0,2\pi)$, $\theta \in [0,\pi]$. The $e_i$'s satisfy the $\mathfrak{so}\left(3\right)$ Poisson algebra and are orthonormal.\footnote{Note that, in spherical coordinates $(\theta,\phi)$, the proper density to be used in the definition \eqref{PoissonBracket} of Poisson brackets is $\sqrt{w\left(\boldsymbol\sigma\right)} = \sin\theta$. Had we, for example, chosen the parametrization
\begin{IEEEeqnarray}{c}
(e_1, e_2, e_3) = (cn\left(\phi|m\right) sn\left(\theta|n\right), sn\left(\phi|m\right)sn\left(\theta|n\right), sn\left(\theta|n\right)), \qquad \{e_i, e_j\} = \epsilon_{ijk} \, e_k, \qquad \int e_i \, e_j\,d^2\sigma = \frac{4\pi}{3} \, \delta_{ij}, \qquad \label{EpsilonDefinition2}
\end{IEEEeqnarray}
for $\phi \in [0,4\mathbb{K}\left(m\right))$, $\theta \in [0,2\mathbb{K}\left(n\right)]$, the corresponding density would have been $\sqrt{w\left(\boldsymbol\sigma\right)} = sn\left(\theta|n\right) dn\left(\theta|n\right)dn\left(\phi|m\right)$.} The ansatz \eqref{SphericalAnsatz1x}--\eqref{SphericalAnsatz3y} essentially splits each one of the two sets of coordinates $x$ and $y$ into three groups:
\begin{IEEEeqnarray}{lllll}
x_{ai} = \tilde{x}_{ai}\left(\tau\right) e_a \qquad \& \qquad y_{bj} = \tilde{y}_{bj}\left(\tau\right) e_b, \qquad i = 1,\ldots,q_a, \quad j = 1,\ldots,s_b, \quad a,b = 1,2,3. \qquad
\end{IEEEeqnarray}
Interestingly, the Gauss law constraint \eqref{GaussLaw1} is immediately satisfied by the above ansatz \eqref{SphericalAnsatz1x}--\eqref{SphericalAnsatz3y}. Just like in the flat-space case (which was worked out in \cite{AxenidesFloratos07}), we put forward the following dielectric top solutions:
\begin{IEEEeqnarray}{lll}
\tilde{\textbf{x}}_1\left(\tau\right) = e^{\Omega_{x1} \tau} \cdot \tilde{\textbf{x}}_{10}, \qquad & \tilde{\textbf{x}}_2\left(\tau\right) = e^{\Omega_{x2} \tau} \cdot \tilde{\textbf{x}}_{20}, \qquad & \tilde{\textbf{x}}_3\left(\tau\right) = e^{\Omega_{x3} \tau} \cdot \tilde{\textbf{x}}_{30} \label{Ansatz1a} \\[6pt]
\tilde{\textbf{y}}_1\left(\tau\right) = e^{\Omega_{y1} \tau} \cdot \tilde{\textbf{y}}_{10}, \qquad & \tilde{\textbf{y}}_2\left(\tau\right) = e^{\Omega_{y2} \tau} \cdot \tilde{\textbf{y}}_{20}, \qquad & \tilde{\textbf{y}}_3\left(\tau\right) = e^{\Omega_{y3} \tau} \cdot \tilde{\textbf{y}}_{30}. \label{Ansatz1b}
\end{IEEEeqnarray}
It is easily verified that the corresponding radii,
\begin{IEEEeqnarray}{lll}
r_{x1}^2 \equiv \tilde{x}_1^2 = \sum_{i=1}^{q_1} \tilde{x}_{10i} \tilde{x}_{10i}, \quad & r_{x2}^2 \equiv \tilde{x}_2^2 = \sum_{j=1}^{q_2} \tilde{x}_{20j} \tilde{x}_{20j}, \quad & r_{x3}^2 \equiv \tilde{x}_3^2 = \sum_{k=1}^{q_3} \tilde{x}_{30k} \tilde{x}_{30k}, \quad \tilde{x}^2 \equiv \sum_{i=1}^{3} r_{xi}^2 \qquad \label{polar1a} \\[6pt]
r_{y1}^2 \equiv \tilde{y}_1^2 = \sum_{i=1}^{s_1} \tilde{y}_{10i} \tilde{y}_{10i}, \quad & r_{y2}^2 \equiv \tilde{y}_2^2 = \sum_{j=1}^{s_2} \tilde{y}_{20j} \tilde{y}_{20j}, \quad & r_{y3}^2 \equiv \tilde{y}_3^2 = \sum_{k=1}^{2_3} \tilde{y}_{30k} \tilde{y}_{30k}, \quad \tilde{y}^2 \equiv \sum_{i=1}^{3} r_{yi}^2, \qquad \label{polar1b}
\end{IEEEeqnarray}
are specified (for all the antisymmetric matrices $\Omega_{x1}$, $\Omega_{x2}$, $\Omega_{x3}$, $\Omega_{y1}$, $\Omega_{y2}$, $\Omega_{y3}$), by means of the conserved angular momenta,
\begin{IEEEeqnarray}{lll}
\left(\ell_{x1}\right)_{ij} \equiv \dot{\tilde{x}}_{1i} \tilde{x}_{1j} - \tilde{x}_{1i} \dot{\tilde{x}}_{1j}, \qquad & \left(\ell_{x2}\right)_{ij} \equiv \dot{\tilde{x}}_{2i} \tilde{x}_{2j} - \tilde{x}_{2i} \dot{\tilde{x}}_{2j}, \qquad & \left(\ell_{x3}\right)_{ij} \equiv \dot{\tilde{x}}_{3i} \tilde{x}_{3j} - \tilde{x}_{3i} \dot{\tilde{x}}_{3j} \qquad \label{polar2a} \\[12pt]
\left(\ell_{y1}\right)_{ij} \equiv \dot{\tilde{y}}_{1i} \tilde{y}_{1j} - \tilde{y}_{1i} \dot{\tilde{y}}_{1j}, \qquad & \left(\ell_{y2}\right)_{ij} \equiv \dot{\tilde{y}}_{2i} \tilde{y}_{2j} - \tilde{y}_{2i} \dot{\tilde{y}}_{2j}, \qquad & \left(\ell_{y3}\right)_{ij} \equiv \dot{\tilde{y}}_{3i} \tilde{y}_{3j} - \tilde{y}_{3i} \dot{\tilde{y}}_{3j}, \qquad \label{polar2b}
\end{IEEEeqnarray}
by minimizing the effective potential of the membrane. We would arrive at the same result, if we had instead plugged the ansatz \eqref{Ansatz1a}--\eqref{Ansatz1b} into the membrane equations of motion \eqref{xEquation}--\eqref{yEquation}. This would lead to a relation between the radii $r_{x1}$, $r_{x2}$, $r_{x3}$, $r_{y1}$, $r_{y2}$, $r_{y3}$ and the matrix elements of $\Omega_{x1}$, $\Omega_{x2}$, $\Omega_{x3}$, $\Omega_{y1}$, $\Omega_{y2}$, $\Omega_{y3}$. In turn, these always combine to form the set of conserved angular momenta $\ell_{x1}$, $\ell_{x2}$, $\ell_{x3}$, $\ell_{y1}$, $\ell_{y2}$, $\ell_{y3}$.
\subsection[Effective potentials]{Effective potentials \label{Subsection:EffectivePotentials}}
\noindent By inserting the spherical ansatz \eqref{SphericalAnsatz1x}--\eqref{SphericalAnsatz3y} into the Hamiltonian \eqref{ppWaveHamiltonian1}, we find the energy of the membrane configurations:
\begin{IEEEeqnarray}{ll}
E = \frac{2\pi T}{3} \Bigg[\dot{\tilde{x}}_1^2 &+ \dot{\tilde{x}}_2^2 + \dot{\tilde{x}}_3^2 + \dot{\tilde{y}}_1^2 + \dot{\tilde{y}}_2^2 + \dot{\tilde{y}}_3^2 + \tilde{x}_1^2 \tilde{x}_2^2 + \tilde{x}_2^2 \tilde{x}_3^2 + \tilde{x}_3^2 \tilde{x}_1^2 + \tilde{y}_1^2 \tilde{y}_2^2 + \tilde{y}_2^2 \tilde{y}_3^2 + \tilde{y}_3^2 \tilde{y}_1^2 + \tilde{x}_1^2 \left(\tilde{y}_2^2 + \tilde{y}_3^2\right) + \nonumber \\
& + \tilde{x}_2^2 \left(\tilde{y}_3^2 + \tilde{y}_1^2\right) + \tilde{x}_3^2 \left(\tilde{y}_1^2 + \tilde{y}_2^2\right) + \frac{\mu^2}{9}\,\tilde{x}^2 + \frac{\mu^2}{36}\,\tilde{y}^2 - 2\mu\,\epsilon_{ijk}\,\tilde{x}_{1i}\tilde{x}_{2j}\tilde{x}_{3k} \Bigg]. \label{EnergySpherical1}
\end{IEEEeqnarray}
Further decomposing the velocities in their tangential and radial/angular parts as,
\begin{IEEEeqnarray}{c}
\dot{\tilde{x}}_1^2 \equiv \dot{\tilde{x}}_{1i}\dot{\tilde{x}}_{1i} = \dot{r}_{x1}^2 + \frac{\ell_{x1}^2}{r_{x1}^2}, \qquad \dot{\tilde{x}}_{2}^2 \equiv \dot{\tilde{x}}_{2i}\dot{\tilde{x}}_{2i} = \dot{r}_{x2}^2 + \frac{\ell_{x2}^2}{r_{x2}^2}, \qquad \dot{\tilde{x}}_{3}^2 \equiv \dot{\tilde{x}}_{3i}\dot{\tilde{x}}_{3i} = \dot{r}_{x3}^2 + \frac{\ell_{x3}^2}{r_{x3}^2} \label{polar3a} \\[6pt]
\dot{\tilde{y}}_{1}^2 \equiv \dot{\tilde{y}}_{1j}\dot{\tilde{y}}_{1j} = \dot{r}_{y1}^2 + \frac{\ell_{y1}^2}{r_{y1}^2}, \qquad \dot{\tilde{y}}_{2}^2 \equiv \dot{\tilde{y}}_{2j}\dot{\tilde{y}}_{2j} = \dot{r}_{y2}^2 + \frac{\ell_{y2}^2}{r_{y2}^2}, \qquad \dot{\tilde{y}}_{3}^2 \equiv \dot{\tilde{y}}_{3j}\dot{\tilde{y}}_{3j} = \dot{r}_{y3}^2 + \frac{\ell_{y3}^2}{r_{y3}^2}, \label{polar3b}
\end{IEEEeqnarray}
and then plugging \eqref{polar1a}--\eqref{polar1b} and \eqref{polar3a}--\eqref{polar3b} into the expression of the membrane energy \eqref{EnergySpherical1}, we find:
\begin{IEEEeqnarray}{ll}
E = \frac{2\pi T}{3} \Bigg[&\dot{r}_{x1}^2 + \dot{r}_{x2}^2 + \dot{r}_{x3}^2 + \dot{r}_{y1}^2 + \dot{r}_{y2}^2 + \dot{r}_{y3}^2 + \frac{\ell_{x1}^2}{r_{x1}^2} + \frac{\ell_{x2}^2}{r_{x2}^2} + \frac{\ell_{x3}^2}{r_{x3}^2} + \frac{\ell_{y1}^2}{r_{y1}^2} + \frac{\ell_{y2}^2}{r_{y2}^2} + \frac{\ell_{y3}^2}{r_{y3}^2} + r_{x1}^2 r_{x2}^2 + \nonumber \\[6pt]
& + r_{x2}^2 r_{x3}^2 + r_{x3}^2 r_{x1}^2 + r_{y1}^2 r_{y2}^2 + r_{y2}^2 r_{y3}^2 + r_{y3}^2 r_{y1}^2 + r_{x1}^2 \left(r_{y2}^2 + r_{y3}^2\right) + r_{x2}^2 \left(r_{y3}^2 + r_{y1}^2\right) + \nonumber \\[6pt]
& + r_{x3}^2 \left(r_{y1}^2 + r_{y2}^2\right) + \frac{\mu^2}{9}\left(r_{x1}^2 + r_{x2}^2 + r_{x3}^2\right) + \frac{\mu^2}{36}\left(r_{y1}^2 + r_{y2}^2 + r_{y3}^2\right) - 2\mu\,\epsilon_{ijk}\tilde{x}_{1i}\tilde{x}_{2j}\tilde{x}_{3k} \Bigg]. \qquad \quad \label{EnergySpherical2}
\end{IEEEeqnarray}
The corresponding effective potential is given by:
\begin{IEEEeqnarray}{ll}
V_{\text{eff}} = \frac{2\pi T}{3} \Bigg[&\frac{\ell_{x1}^2}{r_{x1}^2} + \frac{\ell_{x2}^2}{r_{x2}^2} + \frac{\ell_{x3}^2}{r_{x3}^2} + \frac{\ell_{y1}^2}{r_{y1}^2} + \frac{\ell_{y2}^2}{r_{y2}^2} + \frac{\ell_{y3}^2}{r_{y3}^2} + r_{x1}^2 r_{x2}^2 + r_{x2}^2 r_{x3}^2 + r_{x3}^2 r_{x1}^2 + r_{y1}^2 r_{y2}^2 + r_{y2}^2 r_{y3}^2 + \nonumber \\[6pt]
& + r_{y3}^2 r_{y1}^2 + r_{x1}^2 \left(r_{y2}^2 + r_{y3}^2\right) + r_{x2}^2 \left(r_{y3}^2 + r_{y1}^2\right) + r_{x3}^2 \left(r_{y1}^2 + r_{y2}^2\right) + \frac{\mu^2}{9}\left(r_{x1}^2 + r_{x2}^2 + r_{x3}^2\right) + \nonumber \\[6pt]
& + \frac{\mu^2}{36}\left(r_{y1}^2 + r_{y2}^2 + r_{y3}^2\right) - 2\mu\,\epsilon_{ijk}\tilde{x}_{1i}\tilde{x}_{2j}\tilde{x}_{3k} \Bigg]. \qquad \label{PotentialSpherical1}
\end{IEEEeqnarray}
\indent The effective potential is made up of four basic types of (attractive/repulsive) terms: $\bullet$ (1) kinetic/angular momentum terms (repulsive), $\bullet$ (2) quartic interaction terms (attractive), $\bullet$ (3) mass terms (attractive), and $\bullet$ (4) cubic Myers terms (repulsive). The last two types of terms (i.e.\ mass terms and Myers terms) depend on $\mu$ and are therefore absent from the flat space case ($\mu \rightarrow 0$), which was analyzed in \cite{AxenidesFloratos07}. In either case ($\mu = 0$ or $\mu \neq 0$), it is the balancing between attraction and repulsion which points out where the minima of the effective potential will be. Yet another interesting aspect of these systems is the existence of closed periodic orbits which do not correspond to critical points. In any case however, and because plane-wave backgrounds ($\mu \neq 0$) contain two extra repulsive/attractive terms, the resulting systems are expected to exhibit a much wider variety of dynamical profiles. \\[6pt]
\indent There are basically three ways to combine the three spatial $SO\left(3\right)$ coordinates $x_i$ ($i = 1,2,3$) with the three spherical components $e_i$ in \eqref{EpsilonDefinition1} \cite{AxenidesFloratosKatsinisLinardopoulos20}. This way we obtain our three basic types of membrane configurations (I, II, and III). Types I and II are rotating membranes (tops)\footnote{See appendix \ref{Appendix:ClassicalTopProperty} for more details about the classical top property of our configurations.} that are point-like (or collapsing) in one or two directions in $SO\left(3\right)$. For these configurations the Myers flux term is zero. The third configuration type (III) is the most interesting because it includes all four types of repulsive and attractive terms that we described above and extends to the full geometric background of $SO\left(3\right)\times SO\left(6\right)$. The three types of configurations are introduced below.
\subsubsection[Type I]{Type I configurations}
\noindent Type I configurations have all of their $SO(3)$ coordinates $x_i$ assigned to the $e_1$ spherical component in \eqref{EpsilonDefinition1}, so that $q_1 = 3$ and $q_2 = q_3 = 0$ in \eqref{SphericalAnsatz1x}--\eqref{SphericalAnsatz3x}. This gives
\begin{IEEEeqnarray}{ll}
r_{x} \equiv r_{x1}, \quad r_{x2} = r_{x3} = 0 \quad \& \quad \ell_x \equiv \ell_{x1}, \quad \ell_{x2} = \ell_{x3} = 0,
\end{IEEEeqnarray}
and the Myers flux term becomes zero. The membrane effective potential \eqref{PotentialSpherical1} takes the following form:
\begin{IEEEeqnarray}{ll}
V_{\text{eff}} = \frac{2\pi T}{3} \Bigg[\frac{\ell_{x}^2}{r_{x}^2} &+ \frac{\ell_{y1}^2}{r_{y1}^2} + \frac{\ell_{y2}^2}{r_{y2}^2} + \frac{\ell_{y3}^2}{r_{y3}^2} + r_{y1}^2 r_{y2}^2 + r_{y2}^2 r_{y3}^2 + r_{y3}^2 r_{y1}^2 + r_{x}^2 \left(r_{y2}^2 + r_{y3}^2\right) + \frac{\mu^2r_{x}^2}{9} + \nonumber \\[6pt]
& + \frac{\mu^2}{36}\left(r_{y1}^2 + r_{y2}^2 + r_{y3}^2\right)\Bigg]. \qquad \label{PotentialSpherical2}
\end{IEEEeqnarray}
The effective potential \eqref{PotentialSpherical2} possesses one completely symmetric (single-radius) configuration $r = r_x = r_{y1} = r_{y2} = r_{y3}$, $\ell = \ell_x = \ell_{y1} = \ell_{y2} = \ell_{y3}$. Besides that, there are 5 different axially symmetric (2-radii) configurations, and 4 more configurations with 3 different radii. For each one of these potentials there is a local minimum which corresponds to a stationary top solution with a time-independent radius and non-vanishing total angular momentum.
\subsubsection[Type II]{Type II configurations}
\noindent In type II configurations, two out of the total three $SO(3)$ coordinates $x_i$ point to the direction of $e_1$, while the third $SO(3)$ coordinate $x_3$ points to the direction of $e_3$. Equivalently, $q_1 = 2$, $q_2 = 1$ and $q_3 = 0$ in \eqref{SphericalAnsatz1x}--\eqref{SphericalAnsatz3x}. This leads to
\begin{IEEEeqnarray}{ll}
r_{x3} = 0 \quad \& \quad \quad \ell_{x2} = \ell_{x3} = 0,
\end{IEEEeqnarray}
so that the Myers flux term is again zero, and the effective potential \eqref{PotentialSpherical1} reads:
\begin{IEEEeqnarray}{ll}
V_{\text{eff}} = \frac{2\pi T}{3} \Bigg[\frac{\ell_{x1}^2}{r_{x1}^2} &+ \frac{\ell_{y1}^2}{r_{y1}^2} + \frac{\ell_{y2}^2}{r_{y2}^2} + \frac{\ell_{y3}^2}{r_{y3}^2} + r_{x1}^2 r_{x2}^2 + r_{y1}^2 r_{y2}^2 + r_{y2}^2 r_{y3}^2 + r_{y3}^2 r_{y1}^2 + r_{x1}^2 \left(r_{y2}^2 + r_{y3}^2\right) + \nonumber \\[6pt]
& + r_{x2}^2 \left(r_{y3}^2 + r_{y1}^2\right) + \frac{\mu^2}{9}\left(r_{x1}^2 + r_{x2}^2\right) + \frac{\mu^2}{36}\left(r_{y1}^2 + r_{y2}^2 + r_{y3}^2\right) \Bigg]. \qquad \label{PotentialSpherical3}
\end{IEEEeqnarray}
Type II configurations include one single-radius solution ($r = r_{x1} = r_{x2} = r_{y1} = r_{y2} = r_{y3}$, $\ell = \ell_{x1} = \ell_{y1} = \ell_{y2} = \ell_{y3}$), 13 axially symmetric (2-radius) solutions (tops) and 21 solutions (tops) with 3 different radii.
\paragraph{Example 1} Take for instance a type II configuration where the $SO\left(6\right)$ variables $y_i$ are set to zero:
\begin{IEEEeqnarray}{ll}
x_1 = x\left(\tau\right)\cdot e_1, \quad x_2 = y\left(\tau\right)\cdot e_1, \quad x_3 = z\left(\tau\right)\cdot e_2 \qquad \& \qquad y_i = 0, \quad i = 1,\ldots,6, \label{AnsatzExample1}
\end{IEEEeqnarray}
and time-dependence is of the form \eqref{Ansatz1a}. The effective potential \eqref{PotentialSpherical3} reads,
\begin{IEEEeqnarray}{c}
V_{\text{eff}} = \frac{2\pi T}{3} \Bigg[\frac{\ell^2}{x^2 + y^2} + \left(x^2 + y^2\right) z^2 + \frac{\mu^2}{9}\left(x^2 + y^2 + z^2\right)\Bigg],
\end{IEEEeqnarray}
where we have set $\ell_{x1} = \ell$. The corresponding minimization condition $\nabla V_{\text{eff}} = 0$ leads to
\begin{IEEEeqnarray}{c}
x \, z^2 + \frac{\mu^2 x}{9} - \frac{x \, \ell^2}{\left(x^2+y^2\right)^2} = y \, z^2 + \frac{\mu^2 y}{9} - \frac{y \, \ell^2}{\left(x^2+y^2\right)^2} = z\left(x^2 + y^2\right) + \frac{\mu^2 z}{9} = 0, \qquad
\end{IEEEeqnarray}
which has the following solution
\begin{IEEEeqnarray}{c}
x^2 + y^2 = \frac{3\ell}{\mu} \qquad \& \qquad z = 0. \label{ExampleSolution1}
\end{IEEEeqnarray}
To agree with the form of the ansatz \eqref{Ansatz1a} we can select, for instance,
\begin{IEEEeqnarray}{c}
x\left(\tau\right) = \sqrt{\frac{3\ell}{\mu}}\cos\frac{\mu\,\tau}{3}, \qquad y\left(\tau\right) = \sqrt{\frac{3\ell}{\mu}}\sin\frac{\mu\,\tau}{3}, \qquad z\left(\tau\right) = 0. \label{ExampleSolution2}
\end{IEEEeqnarray}
Alternatively, the ansatz \eqref{AnsatzExample1} could have been directly inserted into the equations of motion \eqref{xEquation}--\eqref{yEquation}:
\begin{IEEEeqnarray}{l}
\ddot{x}\cdot e_1 = -x\, z^2 \cdot e_1 - \frac{\mu^2 x}{9}\cdot e_1 + \mu\,y\,z\cdot e_3 \\[6pt]
\ddot{y}\cdot e_1 = -y\, z^2 \cdot e_1 - \frac{\mu^2 y}{9}\cdot e_1 + \mu\,x\,z\cdot e_3 \\[6pt]
\ddot{z}\cdot e_2 = -z\left(x^2 + y^2\right)\cdot e_2 - \frac{\mu^2 z}{9}\cdot e_2,
\end{IEEEeqnarray}
from which it can be seen that any solution of the type \eqref{Ansatz1a} is bound to satisfy \eqref{ExampleSolution1} as well.
\paragraph{Example 2} Another interesting type II solution is the following:
\begin{IEEEeqnarray}{ll}
x_1 = x\left(\tau\right)\cdot e_1, \quad x_2 = y\left(\tau\right)\cdot e_2, \quad x_3 = 0 \qquad \& \qquad y_i = 0, \quad i = 1,\ldots,6, \label{AnsatzExample2}
\end{IEEEeqnarray}
where again all the $SO\left(6\right)$ variables $y_i$ and the $SO(3)$ coordinate $x_2$ have been set to zero. The effective potential \eqref{PotentialSpherical3} becomes,
\begin{IEEEeqnarray}{c}
V_{\text{eff}} = \frac{2\pi T}{3} \Bigg[x^2 y^2 + \frac{\mu^2}{9}\left(x^2 + y^2\right)\Bigg], \label{YanMillsPotential}
\end{IEEEeqnarray}
so that there is only one trivial critical point at $x = y = 0$, which is obtained by minimizing the effective potential:
\begin{IEEEeqnarray}{c}
x \, y^2 + \frac{\mu^2 x}{9} = y \, x^2 + \frac{\mu^2 y}{9} = 0.
\end{IEEEeqnarray}
On the other hand, potentials of the form \eqref{YanMillsPotential} (which are in fact generalizations of the Yang-Mills potential $x^2 y^2/2$) have a very interesting and rich set of (stable) periodic orbits. A comprehensive study of periodic orbits for potentials of the form \eqref{YanMillsPotential} can be found for example in \cite{ContopoulosHarsoula23, HarsoulaContopoulos24}.
\subsubsection[Type III]{Type III configurations}
\noindent Type III solutions have $q_1 = q_2 = q_3 = 1$, i.e.\ each of the three $SO(3)$ coordinates $x_i$ corresponds to one spherical component $e_i$:
\begin{IEEEeqnarray}{ll}
x_1 = r_{x1} e_1, \quad x_2 = r_{x2} e_2, \quad x_3 = r_{x3} e_3 \quad \& \quad \ell_{x1} = \ell_{x2} = \ell_{x3} = 0, \label{TypeIII}
\end{IEEEeqnarray}
however $r_{x1}$, $r_{x2}$, $r_{x3}$ are not radii anymore, but coordinate components. The effective potential \eqref{PotentialSpherical1} of type III membranes takes the following form:
\begin{IEEEeqnarray}{ll}
V_{\text{eff}} = \frac{2\pi T}{3} \Bigg[\frac{\ell_{y1}^2}{r_{y1}^2} &+ \frac{\ell_{y2}^2}{r_{y2}^2} + \frac{\ell_{y3}^2}{r_{y3}^2} + r_{x1}^2 r_{x2}^2 + r_{x2}^2 r_{x3}^2 + r_{x3}^2 r_{x1}^2 + r_{y1}^2 r_{y2}^2 + r_{y2}^2 r_{y3}^2 + r_{y3}^2 r_{y1}^2 + \nonumber \\[6pt]
& + r_{x1}^2 \left(r_{y2}^2 + r_{y3}^2\right) + r_{x2}^2 \left(r_{y3}^2 + r_{y1}^2\right) + r_{x3}^2 \left(r_{y1}^2 + r_{y2}^2\right) + \frac{\mu^2}{9}\left(r_{x1}^2 + r_{x2}^2 + r_{x3}^2\right) + \nonumber \\[6pt]
& + \frac{\mu^2}{36}\left(r_{y1}^2 + r_{y2}^2 + r_{y3}^2\right) - 2\mu r_{x1} r_{x2} r_{x3} \Bigg]. \qquad \label{PotentialSpherical4}
\end{IEEEeqnarray}
Once more, the various radii (and the respective angular momenta) can be combined into groups of one, two and three different values. We end up with different top configurations, one of which corresponds to a totally symmetric (single-radius) top, 9 to axially symmetric (2-radius) tops and 10 to tops with 3 different radii.
\subsection[Type III solutions]{Type III solutions \label{Subsection:ParticularSolutions}}
\noindent From now on we focus exclusively on type III configurations. The effective potential of these configurations contain all sorts of possible attractive and repulsive terms (which we enumerated in section \ref{Subsection:EffectivePotentials} above), including the repulsive cubic (or Myers) terms. So these solutions enjoy a much broader phase space than the other two types (I and II) and are naturally expected to be more interesting. In \cite{AxenidesFloratosLinardopoulos17a} the following $SO\left(3\right) \times SO\left(3\right) \times SO\left(3\right) \subset SO\left(3\right) \times SO\left(6\right)$ invariant ansatz was proposed:\footnote{For spherical membrane topologies, the functions which are appropriate for describing their internal degrees of freedom are the well-known spherical harmonics $Y_{jm}\left(\theta,\phi\right)$ (for $j = 0,1,\ldots$, $\left|m\right| = 0,1,\ldots j$). Spherical harmonics satisfy the infinite-dimensional Lie algebra SDiff$\left(\text{S}^2\right)$ \cite{Hoppe82}:
\begin{IEEEeqnarray}{c}
\left\{Y_{j_1 m_1},Y_{j_2 m_2}\right\} = f_{j_1 m_1, j_2 m_2}^{j_3 m_3} Y_{j_3 m_3}.
\end{IEEEeqnarray}
They are homogeneous and harmonic polynomials of the spherical coordinate system $\left\{e_i\right\}$ in \eqref{EpsilonDefinition1}. The spatial sets of coordinates $x$ ($SO(3)$) and $y$ ($SO(6)$) can be expanded in spherical harmonics as
\begin{IEEEeqnarray}{c}
x_i = \sum_{j,m} x_{i}^{jm}\left(\tau\right) Y_{jm}\left(\theta,\phi\right), \qquad y_i = \sum_{j,m} y_{i}^{jm}\left(\tau\right) Y_{jm}\left(\theta,\phi\right),
\end{IEEEeqnarray}
which leads to an infinite dimensional system of coupled second order ODEs for the set of unknown mode functions $x_{i}^{jm}\left(\tau\right)$ and $y_{i}^{jm}\left(\tau\right)$. The mode functions and their time derivatives should satisfy the Gauss-law constraint \eqref{GaussLaw1} at all times. Then it follows that the only finite subalgebra of SDiff$\left(\text{S}^2\right)$ which can be used to reduce the above infinite system of equations to a finite system is $SO\left(3\right)$ \cite{Banyaga78}.}
\begin{IEEEeqnarray}{ll}
x_i = \tilde{u}_i\left(\tau\right) e_i, \qquad & y_j = \tilde{v}_j\left(\tau\right) e_j, \qquad y_{j+3} = \tilde{w}_j\left(\tau\right) e_j, \qquad i,j = 1,2,3. \qquad \quad \label{Ansatz2}
\end{IEEEeqnarray}
This ansatz automatically satisfies the Gauss-law constraint \eqref{GaussLaw1}. The reduced system for $(\tilde{u}_i, \tilde{v}_i, \tilde{w}_i)$ will turn out to be a very diverse dynamical system with stable and unstable solutions which correspond to spinning and bouncing M2-branes of spherical topologies. The ansatz \eqref{Ansatz2} corresponds to the following Hamiltonian:
\begin{IEEEeqnarray}{l}
H = \frac{2\pi T}{3}\left(\tilde{p}_u^2 + \tilde{p}_v^2 + \tilde{p}_w^2\right) + U, \label{Hamiltonian}
\end{IEEEeqnarray}
which is obtained by integrating out the worldvolume variables $\theta$ and $\phi$. The expression for the potential energy $U$ reads:
\begin{IEEEeqnarray}{ll}
U = \frac{2\pi T}{3}\bigg[&\tilde{u}_1^2 \tilde{u}_2^2 + \tilde{u}_2^2 \tilde{u}_3^2 + \tilde{u}_3^2 \tilde{u}_1^2 + \tilde{r}_1^2 \tilde{r}_2^2 + \tilde{r}_2^2 \tilde{r}_3^2 + \tilde{r}_3^2 \tilde{r}_1^2 + \tilde{u}_1^2 \left(\tilde{r}_2^2 + \tilde{r}_3^2\right) + \tilde{u}_2^2 \left(\tilde{r}_3^2 + \tilde{r}_1^2\right) + \tilde{u}_3^2 \left(\tilde{r}_1^2 + \tilde{r}_2^2\right) + \nonumber \\
& \frac{\mu^2}{9}\left(\tilde{u}_1^2 + \tilde{u}_2^2 + \tilde{u}_3^2\right) + \frac{\mu^2}{36}\left(\tilde{r}_1^2 + \tilde{r}_2^2 + \tilde{r}_3^2\right) - 2\mu \tilde{u}_1 \tilde{u}_2 \tilde{u}_3 \bigg], \quad \tilde{r}_j^2 \equiv \tilde{v}_j^2 + \tilde{w}_j^2, \ j = 1,2,3.
\end{IEEEeqnarray}
\indent Because the Hamiltonian \eqref{Hamiltonian} has a manifest $SO(2)\times SO(2)\times SO(2)$ symmetry in the $SO(6)$ variables $\tilde{v}_i$ and $\tilde{w}_i$, any solution is required to preserve three $SO(2)$ angular momenta $\ell_i$ ($i = 1,2,3$). The kinetic terms of \eqref{Hamiltonian} take the following forms:
\begin{IEEEeqnarray}{l}
\tilde{p}_v^2 + \tilde{p}_w^2 = \sum_{i=1}^3\left(\dot{\tilde{r}}_i^2 + \frac{\ell_i^2}{\tilde{r}_i^2}\right),
\end{IEEEeqnarray}
which leads to the effective potential
\begin{IEEEeqnarray}{ll}
V_{\text{eff}} = U + \frac{2\pi T}{3}\left(\frac{\ell_1^2}{\tilde{r}_1^2} + \frac{\ell_2^2}{\tilde{r}_2^2} + \frac{\ell_3^2}{\tilde{r}_3^2}\right). \label{EffectivePotential}
\end{IEEEeqnarray}
\indent To date, a wide range of solutions of the BMN matrix model \eqref{BMN_MatrixModel} is known, for both finite values of the matrix size $N$, as well as in the classical ($N\rightarrow \infty$) limit. The latter is known to give rise to a theory of membranes in the plane-wave background \eqref{MaximallySupersymmetricBackground1}--\eqref{MaximallySupersymmetricBackground2}. In \cite{Bak02a, Mikhailov02b, Park02b, BakKimLee05, HoppeLee07} various BPS solutions of varying topologies were found, while spinning non-BPS solutions were spelled out in \cite{ArnlindHoppe03b, ArnlindHoppeTheisen04, BerensteinDzienkowskiLashof-Regas15, Hoppe15}. In what follows we will follow closely \cite{AxenidesFloratosLinardopoulos17a}, where a number of interesting pulsating and (from the extrema of the effective potential \eqref{EffectivePotential}) spinning membrane solutions were identified.
\subsubsection[Static dielectric membranes in $SO\left(3\right)$]{Static dielectric membranes in $SO\left(3\right)$ \label{SubSection:StaticDielectricMembrane}}
\begin{figure}[H]
\begin{center}
\includegraphics[scale=0.4]{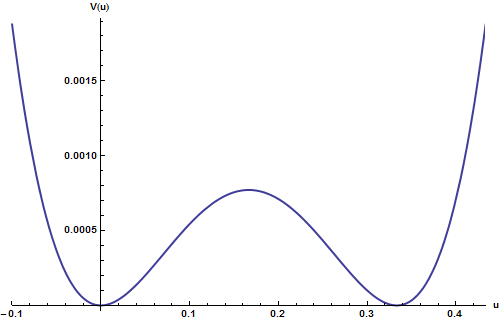} \qquad \includegraphics[scale=0.4]{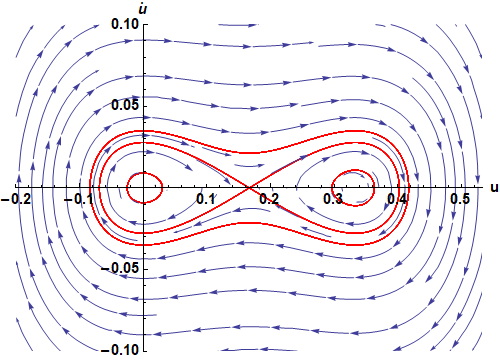}
\caption{Potential (left) and phase portrait (right) of the spherically symmetric membrane.} \label{Graph:PotentialPhasePortrait}
\end{center}
\end{figure}

\noindent We begin by revisiting static dielectric membranes in $SO\left(3\right)$, an example which has been studied extensively in the literature, as we have already mentioned in the introduction. These membranes are obtained by setting all the $SO\left(6\right)$ coordinates of type III configurations equal to zero. For simplicity, we switch to dimensionless time $t \equiv \mu\tau$ and adopt the following notation:
\begin{IEEEeqnarray}{ll}
x_i = \mu u_i e_i, \quad i = 1,2,3 \qquad \& \qquad y_i = \mu v_i = 0, \quad i = 1,\ldots 6. \label{StaticMembrane1}
\end{IEEEeqnarray}
The equations of motion \eqref{xEquation}--\eqref{yEquation} then become:
\begin{IEEEeqnarray}{llc}
\ddot{u}_1 + \left(u_2^2 + u_3^2 +\frac{1}{9}\right) u_1 =& u_2 u_3& \qquad \& \qquad \ddot{v}_i = 0, \quad i=1,\ldots, 6 \label{StaticMembrane2a} \\[6pt]
\ddot{u}_2 + \left(u_1^2 + u_3^2 +\frac{1}{9}\right) u_2 =& u_1 u_3& \label{StaticMembrane2b} \\[6pt]
\ddot{u}_3 + \left(u_1^2 + u_2^2 +\frac{1}{9}\right) u_3 =& u_1 u_2.& \label{StaticMembrane2c}
\end{IEEEeqnarray}
The corresponding dynamics is fully specified in terms of the Hamiltonian
\begin{IEEEeqnarray}{c}
H = \frac{4\pi T\mu^4}{3} \cdot \mathcal{H}, \quad \mathcal{H} \equiv \frac{1}{2}\Bigg[p_1^2 + p_2^2 + p_3^2 + u_1^2 u_2^2 + u_2^2 u_3^2 + u_1^2 u_3^2 + \frac{1}{9}\left(u_1^2 + u_2^2 + u_3^2\right) - 2 u_1 u_2 u_3\Bigg], \qquad \label{StaticHamiltonian}
\end{IEEEeqnarray}
and the Hamilton equations of motion:
\begin{IEEEeqnarray}{cl}
p_i = \dot{u}_{i}, \qquad \dot{p}_i = - \frac{\partial{\mathcal{H}}}{\partial{u}_i}, \label{HamiltonEquations}
\end{IEEEeqnarray}
which evidently imply the Lagrangian equations of motion \eqref{StaticMembrane2a}--\eqref{StaticMembrane2c}. The effective potential energy of the static membrane \eqref{StaticMembrane1} is given by
\begin{IEEEeqnarray}{c}
V_{\text{eff}} = \frac{2\pi T \mu^4}{3}\Bigg[\left(u_1^2 u_2^2 + u_2^2 u_3^2 + u_1^2 u_3^2\right) + \frac{1}{9}\left(u_1^2 + u_2^2 + u_3^2\right) - 2 u_1 u_2 u_3\Bigg]. \label{StaticPotential}
\end{IEEEeqnarray}
This potential turns out to be a special case of the so-called generalized 3-dimensional H\'{e}non-Heiles potential, which was introduced in \cite{EfstathiouSadovskii04}:
\begin{IEEEeqnarray}{c}
V_{\text{HH}} = \frac{1}{2}\left(u_1^2 + u_2^2 + u_3^2\right) + K_3 \, u_1 u_2 u_3 + K_0\left(u_1^2 + u_2^2 + u_3^2\right)^2 + K_4\left(u_1^4 + u_2^4 + u_3^4\right). \label{HenonHeiles3d}
\end{IEEEeqnarray}
For $K_3 = -9$, $K_0 = -K_4 = 9/4$, \eqref{HenonHeiles3d} obviously reduces to the effective potential \eqref{StaticPotential}.
\paragraph{Extrema} The extrema of \eqref{StaticPotential} solve the equilibrium conditions:
\begin{IEEEeqnarray}{ll}
\partial_i V_{\text{eff}} = 0 \Rightarrow & \left(u_2^2 + u_3^2 + \frac{1}{9}\right)u_1 = u_2 u_3 \\[6pt]
& \left(u_3^2 + u_1^2 + \frac{1}{9}\right)u_2 = u_3 u_1 \\[6pt]
& \left(u_1^2 + u_2^2 + \frac{1}{9}\right)u_3 = u_1 u_2.
\end{IEEEeqnarray}
Here are the corresponding roots:
\begin{IEEEeqnarray}{c}
\textbf{u}_0 = 0, \qquad \textbf{u}_{1/6} = \frac{1}{6}\cdot\left(\pm 1,\pm 1,\pm 1\right), \qquad \textbf{u}_{1/3} = \frac{1}{3}\cdot\left(\pm 1,\pm 1,\pm 1\right), \label{So3Extrema1}
\end{IEEEeqnarray}
which are nine in total because the product of their components must be non-negative. The effective potential \eqref{StaticPotential} shares the symmetry of a tetrahedron $T_d$ which is generated by the 4 extremal points $\textbf{u}_{1/3}$ and $\textbf{u}_{1/6}$. Two minima are degenerate, namely $\textbf{u}_0$ (a point-like membrane) and $\textbf{u}_{1/3}$ (the Myers dielectric sphere), while $\textbf{u}_{1/6}$ is a saddle point. The value of the effective potential at the extremal points is
\begin{IEEEeqnarray}{c}
V_{\text{eff}}\left(0\right) = V_{\text{eff}}\left(\frac{1}{3}\right) = 0, \qquad V_{\text{eff}}\left(\frac{1}{6}\right) = \frac{2\pi T \mu^{4}}{6^4}. \label{So3Extrema2}
\end{IEEEeqnarray}
It is easy to show that the corresponding Hessian matrix is positive-definite for $\textbf{u}_0$ and $\textbf{u}_{1/3}$ and indefinite for $\textbf{u}_{1/6}$. Therefore the former are minima of the potential (and in fact global minima), while the latter is a saddle point. This result will be confirmed in section \ref{Section:LOperturbations} below by means of a detailed analysis of radial (section \ref{SubSubSection:RadialPerturbations1}) and angular/mutlipole perturbations (section \ref{SubSubSection:AngularPerturbations1}), to leading order (LO) in perturbation theory. Next-to-leading order (NLO) perturbations will be studied right after, in section \ref{Section:NLOperturbations}. \\[6pt]
\indent When the $u_i$ in \eqref{StaticMembrane1} are not all equal, the corresponding equations of motion \eqref{StaticMembrane2a}--\eqref{StaticMembrane2c} are so involved that the exact (time-dependent) solutions can only be determined by numerical methods. There exists however a special configuration for which an analytic solution of the system \eqref{StaticMembrane2a}--\eqref{StaticMembrane2c} is possible. This case comes about when all the $SO\left(3\right)$ membrane coordinates $u_i$ are equal.
\paragraph{Spherically symmetric membrane} Specifically, when $u_1 = u_2 = u_3$ in \eqref{StaticMembrane1}, the equations of motion reduce to those corresponding to the double-well potential which is exactly solvable (cf.\ \cite{BrizardWestland17}). The analytic solutions are pulsating membranes of a spherical topology which oscillate around the lobes (either one or both) of the double-well potential (see figure \ref{Graph:PotentialPhasePortrait}).\footnote{The interested reader may find more integrable membrane configurations, as well as more information about classical membrane integrability in \cite{Hoppe90, Hoppe21a, Hoppe21b, Hoppe21c, Hoppe21d, Hoppe22b}.}
\vspace{-0.4cm}\noindent Setting $p = p_1 = p_2 = p_3$ and $u = u_1 = u_2 = u_3$ for the membrane momenta and coordinates, the corresponding Hamiltonian \eqref{StaticHamiltonian} becomes:
\begin{IEEEeqnarray}{c}
H = 2\pi T\mu^4 \left[p^2 + u^2\left(u - \frac{1}{3}\right)^2\right], \qquad \label{So3Hamiltonian}
\end{IEEEeqnarray}
which implies the following Hamilton's equations of motion (using dimensionless time $t \equiv \mu\tau$):
\begin{IEEEeqnarray}{c}
\dot{u} = p, \qquad \dot{p} = -u\left(2u^2 - u + \frac{1}{9}\right). \qquad \label{So3Equations}
\end{IEEEeqnarray}
\indent The phase diagram of the dynamical system \eqref{So3Equations} has been plotted in figure \ref{Graph:PotentialPhasePortrait}. We may distinguish 3 main types of trajectories: $\bullet$ (1) small-energy oscillations ($\mathcal{E} \equiv E/ 2\pi T\mu^4 < 6^{-4} \equiv \mathcal{E}_{c}$) around each of the two stable global minima ($u_0 = 0, 1/3$), $\bullet$ (2) large-energy oscillations ($\mathcal{E} > \mathcal{E}_{c}$) around the local maximum ($u_0 = 1/6$), and $\bullet$ (3) two homoclinic trajectories which cross the unstable critical point at $u_0 = 1/6$ with an energy equal to the height of the potential (i.e.\ the critical energy, $\mathcal{E} = \mathcal{E}_{c}$).
\begin{figure}[H]
\begin{center}
\includegraphics[scale=0.31]{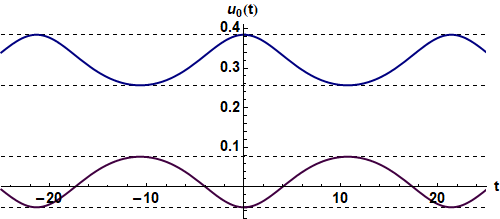} \includegraphics[scale=0.31]{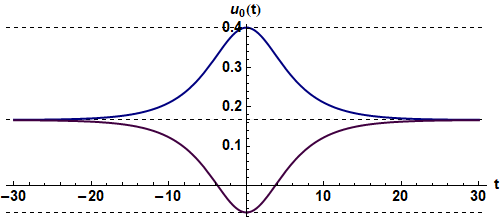} \includegraphics[scale=0.31]{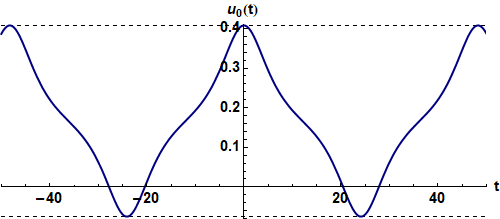}
\caption{Plots of \eqref{OrbitSO3} for $\mathcal{E} < 1/6^4$ (left), $\mathcal{E} = 1/6^4$ (center), and $\mathcal{E} > 1/6^4$ (right).} \label{Graph:OrbitsSO3}
\end{center}
\end{figure}
\vspace{-0.4cm}\indent The trajectories can be determined from the conserved energy integral and the initial conditions:
\begin{IEEEeqnarray}{l}
\dot{u}_0\left(0\right) = 0, \qquad u_0\left(0\right) = \frac{1}{6} \pm \sqrt{\frac{1}{6^2} + \sqrt{\mathcal{E}}}, \qquad
\end{IEEEeqnarray}
\noindent where the positive/negative sign should be taken, depending on whether the motion takes place in the right/left well of the double-well potential. We find:
\begin{IEEEeqnarray}{ll}
u_0\left(t\right) = &\frac{1}{6} \pm \sqrt{\frac{1}{6^2} + \sqrt{\mathcal{E}}} \cdot cn\left[\sqrt{2\sqrt{\mathcal{E}}}\cdot t\Bigg|\frac{1}{2}\left(1 + \frac{1}{36\sqrt{\mathcal{E}}}\right)\right]. \qquad \ \label{OrbitSO3}
\end{IEEEeqnarray}
When the energy is greater than the critical energy ($\mathcal{E} \geq \mathcal{E}_{c}$), only the positive sign should be kept in \eqref{OrbitSO3}. At the critical energy $\mathcal{E} = \mathcal{E}_{c}$, the solution \eqref{OrbitSO3} becomes the homoclinic trajectory:
\begin{IEEEeqnarray}{c}
u_0\left(t\right) = \frac{1}{6} \pm \frac{1}{3 \sqrt{2}} \cdot \text{sech}\left(\frac{t}{3 \sqrt{2}}\right). \qquad \label{HomoclinicOrbitSO3}
\end{IEEEeqnarray}
\indent The plots of the solutions \eqref{OrbitSO3}--\eqref{HomoclinicOrbitSO3} for different energies $\mathcal{E}$ can be found in figure \ref{Graph:OrbitsSO3} above. The rightmost plot corresponds to single-well oscillations around point-like membrane configurations, whereas the leftmost plot corresponds to oscillations around the pointlike membrane and the Myers sphere. The middle plot is the homoclinic orbit \eqref{HomoclinicOrbitSO3}, which takes place at the critical energy $\mathcal{E} = \mathcal{E}_{c}$. Because of the potential barrier, oscillations around the Myers sphere with $\mathcal{E} < \mathcal{E}_c$ cannot make the membrane collapse to a point (as e.g.\ when $\mathcal{E} > \mathcal{E}_c$). For $u < 0$, the membrane reverses its orientation.
\begin{figure}[H]
\begin{center}
\includegraphics[scale=0.31]{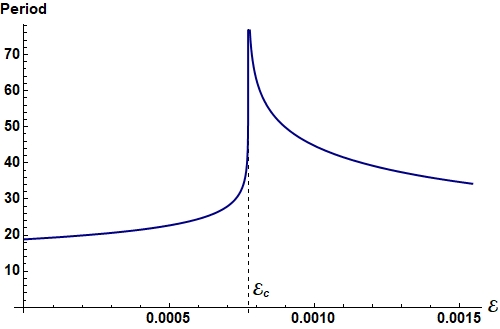}
\caption{Membrane oscillation period \eqref{PeriodSO3} as a function of the energy.} \label{Graph:PeriodSO3}
\end{center}
\end{figure}
\vspace{-0.4cm}\noindent We may also compute the period of membrane oscillations as a function of its energy $\mathcal{E}$. The period is given by the complete elliptic integral of the first kind:
\begin{IEEEeqnarray}{ll}
T\left(\mathcal{E}\right) = 2\sqrt{\frac{2}{\sqrt{\mathcal{E}}}} \cdot \textbf{K}\left(\frac{1}{2}\left(1 + \frac{1}{36\sqrt{\mathcal{E}}}\right)\right), \label{PeriodSO3}
\end{IEEEeqnarray}
and it has been plotted in figure \ref{Graph:PeriodSO3}. The homoclinic trajectory \eqref{HomoclinicOrbitSO3} (at $\mathcal{E} = \mathcal{E}_c$) has infinite period.
\subsubsection[\texorpdfstring{Axially symmetric tops in $SO\left(3\right) \times SO\left(6\right)$}{Axially symmetric tops in $SO\left(3\right)xSO\left(6\right)$}]{Axially symmetric tops in $SO\left(3\right)\times SO\left(6\right)$ \label{SubSection:AxiallySymmetricMembrane}}
\begin{figure}[H]
\begin{center}
\includegraphics[scale=0.4]{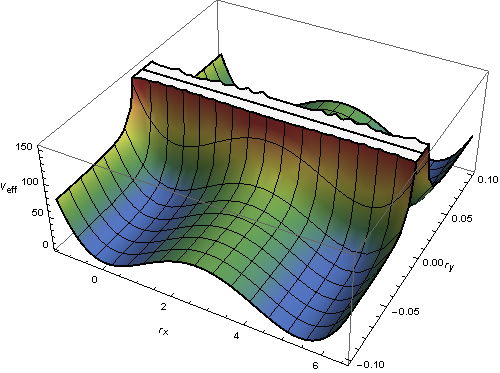} \qquad \includegraphics[scale=0.4]{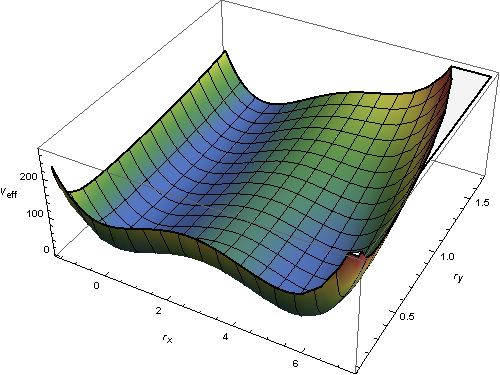}
\caption{Effective potential \eqref{AxiallySymmetricPotential1} of the top \eqref{Ansatz3} as a function of the radii ($\ell_y = 0.1$, $\mu = 16$). In the left figure, $-0.1 \leq r_y \leq +0.1$, while in the right figure, $0 \leq r_y \leq 1.7$.} \label{Graph:EffectivePotential13D}
\end{center}
\end{figure}

\begin{figure}[H]
\begin{center}
\includegraphics[scale=0.4]{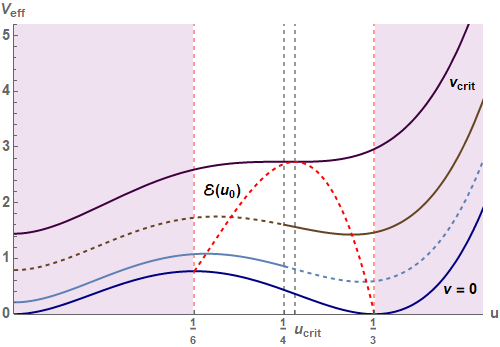} \qquad \includegraphics[scale=0.49]{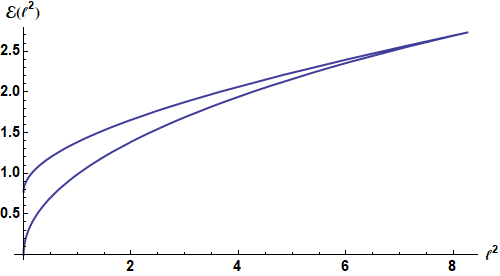}
\caption{$V_{\text{eff}}$ \eqref{AxiallySymmetricPotential1} for various $v$'s and $\ell$'s (left) and dispersion relation $\mathcal{E} = \mathcal{E}\left(\ell^2\right)$ (right).} \label{Graph:EffectivePotentialDispersionRelation}
\end{center}
\end{figure}

\begin{figure}[H]
\begin{center}
\includegraphics[scale=0.5]{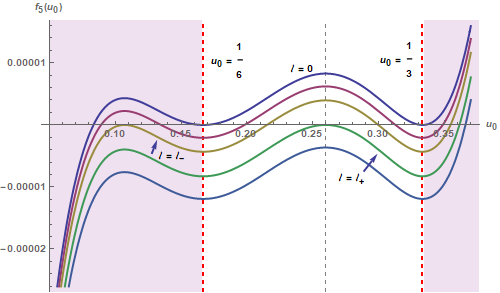}
\caption{Extremal function $f_5\left(u_0\right)$ for various values of the angular momentum $\ell$.} \label{Graph:ExtremalFunction}
\end{center}
\end{figure}

\begin{figure}[H]
\begin{center}
\includegraphics[scale=0.4]{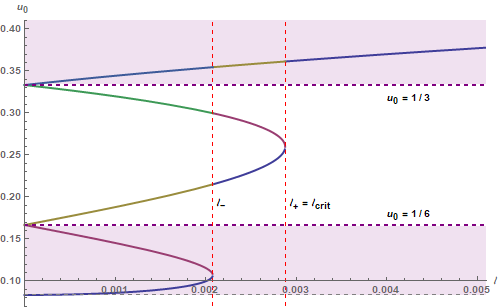} \qquad \includegraphics[scale=0.4]{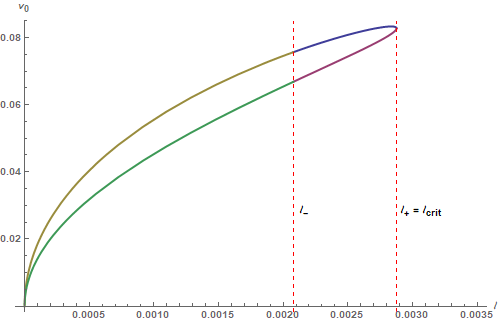}
\caption{Extrema $(u_0,v_0)$ of the axially symmetric potential \eqref{AxiallySymmetricPotential2} as a function of the angular momentum $\ell$.} \label{Graph:RootFunction}
\end{center}
\end{figure}

\begin{figure}[H]
\begin{center}
\includegraphics[scale=0.4]{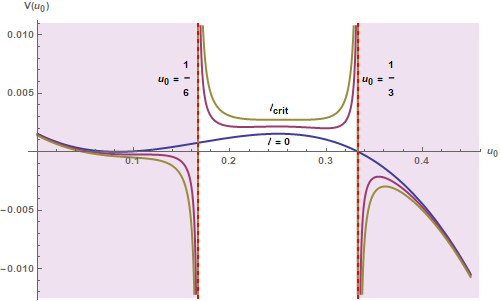} \qquad \includegraphics[scale=0.4]{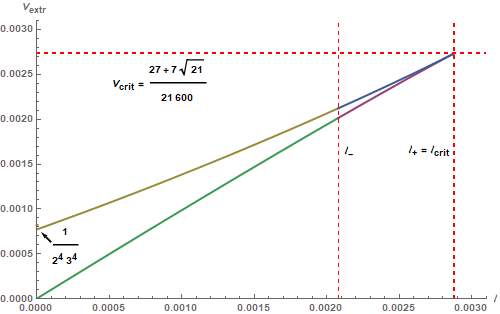}
\caption{The axially symmetric potential \eqref{AxiallySymmetricPotential2} as a function of $u_0$ for various angular momenta $\ell$ (left). The figure on the right is a plot of the two extrema of the axially symmetric potential \eqref{AxiallySymmetricPotential2} as a function of the angular momentum $\ell$.} \label{Graph:Potential}
\end{center}
\end{figure}

\noindent The second major type III configurations that will be treated in this paper is an axially symmetric configuration which lives inside the full geometric background of $SO\left(3\right)\times SO\left(6\right)$. The binary configuration consists of a static dielectric membrane in $SO\left(3\right)$, coupled to an uncharged rigidly spinning (non-collapsed in general) top in $SO\left(6\right)$:
\begin{IEEEeqnarray}{ll}
r_x = r_{x1} = r_{x2} = r_{x3}, \qquad r_y = r_{y1} = r_{y2} = r_{y3}, \qquad \ell_y = \ell_{y1} = \ell_{y2} = \ell_{y3}, \label{Ansatz3}
\end{IEEEeqnarray}
which corresponds to the effective potential,
\begin{IEEEeqnarray}{c}
V_{\text{eff}} = 2\pi T \left[\frac{\ell_{y}^2}{r_{y}^2} + r_{x}^4 + r_{y}^4 + 2r_{x}^2 r_{y}^2 + \frac{\mu^2 r_x^2}{9} + \frac{\mu^2 r_y^2}{36} - \frac{2\mu r_x^3}{3}\right]. \label{AxiallySymmetricPotential1}
\end{IEEEeqnarray}
The effective potential \eqref{AxiallySymmetricPotential1} as a function of the radii $r_x$ and $r_y$ has been plotted in figure \ref{Graph:EffectivePotential13D} for fixed values of the angular momentum $\ell_y$ and the mass parameter $\mu$.
\vspace{-0.4cm}\indent The ansatz \eqref{Ansatz3}--\eqref{AxiallySymmetricPotential1} presupposes an $s_1 = s_2 = s_3 = 2$ split of the six $SO\left(6\right)$ coordinates $y_{bj}$ in \eqref{SphericalAnsatz1y}--\eqref{SphericalAnsatz3y} (guaranteeing $\ell_{yi} \neq 0$ for all $i = 1,2,3$), otherwise $\ell_{yi} = 0$ for some of the angular momenta in $SO\left(6\right)$. Our present treatment however, will turn out to be independent of the specific values of $s_i$. For $\mu \neq 0$ it is also very convenient to work with the dimensionless quantities,
\begin{IEEEeqnarray}{ll}
u \equiv \frac{r_x}{\mu}, \qquad v \equiv \frac{r_y}{\mu}, \qquad \ell \equiv \frac{\ell_y}{\mu^3}, \qquad \label{Dimensionless}
\end{IEEEeqnarray}
in terms of which the effective potential \eqref{AxiallySymmetricPotential1} of the axially symmetric membrane \eqref{Ansatz3} becomes:
\begin{IEEEeqnarray}{ll}
V \equiv \frac{V_{\text{eff}}}{2 \pi T \mu^4} = u^4 + 2 u^2 v^2 + v^4 + \frac{u^2}{9} + \frac{v^2}{36} - \frac{2 u^3}{3} + \frac{\ell ^2}{v^2}, \qquad \label{AxiallySymmetricPotential2}
\end{IEEEeqnarray}
and corresponds to the Hamiltonian:
\begin{IEEEeqnarray}{ll}
\frac{H}{2 \pi T \mu^4} = p_u^2 + p_v^2 + V. \label{AxiallySymmetricHamiltonian}
\end{IEEEeqnarray}
\paragraph{Example} A type III configuration of the above form \eqref{Ansatz3} was studied in \cite{AxenidesFloratosLinardopoulos17a, AxenidesFloratosLinardopoulos17b}. It consists of a membrane that is static in the $SO(3)$ sector and is rigidly rotating in $SO(6)$:\footnote{See also \cite{HarmarkSavvidy00} for a similar solution in the context of D0-brane matrix mechanics. Of course, the prototype solution first appeared in \cite{CollinsTucker76}.}
\begin{IEEEeqnarray}{ll}
\tilde{u}_i = \mu u\left(t\right), \quad \tilde{v}_j = \mu v\left(t\right)\cos\left(\omega t + \varphi_k\right), \quad \tilde{w}_j = \mu v\left(t\right)\sin\left(\omega t + \varphi_k\right), \quad i,j= 1,2,3.\footnote{Note the similarity between the ansatz \eqref{AnsatzSO3xSO6} and the definition of cylindrical coordinates with $(z,\rho) = \mu\cdot(u, v)$.} \qquad \label{AnsatzSO3xSO6}
\end{IEEEeqnarray}
Here are the corresponding equations of motion (for $t \equiv \mu\tau$):
\begin{IEEEeqnarray}{ll}
\ddot{u} = -u \left[2 u^2 - u + \frac{1}{9} + 2 v^2\right], \qquad \ddot{v} = -\frac{1}{v^3} \left[2v^6 + \left(\frac{1}{36} + 2u^2\right)v^4 -\ell^2\right]. \qquad
\end{IEEEeqnarray}
\paragraph{Extrema} The extrema of the potential \eqref{AxiallySymmetricPotential2} can be found by solving the following system:
\begin{IEEEeqnarray}{ll}
\partial_u V = 2u_0 \left[2 u_0^2 - u_0 + \frac{1}{9} + 2 v_0^2\right] = 0 \quad \& \quad \partial_v V = \frac{2}{v_0^3} \left[2v_0^6 + \left(\frac{1}{36} + 2u_0^2\right)v_0^4 -\ell^2\right] = 0. \qquad \label{AxiallySymmetricExtrema1}
\end{IEEEeqnarray}
The first equation in \eqref{AxiallySymmetricExtrema1} is satisfied whenever $u_0 = 0$ or whenever the quantity
\begin{IEEEeqnarray}{ll}
v_0^2 = -\frac{1}{18} \left(18 u_0^2 -9u_0+ 1\right) = - \frac{1}{18} \left(3u_0 - 1\right)\left(6u_0 - 1\right) > 0, \label{AxiallySymmetricExtrema2}
\end{IEEEeqnarray}
is positive. Before we go on to examine these two setups, let us briefly discuss the case $\ell = 0$. Setting $\ell = 0$ in the effective potential \eqref{AxiallySymmetricPotential2} and differentiating, we find that the first equation in \eqref{AxiallySymmetricExtrema1} remains the same, whereas the second one becomes:
\begin{IEEEeqnarray}{ll}
\partial_v V\Big|_{\ell = 0} = 4v_0 \left[u_0^2 + v_0^2 + \frac{1}{72}\right] = 0. \qquad
\end{IEEEeqnarray}
We therefore obtain the same extrema that we found in the previous section for the static spherically symmetric (i.e.\ $u_1 = u_2 = u_3 = u_0$) membrane in $SO\left(3\right)$ (in dimensionless units):
\begin{IEEEeqnarray}{ll}
u_0 \ \in \left\{0, \ \frac{1}{3}, \ \frac{1}{6}\right\} \quad \& \quad v_0 = 0 \qquad \left(\ell = 0\right).
\end{IEEEeqnarray}
Now let us see what happens when $\ell \neq 0$. As we mentioned earlier there are two main setups, depending on whether the $SO\left(3\right)$ component is a point-like membrane or not. Let us examine each one of them separately. \\[6pt]
$\bullet \ u = 0$: This setup corresponds to a membrane which has collapsed in $SO\left(3\right)$, while its counterpart in $SO\left(6\right)$ is an uncharged Euler top with effective potential,
\begin{IEEEeqnarray}{ll}
V = v^4 + \frac{v^2}{36} + \frac{\ell ^2}{v^2}, \qquad \label{CollapsedEulerTop}
\end{IEEEeqnarray}
very similar to the tops that we encountered in categories I and II above. These (uncharged) tops can be studied along the lines of the paper \cite{AxenidesFloratos07} where we refer the interested reader for more details. \\[6pt]
$\bullet \ u \neq 0$: In this case, equation \eqref{AxiallySymmetricExtrema2} implies the following intervals of allowed values for $u_0$ and $v_0$:
\begin{IEEEeqnarray}{ll}
\frac{1}{6} \leq u_0 \leq \frac{1}{3} \quad \& \quad 0 \leq v_0 \leq \frac{1}{12} \equiv v_{\text{max}}. \qquad \label{ExtremalBounds1}
\end{IEEEeqnarray}
Plugging the solution \eqref{AnsatzSO3xSO6} into the second equation of motion in \eqref{AxiallySymmetricExtrema1}, we get for $\ell \neq 0$, $\dot{\varphi} = \omega$ (constant) and $\ddot{v} = 0$:
\begin{IEEEeqnarray}{l}
\omega^2 = 2u_0^2 + 2v_0^2 + \frac{1}{36} = u_0 - \frac{1}{12} \qquad \& \qquad \ell = \omega v_0^2, \qquad \label{AngularVelocity}
\end{IEEEeqnarray}
so that by inserting \eqref{AxiallySymmetricExtrema2} into \eqref{AngularVelocity}, we can write the conserved angular momentum in terms of $u_0$:
\begin{IEEEeqnarray}{ll}
\ell^2 = \left(u_0 - \frac{1}{12}\right)\left(u_0 - \frac{1}{6}\right)^2\left(\frac{1}{3} - u_0\right)^2. \qquad \label{BinaryAngularMomentum}
\end{IEEEeqnarray}
The membrane energy \eqref{AxiallySymmetricHamiltonian}, which is positive in the interval \eqref{ExtremalBounds1}, becomes:
\begin{IEEEeqnarray}{ll}
\mathcal{E} = \frac{5}{3}\left(u_0 - \frac{1}{12}\right)\left(u_0 - \frac{2}{15}\right)\left(\frac{1}{3} - u_0\right). \qquad \label{BinaryEnergy}
\end{IEEEeqnarray}
\indent The energy \eqref{BinaryEnergy} has been plotted with a red dashed line on the left diagram of figure \ref{Graph:EffectivePotentialDispersionRelation} which also contains a plot of the effective potential \eqref{AxiallySymmetricPotential1} for various values of $v_0$. The dispersion relation $\mathcal{E} = \mathcal{E}\left(\ell^2\right)$ has been plotted on the right diagram of figure \ref{Graph:EffectivePotentialDispersionRelation}.
\vspace{-0.4cm}\indent For all the $u_0$'s that lie in the interval \eqref{ExtremalBounds1}, the variable $v_0$ in \eqref{AxiallySymmetricExtrema2} is real. This allows us to eliminate it from the second equation in \eqref{AxiallySymmetricExtrema1}:
\begin{IEEEeqnarray}{ll}
f_5\left(u_0\right) \equiv u_0^5 - \frac{13 u_0^4}{12} + \frac{4u_0^3}{9} - \frac{37 u_0^2}{432} + \frac{5u_0}{648} - \frac{1}{3888} - \ell ^2 = 0. \label{AxiallySymmetricExtrema3}
\end{IEEEeqnarray}
The extremal function $f_5$ in \eqref{AxiallySymmetricExtrema3} has been plotted as a function of $u_0$, for various values of the angular momentum $\ell$, in figure \ref{Graph:ExtremalFunction}. Since \eqref{AxiallySymmetricExtrema3} is a fifth-degree polynomial equation, it always has exactly 5 roots which, depending on the value of the angular momentum $\ell$, can all be real or not. Because relation \eqref{AxiallySymmetricExtrema2} holds, the real roots of $f_5$ can give rise to real extrema of $v$, only when $u_0$ lies within the interval \eqref{ExtremalBounds1}. These bounds on the allowed values of $u_0$ have been denoted with red dashed lines in figure \ref{Graph:ExtremalFunction}.
\vspace{-0.4cm}\indent The resulting picture is the following. For $\ell = 0$, the function $f_5$ in \eqref{AxiallySymmetricExtrema3} has exactly five real roots but only two of them satisfy \eqref{ExtremalBounds1}. These (double) roots are the two minima at $u_0 = 1/6$ and $u_0 = 1/3$ which were found above by setting $\ell = 0$ in the effective potential \eqref{AxiallySymmetricPotential2}. As we increase the angular momentum, it is clear from the form of $f_5$ in \eqref{AxiallySymmetricExtrema3} that the $\ell = 0$ curve will gradually start moving below the horizontal axis, making the two allowed real roots approach each other. Meanwhile, no additional real root enters the interval \eqref{ExtremalBounds1}. The two local maxima of the function $f_5$ vanish for
\begin{IEEEeqnarray}{ll}
\ell_{\pm} = \sqrt{\frac{102 \pm 7\sqrt{21}}{16\,200\,000}}.
\end{IEEEeqnarray}
\indent For $\ell = \ell_-$, the first (from the left) local maximum of $f_5$ touches the $u_0$-axis and the total number of (different) real roots of $f_5$ is reduced to four (or better, there are still five real roots, but one is double). For $\ell>\ell_-$, the function $f_5$ cannot have more than three real roots (and two or none allowed within the interval \eqref{ExtremalBounds1}). The second maximum of the function $f_5$ touches the $u_0$-axis when $\ell = \ell_+$, and the two allowed double roots coalesce into a single double (allowed) root located at:
\begin{IEEEeqnarray}{ll}
u_{\text{crit}} = \frac{1}{60} \left(11+\sqrt{21}\right) \approx 0.25971, \quad v_{\text{crit}} = \frac{1}{30} \sqrt{2\sqrt{21} - 3} \approx 0.0827657. \label{AxiallySymmetricExtrema4}
\end{IEEEeqnarray}
This explains the cusp in the dispersion relation $\mathcal{E} = \mathcal{E}\left(\ell^2\right)$. The critical value of the angular momentum, above which the potentials \eqref{AxiallySymmetricPotential1}--\eqref{AxiallySymmetricPotential2} have no real minima is found by setting the second maximum of $f_5$ in \eqref{AxiallySymmetricExtrema3} equal to zero:
\begin{IEEEeqnarray}{ll}
\ell_{\text{crit}} = \ell_+ \approx 0.00287688. \label{CriticalAngularMomentum}
\end{IEEEeqnarray}
The real roots of the function $f_5$ (that is the extrema of the axially symmetric potential \eqref{AxiallySymmetricPotential2} w.r.t.\ $u$) as a function of the angular momentum $\ell$ have been plotted on the left diagram of figure \ref{Graph:RootFunction}.
\vspace{-0.4cm}\indent The right diagram of figure \ref{Graph:RootFunction} is the plot of $v_0$ in \eqref{AxiallySymmetricExtrema2} as a function of the angular momentum $\ell$. Different colors (in both diagrams) correspond to (maximum five) different real roots of the function $f_5$ in \eqref{AxiallySymmetricExtrema3}. Note that although the allowed roots are not the same for all the $\ell$'s, their number is never greater than two. \\[6pt]
\indent For (nonzero) $u_0$ within the allowed range \eqref{ExtremalBounds1} we may directly plug the value of $v_0$ that is specified in equation \eqref{AxiallySymmetricExtrema2} into the axially symmetric potential \eqref{AxiallySymmetricPotential2}, finding:
\begin{IEEEeqnarray}{ll}
V = \frac{1}{648} - \frac{u_0}{24}\left(4u_0 - 1\right)^2 - \frac{18\ell^2}{1 + 9u_0\left(2u_0 -1\right)}. \label{AxiallySymmetricPotential3}
\end{IEEEeqnarray}
A plot of the potential \eqref{AxiallySymmetricPotential3} as a function of the variable $u_0$, for various values of the angular momentum $\ell$ can be found in figure \ref{Graph:Potential}. Figure \ref{Graph:Potential} also contains the plot of the two extrema of the potential \eqref{AxiallySymmetricPotential3} in terms of the angular momentum $\ell$. This graph is obtained by plugging the allowed extremal values of $u_0\left(\ell\right)$ and $v_0\left(\ell\right)$ (plotted in figure \ref{Graph:RootFunction}) into the axially symmetric potential \eqref{AxiallySymmetricPotential3}.
\vspace{-0.4cm}\indent To determine the type of each extremum of the axially symmetric potential \eqref{AxiallySymmetricPotential2}, we compute the corresponding Hessian matrix by also taking into account the extremization conditions \eqref{AxiallySymmetricExtrema2}--\eqref{BinaryAngularMomentum}:
\begin{IEEEeqnarray}{l}
\partial^2_{\{u,v\}} V\Big|_{u_0,v_0} = \left(\begin{array}{cc} 2u_0\left(4u_0-1\right) & 8u_0v_0 \\ 8u_0v_0 & -8u_0^2 + 12u_0 - 10/9\end{array}\right). \qquad \ \label{HessianMatrix}
\end{IEEEeqnarray}
The eigenvalues of the Hessian \eqref{HessianMatrix} point out that there are two types of extrema: a continuum of saddle points between $1/6 \leq u_0 \leq u_{\text{crit}}$ and a continuum of minima between $u_{\text{crit}} < u_0 \leq 1/3$. See figure \ref{Graph:HessianEigenvalues} below.
\begin{figure}[H]
\begin{center}
\includegraphics[scale=0.4]{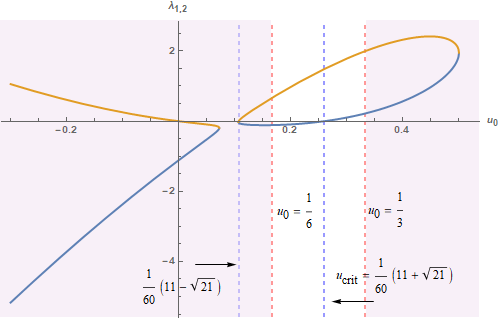}
\caption{Eigenvalues of the Hessian matrix \eqref{HessianMatrix}.} \label{Graph:HessianEigenvalues}
\end{center}
\end{figure}
\vspace{-0.4cm}\indent A convenient way to classify the extrema of the spherically symmetric potential \eqref{AxiallySymmetricPotential2} is provided by the following set of variables $u_{\pm}$ which are obtained by inverting \eqref{AxiallySymmetricExtrema2}:
\begin{IEEEeqnarray}{ll}
u_{\pm} = \frac{1}{4} \pm \sqrt{v_{\text{max}}^2 - v_0^2}. \label{MaximaMinima}
\end{IEEEeqnarray}
\indent It is easy to see that $u_-$ parametrizes the set of saddle points between $1/6 \leq u_0 \leq 1/4$, while $u_+$ parametrizes the (remaining) series of saddle points between $1/4 \leq u_0 \leq u_{\text{crit}}$ and the set of minima between $u_{\text{crit}} < u_0 \leq 1/3$. When the $SO(6)$ coordinate $v$ becomes zero, the former set of points (which is described by $u_-$) is reduced to a single unstable point, namely the saddle point $\textbf{u}_{1/6}$ of the double-well potential, while the latter set of points (which is parametrized by $u_+$) reduces to the Myers minimum $\textbf{u}_{1/3}$. \\[6pt]
\indent Obviously, when $v_0 > 0$ the degeneracy of the double-well potential at the minima $u_0 =0, 1/3$ (see \eqref{So3Extrema2} for the exact values of the potential) is lifted, and the corresponding extrema at $\textbf{u}_{1/6}$ and $\textbf{u}_{1/3}$ increase in value towards $u_0 = 1/4$, with the corresponding energy difference given by
\begin{IEEEeqnarray}{ll}
\mathcal{E}_+ - \mathcal{E}_- = \frac{10}{3}\left(v_0^2 - \frac{1}{360}\right)\sqrt{v_{\text{max}}^2 - v_0^2}, \qquad \label{PotentialDifference}
\end{IEEEeqnarray}
where $\mathcal{E}_{\pm} \equiv \mathcal{E}\left(u_{\pm}\right)$. It is further interesting to notice from \eqref{PotentialDifference} that the minima at $u_+$ are energetically preferred only inside the interval $0 \leq v_0 \leq 1/(6\sqrt{10}) < v_{\text{crit}}$, while inside the interval $1/(6\sqrt{10}) < v_0 \leq v_{\text{crit}}$ the minima at $u_+$ have greater energies compared to the saddle points at $u_-$ and so they are less favored energetically. Right at the point $(v = v_{\text{max}}, u_+ = u_- = 1/4)$ the energy difference in \eqref{PotentialDifference} becomes zero and the two sets of saddle extrema at $u_{\pm}$ merge with each other. It is also worth mentioning that beyond the range of extrema of $u$ and $v$ (just described) there is no equilibration between the applied forces on the membrane and its motion can easily become chaotic.\footnote{See e.g.\ \cite{AsanoKawaiYoshida15} for a study of chaos in the dynamical system that emerges when $\ell = 0$.}
\section[LO stability analysis]{LO stability analysis \label{Section:LOperturbations}}
\noindent Having classified the (spherical) membrane configurations that arise from the ansätze \eqref{SphericalAnsatz1x}--\eqref{SphericalAnsatz3y}, we are now ready to study their stability. As we have already noted, type III configurations are much more interesting because they combine 4 different types of attractive and repulsive terms (including a Myers flux term); these terms were described in section \ref{Subsection:EffectivePotentials}. Therefore, from now on, we will only focus on the two simple configurations of type III that we introduced in sections \ref{SubSection:StaticDielectricMembrane} and \ref{SubSection:AxiallySymmetricMembrane} above, namely the static dielectric membrane in $SO(3)$, and the $SO(3)\times SO(6)$ axially symmetric top. \\[6pt]
\indent With regard to the stability of these solutions, it seems that there are three different (but complementary as it will turn out) approaches. To leading order (LO) in perturbation theory, we can either perform radial perturbations \cite{AxenidesFloratosLinardopoulos17a}, or angular/multipole perturbations \cite{AxenidesFloratosLinardopoulos17b}. All the same, we may also process angular/multipole perturbations at the next-to-leading order (NLO) \cite{AxenidesFloratosKatsinisLinardopoulos21}. In the present section we will take on the study of LO perturbations and leave the analysis of NLO perturbations for the next section \ref{Section:NLOperturbations}. But let us get right to it, starting from LO perturbations in the radial direction.
\subsection[Radial stability]{Radial stability \label{SubSection:StabilityAnalysis}}
\subsubsection[$SO(3)$ sector]{$SO(3)$ sector \label{SubSubSection:RadialPerturbations1}}
\noindent The $SO(3)$ potential \eqref{StaticPotential} has 9 fixed points which have been spelled out in \eqref{So3Extrema1}. As we have already argued, the corresponding Hessian is positive-definite for $\textbf{u}_0$ and $\textbf{u}_{1/3}$, and indefinite for $\textbf{u}_{1/6}$. Therefore the former extremal points are minima while the latter is a saddle point. \\[6pt]
\indent This conclusion is confirmed by perturbation theory. By perturbing the classical equations of motion of the membrane we are led to a linearized system which can then be transformed to an eigenvalue/eigenvector problem. First off, the full set of equations of motion, for any type III configuration reads:
\begin{IEEEeqnarray}{ll}
\ddot{u}_1 + \left(u_2^2 + u_3^2 + \frac{r_{y2}^2}{\mu^2} + \frac{r_{y3}^2}{\mu^2} + \frac{1}{9}\right) u_1 = u_2 u_3,& \quad \ddot{v}_i + \left(\frac{r_{y2}^2}{\mu^2} + \frac{r_{y3}^2}{\mu^2} + u_2^2 + u_3^2 + \frac{1}{36}\right) v_i = 0 \label{EquationsOfMotion1} \qquad \\[6pt]
\ddot{u}_2 + \left(u_3^2 + u_1^2 + \frac{r_{y3}^2}{\mu^2} + \frac{r_{y1}^2}{\mu^2} + \frac{1}{9}\right) u_2 = u_3 u_1,& \quad \ddot{v}_j + \left(\frac{r_{y3}^2}{\mu^2} + \frac{r_{y1}^2}{\mu^2} + u_3^2 + u_1^2 + \frac{1}{36}\right) v_j = 0 \label{EquationsOfMotion2} \qquad \\[6pt]
\ddot{u}_3 + \left(u_1^2 + u_2^2 + \frac{r_{y1}^2}{\mu^2} + \frac{r_{y2}^2}{\mu^2} + \frac{1}{9}\right) u_3 = u_1 u_2,& \quad \ddot{v}_k + \left(\frac{r_{y1}^2}{\mu^2} + \frac{r_{y2}^2}{\mu^2} + v_1^2 + v_2^2 + \frac{1}{36}\right) v_k = 0, \qquad \label{EquationsOfMotion3}
\end{IEEEeqnarray}
where we have set $t \equiv \mu \tau$ and
\begin{IEEEeqnarray}{lll}
x_i = \mu u_i e_i, \quad i=1,2,3 \qquad \& \qquad & y_i = \mu v_i e_1, \quad & i = 1,\ldots,s_1 \label{TypeIIIAnsatz1} \\[6pt]
& y_j = \mu v_j e_2, \quad & j = s_1 + 1, \ldots, s_1 + s_2 \label{TypeIIIAnsatz2} \\[6pt]
& y_k = \mu v_k e_3, \quad & k = s_1 + s_2 + 1, \ldots, s_1 + s_2 + s_3. \label{TypeIIIAnsatz3}
\end{IEEEeqnarray}
By inserting the (radial) perturbations
\begin{IEEEeqnarray}{c}
u_i = u_i^0 + \delta u_i\left(t\right), \quad i = 1, 2, 3, \qquad \& \qquad v_j = \delta v_j\left(t\right), \quad j = 1,\ldots,6, \label{RadialPerturbations}
\end{IEEEeqnarray}
into the system of equations \eqref{EquationsOfMotion1}--\eqref{EquationsOfMotion3}, we obtain the following system of fluctuation equations
\begin{IEEEeqnarray}{c}
\delta\ddot{\textbf{u}} = -\left[\begin{array}{ccc} 2u_0^2 + \frac{1}{9} & 2u_1^0u_2^0 - u_3^0 & 2u_1^0u_3^0 - u_2^0 \\ 2u_2^0u_1^0 - u_3^0 & 2u_0^2 + \frac{1}{9} & 2u_2^0u_3^0 - u_1^0 \\ 2u_3^0u_1^0 - u_2^0 & 2u_3^0u_2^0 - u_1^0 & 2u_0^2 + \frac{1}{9} \end{array}\right] \cdot \delta\textbf{u} \quad \& \quad \delta\ddot{\textbf{v}} = -\left(2u_0^2 + \frac{1}{36}\right) \cdot \delta\textbf{v}, \qquad \label{LinearizedSystem1}
\end{IEEEeqnarray}
where $u_i^0$ is the set of extremal points \eqref{So3Extrema1}, and we have defined
\begin{IEEEeqnarray}{c}
u_0^2 \equiv \left(u_1^0\right)^2 = \left(u_2^0\right)^2 = \left(u_3^0\right)^2,
\end{IEEEeqnarray}
for the common value of the square of each extremum's components. Plugging the particular solution
\begin{IEEEeqnarray}{c}
\left[\begin{array}{c} \delta \textbf{u} \\ \delta \textbf{v}\end{array}\right] = e^{\lambda t} \, \boldsymbol{\xi},
\end{IEEEeqnarray}
into the linearized system \eqref{LinearizedSystem1} we can transform it into an eigenvalue/eigenvector problem which we may subsequently solve. The negative eigenvalues $r = \lambda^2 < 0$ correspond to stable directions, whereas the positive eigenvalues $r = \lambda^2 > 0$ lead to stable/unstable directions (depending on the sign of the real eigenvalue $\lambda$). For the nine extremal points \eqref{So3Extrema1} we find the following set of eigenvalues:
\vspace{-.4cm}\begin{table}[H]
\begin{center}
\begin{eqnarray}
\begin{array}{|c|c|c|c|}
\hline &&& \\
\text{extremum}& \text{location} &\text{eigenvalues } r = \lambda^2 \text{ (degeneracy)} & \text{stability} \\[6pt]
\hline &&& \\
\textbf{u}_0 & 0 & -\frac{1}{9} \, \left(3\right), \ -\frac{1}{36} \, \left(6\right) & \text{center (stable)} \\[12pt]
\textbf{u}_{1/6} & \left(\pm \frac{1}{6},\pm \frac{1}{6},\pm \frac{1}{6}\right) & \frac{1}{18} \, \left(1\right), \ -\frac{5}{18} \, \left(2\right), \ -\frac{1}{12} \, \left(6\right) & \text{saddle point} \\[12pt]
\textbf{u}_{1/3} & \left(\pm \frac{1}{3},\pm \frac{1}{3},\pm \frac{1}{3}\right) & -\frac{1}{9} \, \left(1\right), \ -\frac{4}{9} \, \left(2\right), \ -\frac{1}{4} \, \left(6\right) & \text{center (stable)} \\[6pt]
\hline
\end{array} \nonumber
\end{eqnarray}
\end{center}
\caption{Radial spectrum and stability of the static dielectric membrane in $SO(3)$. \label{Table:RadialPerturbations}}
\end{table}
\noindent This provides further justification to our previous conclusion (section \ref{SubSection:StaticDielectricMembrane}) that there are 2 stable degenerate global minima ($\textbf{u}_0$ and $\textbf{u}_{1/3}$) and a single saddle (local maximum) point ($\textbf{u}_{1/6}$) between them.
\subsubsection[\texorpdfstring{$SO(3) \times SO(6)$ sector}{$SO(3)xSO(6)$ sector}]{$SO(3)\times SO(6)$ sector \label{SubSubSection:RadialPerturbations2}}
\noindent We will now examine the radial stability of the two allowed extremal points of the axially symmetric potential \eqref{AxiallySymmetricPotential2} that we found in the previous section (see \ref{SubSection:AxiallySymmetricMembrane}). For simplicity, we concentrate on the radial modes, that is we ignore the fluctuations that are induced on the angular momentum. A more thorough analysis (which takes into account the fluctuations of the angular momentum) appears in appendix \ref{Appendix:RadialPerturbations}, where we perturb a specific axially symmetric configuration in $SO\left(3\right)\times SO\left(6\right)$. \\[6pt]
\indent The Lagrangian of the axially symmetric membrane \eqref{Ansatz3} is given in terms of the dimensionless variables \eqref{Dimensionless} and time $t \equiv \mu\tau$, by the following expression:
\begin{IEEEeqnarray}{ll}
\mathcal{L} \equiv \frac{L}{2\pi T \mu^4} = \dot{u}^2 + \dot{v}^2 - \left[u^4 + 2 u^2 v^2 + v^4 + \frac{u^2}{9} + \frac{v^2}{36} - \frac{2 u^3}{3} + \frac{\ell ^2}{v^2}\right]. \qquad \label{AxiallySymmetricLagrangian}
\end{IEEEeqnarray}
Here are the corresponding equations of motion:
\begin{IEEEeqnarray}{ll}
\ddot{u} + 2u^3 + 2uv^2 + \frac{u}{9} - u^2 = 0 \quad \& \quad \ddot{v} + 2u^2v + 2v^3 + \frac{v}{36} - \frac{\ell^2}{v^3} = 0. \label{AxiallySymmetricMembrane1}
\end{IEEEeqnarray}
We now introduce the perturbation\footnote{Alternatively, we could directly compute the eigenvalues of the second derivative matrix (Hessian). See section \ref{SubSection:AxiallySymmetricMembrane} above.}
\begin{IEEEeqnarray}{l}
u = u_0 + \delta u\left(t\right) \qquad \& \qquad v = v_0 + \delta v\left(t\right), \label{AxialPerturbations1}
\end{IEEEeqnarray}
where $u_0$ and $v_0$ are the extrema of \eqref{AxiallySymmetricPotential2} that satisfy the potential minimization equations \eqref{AxiallySymmetricExtrema1}. By plugging the perturbations \eqref{AxialPerturbations1} into the equations of motion \eqref{AxiallySymmetricMembrane1} and using the extremization conditions \eqref{AxiallySymmetricExtrema1}, we are led to the following system of linearized equations:
\begin{IEEEeqnarray}{ll}
\delta\ddot{u} + u_0\left(4u_0 - 1\right)\delta u + 4u_0v_0 \delta v = 0 \qquad \& \qquad \delta\ddot{v} + 4u_0v_0 \delta u - \left(4u_0^2 - 6u_0 + \frac{5}{9}\right)\delta v = 0 \label{LinearizedSystem2}. \qquad
\end{IEEEeqnarray}
The system of equations \eqref{LinearizedSystem2} can be written in the following matrix form,
\begin{IEEEeqnarray}{c}
\left[\begin{array}{c} \delta \ddot{u} \\ \delta \ddot{v}\end{array}\right] = \left[\begin{array}{cc} u_0\left(1 - 4u_0\right) & -4u_0v_0 \\ -4u_0v_0 & 4u_0^2 - 6u_0 + \frac{5}{9} \end{array}\right]\cdot\left[\begin{array}{c} \delta u \\ \delta v\end{array}\right],
\end{IEEEeqnarray}
so that by further expressing the general solution as
\begin{IEEEeqnarray}{c}
\left[\begin{array}{c} \delta u \\ \delta v\end{array}\right] = e^{\lambda t} \, \boldsymbol{\xi}, \label{Eigenvalues1}
\end{IEEEeqnarray}
we find the following two eigenvalues:
\begin{IEEEeqnarray}{c}
r_{\pm} = \lambda_{\pm}^2 = \frac{5}{18} - \frac{5u_0}{2} \pm \sqrt{-20u_0^3 + \frac{163u_0^2}{12} - \frac{35u_0}{18} + \frac{5^2}{18^2}}. \label{AxiallySymmetricEigenvalues1}
\end{IEEEeqnarray}
\begin{figure}[H]
\begin{center}
\includegraphics[scale=0.4]{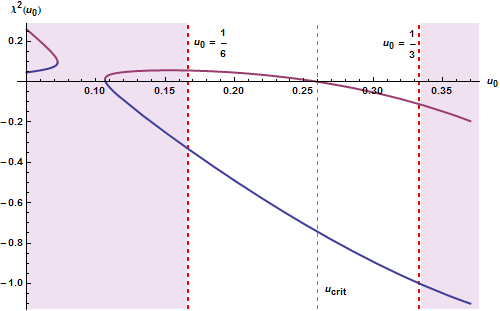}
\caption{Eigenvalues \eqref{AxiallySymmetricEigenvalues1} of radial perturbations \eqref{AxialPerturbations1} as a function of the extremal value $u_0$.} \label{Graph:Eigenvalues1}
\end{center}
\end{figure}
\vspace{-0.4cm} \indent We have plotted the two eigenvalues \eqref{AxiallySymmetricEigenvalues1} as a function of the extremal value $u_0$ in figure \ref{Graph:Eigenvalues1}. In the region of allowed $u_0$'s, given by \eqref{ExtremalBounds1}, $\lambda_-$ is always a purely imaginary eigenvalue that corresponds to a stable direction. On the other hand, $\lambda_+$ is purely imaginary only when $u_0$ is greater than the critical value \eqref{AxiallySymmetricExtrema4} (i.e.\ in the interval $u_{\text{crit}} < u_0 < 1/3$) and it is purely real (positive/negative) when $u_0$ is less than $u_{\text{crit}}$ (i.e.\ in the interval $1/6 < u < u_{\text{crit}}$). The former interval corresponds to a stable direction, whereas the latter is a stable/unstable direction, depending on the sign of the real eigenvalue $\lambda_+$. Because the two extrema coalesce right at the critical point $u_{\text{crit}} \approx 0.25971$ (as we have described in section \ref{SubSection:AxiallySymmetricMembrane} above), we infer that the rightmost extremum ($u_0 > u_{\text{crit}}$) is always stable, whereas the leftmost extremum ($u_0 < u_{\text{crit}}$) may have one unstable direction.
\subsection[Angular stability]{Angular stability \label{SubSection:AngularPerturbations}}
\noindent Having completed an in-depth study of the radial spectrum of static dielectric membranes \eqref{StaticMembrane1} in $SO\left(3\right)$ and axially symmetric dielectric membranes \eqref{AnsatzSO3xSO6} in $SO\left(3\right)\times SO\left(6\right)$, we will now embark on the study of stability of their angular modes. Our analysis will follow the recent publication \cite{AxenidesFloratosLinardopoulos17b} and will effectively parallel the one which was carried out for the BMN matrix model \cite{DasguptaJabbariRaamsdonk02}. However, our overall approach will be very different. We will employ a method which was by and large introduced in \cite{AxenidesFloratosPerivolaropoulos00, AxenidesFloratosPerivolaropoulos01} for the study of stability of membrane configurations inside a flat Minkowski background.\footnote{Essentially, our analysis follows the work of Lyapunov (1892). Details can be found in many textbooks, see e.g.\ \cite{Lefschetz57}.} Interestingly, these flat space results can be recovered in the zero flux limit ($\mu \rightarrow 0$).
\subsubsection[$SO(3)$ sector]{$SO(3)$ sector \label{SubSubSection:AngularPerturbations1}}
\noindent As a warmup, let us begin by examining the stability of the static $SO\left(3\right)$ dielectric membrane ansatz \eqref{StaticMembrane1} under linearized multipole perturbations. Consider the perturbation,
\begin{IEEEeqnarray}{c}
x_i = x_i^0 + \delta x_i, \quad i = 1, 2, 3 \qquad \& \qquad y_i = \delta y_i, \quad i = 1,\ldots 6, \label{AngularPerturbations1}
\end{IEEEeqnarray}
where the tree-level ansatz $x_i^0 = \mu u_0 e_i$ ($i = 1,2,3$) is given in terms of the spherically symmetric critical points \eqref{So3Extrema1}. By plugging the perturbations \eqref{AngularPerturbations1} into the equations of motion \eqref{xEquation}--\eqref{yEquation} we are led to:
\begin{IEEEeqnarray}{l}
\delta\ddot{x}_i = \left\{\left\{\delta x_i,x_j^0\right\},x_j^0\right\} + \left\{\left\{x_i^0,\delta x_j\right\},x_j^0\right\} + \left\{\left\{x_i^0,x_j^0\right\},\delta x_j\right\} - \frac{\mu^2}{9}\,\delta x_i^0 + \mu\epsilon_{ijk}\left\{\delta x_j,x_k^0\right\} \qquad \label{PerturbationEquation1} \\[6pt]
\delta\ddot{y}_i = \left\{\left\{\delta y_i,x_j^0\right\},x_j^0\right\} - \frac{\mu^2}{36}\,\delta y_i^0. \label{PerturbationEquation2}
\end{IEEEeqnarray}
The corresponding Gauss law constraint \eqref{GaussLaw1} becomes:
\begin{IEEEeqnarray}{c}
\left\{\delta\dot{x}_i, x_i^0\right\} = 0, \label{GaussLaw2}
\end{IEEEeqnarray}
since $\dot{x}_i^0 = {y}_i^0 = \dot{y}_i^0 = 0$. We will show below that if the Gauss law \eqref{GaussLaw2} is satisfied at one particular instant in time (typically at $\tau = 0$), then it will be valid for all times $\tau$. \\[6pt]
\indent In order to examine the angular stability of the static dielectric membrane \eqref{StaticMembrane1} in $SO(3)$, we need to expand the $SO(3)$ perturbations $\delta x$ and the $SO(6)$ perturbations $\delta y$ in spherical harmonics:
\begin{IEEEeqnarray}{lll}
\delta x_i = \mu\cdot\sum_{j,m}\eta_i^{jm}\left(\tau\right) Y_{jm}\left(\theta,\phi\right), \qquad &\eta_i^{jm}\left(0\right) = 0, \qquad &i = 1,2,3 \label{AngularPerturbations2a} \\[6pt]
\delta y_i = \mu\cdot\sum_{j,m}\theta_i^{jm}\left(\tau\right) Y_{jm}\left(\theta,\phi\right), \qquad &\theta_i^{jm}\left(0\right) = 0, \qquad &i = 1,\ldots,6. \label{AngularPerturbations2b}
\end{IEEEeqnarray}
Because the fluctuations $\left(\delta x, \delta y\right)$ in \eqref{AngularPerturbations2a}--\eqref{AngularPerturbations2b} must be real, the corresponding fluctuation modes $(\eta_i^{jm}\left(t\right), \theta_i^{jm}\left(t\right))$ satisfy the following reality conditions:
\begin{IEEEeqnarray}{ll}
\eta_i^{jm*}\left(t\right) = (-1)^{m} \eta_i^{j(-m)}\left(t\right), \qquad \theta_i^{jm*}\left(t\right) = (-1)^{m} \theta_i^{j(-m)}\left(t\right). \label{RealityCondition}
\end{IEEEeqnarray}
By using the property of spherical harmonics
\begin{IEEEeqnarray}{c}
\left\{e_i,Y_{jm}\left(\theta,\phi\right)\right\} = -i \hat{J}_i^{\,(j)} Y_{jm}\left(\theta,\phi\right) = -i \sum_{m'}\left(J_i\right)_{m'm}^{(j)} Y_{jm'}\left(\theta,\phi\right), \label{SphericalHarmonics1}
\end{IEEEeqnarray}
where the operator $\hat{J}_i$ provides a representation of angular momentum in the spherical coordinate system and the matrix $\left(J_i\right)_{mm'}$ provides a $2j+1$ dimensional representation of the group $\mathfrak{su}\left(2\right)$, we can derive the equations of motion for the fluctuation modes $\eta_i$ and $\theta_i$:
\begin{IEEEeqnarray}{c}
\ddot\eta_i + \omega_3^2 \eta_i = u_0^2 T_{ik} \eta_k + u_0 Q_{ik} \eta_k \qquad \& \qquad \ddot\theta_i + \omega_6^2 \theta_i = 0. \label{AngularPerturbations3}
\end{IEEEeqnarray}
\indent Note that we are using dimensionless time $t \equiv \mu\tau$ and we have employed the Einstein summation convention for the repeated indices. In order to make our expressions more transparent, we will generally omit the angular momentum indices $(j,m)$, unless they are absolutely necessary. We have also defined the following quantities:
\begin{IEEEeqnarray}{c}
\omega_3^2 \equiv u_0^2 \, \textbf{J}^2 + \frac{1}{9}, \quad \omega_6^2 \equiv u_0^2\, \textbf{J}^2 + \frac{1}{36} \quad \& \quad T_{ik} \equiv J_i J_k - 2i\epsilon_{ikl}J_l, \quad Q_{ik} \equiv i\epsilon_{ikl}J_l, \qquad \label{Definitions1}
\end{IEEEeqnarray}
where $\textbf{J}^2 \equiv j(j+1)$. Moreover, we can write \eqref{AngularPerturbations3} in a compact form as follows
\begin{IEEEeqnarray}{c}
\ddot{H} + \mathcal{K} \cdot H = 0, \qquad \mathcal{K} \equiv \omega_3^2 - u_0^2 T - u_0 Q \qquad \& \qquad \ddot\Theta + \omega_6^2 \Theta = 0, \label{AngularPerturbations4}
\end{IEEEeqnarray}
where the index structure of the vectors/matrices $H$, $\Theta$, $Q$ and $T$ is given by $H \equiv \left(\eta_i\right)$, $\Theta \equiv \left(\theta_i\right)$, $Q \equiv \left(Q_{ik}\right)$ and $T \equiv \left(T_{ik}\right)$. This way, we have obtained a $3\times(2j + 1)$ dimensional representation of the fluctuation modes $\eta_i^{jm}$, $\theta_i^{jm}$ and the matrices $Q_{ik}$ and $T_{ik}$:\\[-18pt]

\footnotesize\begin{IEEEeqnarray}{l}
H = \begin{pmatrix} \eta_x^{jm} \\ \eta_y^{jm} \\ \eta_z^{jm} \end{pmatrix}, \ \Theta = \begin{pmatrix} \theta_x^{jm} \\ \theta_y^{jm} \\ \theta_z^{jm} \end{pmatrix}, \quad Q = i\begin{pmatrix} 0 & J_z & -J_y \\ -J_z & 0 & J_x \\ J_y & -J_x & 0 \end{pmatrix}, \
T = \begin{pmatrix} J_x^2 & J_x J_y - 2i J_z & J_x J_z + 2i J_y \\ J_y J_x + 2i J_z & J_y^2 & J_y J_z - 2i J_x \\ J_z J_x - 2i J_y & J_z J_x + 2i J_x & J_z^2 \end{pmatrix}. \qquad \label{Definitions2}
\end{IEEEeqnarray}\normalsize
\indent As it will become apparent in the next section, where we will perform the NLO stability analysis for the static dielectric membrane \eqref{StaticMembrane1} in $SO(3)$, the $2j + 1$ and the $3\times(2j + 1)$ dimensional representations \eqref{AngularPerturbations3} and \eqref{AngularPerturbations4} are practically different. Subtle details of our analysis at the NLO will turn out to depend on which of the two representations we use.
\paragraph{Eigenvalues} To obtain the spectrum of fluctuations for the membrane solution \eqref{StaticMembrane1}, we must solve the fluctuation equations \eqref{AngularPerturbations3} and \eqref{AngularPerturbations4}. We may readily transform \eqref{AngularPerturbations3} and \eqref{AngularPerturbations4} into an eigenvalue/eigenvector problem by setting
\begin{IEEEeqnarray}{c}
\left[\begin{array}{c} H \\ \Theta \end{array}\right] = e^{i\lambda t} \left[\begin{array}{c} \boldsymbol{\xi}_1 \\ \boldsymbol{\xi}_2 \end{array}\right],
\end{IEEEeqnarray}
which leads us to the following linear system of equations (by also setting $\boldsymbol{\xi}_{a} \equiv (\xi_{a,i})$, for $a = 1,2$):
\begin{IEEEeqnarray}{c}
\left(-\lambda^2 + \omega_3^2 - u_0^2 \, T - u_0 \, Q\right) \cdot \boldsymbol{\xi}_1 = 0 \qquad \& \qquad \left(-\lambda^2 + \omega_6^2\right) \cdot \boldsymbol{\xi}_2 = 0. \label{AngularPerturbations5}
\end{IEEEeqnarray}
\indent To solve the system of equations \eqref{AngularPerturbations5}, we will follow a method which was introduced in \cite{AxenidesFloratosPerivolaropoulos00, AxenidesFloratosPerivolaropoulos01}. We need the $(2j+1)\times(2j+1)$ matrices
\begin{IEEEeqnarray}{c}
P_{ik} \equiv \frac{1}{j\left(j+1\right)} \, J_i J_k \quad \& \quad R_{ik}^{\pm} \equiv \frac{1}{2j +1} \Big[\frac{1}{2}\cdot\left(2j+1\mp1\right)\cdot\left(\delta_{ik}\times I - P_{ik}\right) \pm \left(\delta_{ik}\times I - Q_{ik}\right)\Big], \qquad \label{AngularEigenvalueProblem1}
\end{IEEEeqnarray}
which are orthogonal projectors (i.e.\ Hermitian and idempotent) that form a complete set:
\begin{IEEEeqnarray}{l}
P_{il}\cdot P_{lk} = P_{ik}, \quad R^{\pm}_{il} \cdot R^{\pm}_{lk} = R^{\pm}_{ik}, \quad P_{il} \cdot R^{\pm}_{lk} = R^{+}_{il} \cdot R^{-}_{lk} = 0, \quad P_{ik} + R_{ik}^{+} + R_{ik}^{-} = \delta_{ik} \times I, \qquad \label{AngularEigenvalueProblem2}
\end{IEEEeqnarray}
where $i, k, l = 1, 2, 3$ are spatial $SO(3)$ indices and $I$ is the $(2j+1)\times(2j+1)$ unit matrix. Note that the identities \eqref{AngularEigenvalueProblem2} are also valid in the $3(2j+1)\times3(2j+1)$ representation. As it turns out, the matrices $T$ and $Q$ are Hermitian, as well as commuting. Moreover, they can be expressed by means of the projection operators $P$, $R_{\pm}$ as follows:
\begin{IEEEeqnarray}{c}
T = \left[j\left(j+1\right) - 2\right]P + 2jR_+ - 2\left(j+1\right)R_- \qquad \& \qquad Q = P - j R_+ + \left(j+1\right)R_-, \label{AngularEigenvalueProblem3}
\end{IEEEeqnarray}
where this time we have omitted the spatial indices $i,k$ for simplicity. The advantage of the decomposition \eqref{AngularEigenvalueProblem3} is that we can directly read off the eigenvalues of the matrices $T$ and $Q$ (along with the corresponding multiplicities) in each of the projective spaces $P,R_{\pm}$. Then the eigenvalues $\lambda$ of the fluctuation equations \eqref{AngularPerturbations4} can be found by solving the system of equations \eqref{AngularPerturbations5}. This system is more easily solved by expressing the first equation in terms of the projectors $P,R$ in \eqref{AngularEigenvalueProblem1}:
\begin{IEEEeqnarray}{c}
\left(\omega_3^2 - \lambda^2\right)\boldsymbol{\xi}_1 = \Bigg[\left(u_0^2\left[j\left(j+1\right) - 2\right] + u_0\right)P + j u_0 \left(2u_0 - 1\right)R_+ - \left(j+1\right)u_0\left(2u_0 - 1\right)R_-\Bigg]\boldsymbol{\xi}_1. \qquad \ \label{AngularPerturbations6}
\end{IEEEeqnarray}
The $SO(6)$ eigenvalues $\lambda_{\theta}$ can be read-off from the second equation in \eqref{AngularPerturbations5}. Multiplying both sides of \eqref{AngularPerturbations6} with the projection operators $P$, $R_{\pm}$, we obtain the following set of eigenvalues:
\begin{IEEEeqnarray}{ll}
\lambda_P^2 = 2(u_0 - \frac{1}{3})(u_0 - \frac{1}{6}), \qquad & \lambda_+^2 = j\left(j-1\right)u_0^2 + j u_0 + \frac{1}{9} \label{Eigenfrequency1} \\[6pt]
\lambda_{\theta}^2 = u_0^2 j\left(j+1\right) + \frac{1}{36}, \qquad & \lambda_-^2 = \left(j+1\right)\left(j+2\right)u_0^2 - \left(j+1\right)u_0 + \frac{1}{9}. \qquad \label{Eigenfrequency2}
\end{IEEEeqnarray}
\indent The degeneracies of the eigenvalues \eqref{Eigenfrequency1}--\eqref{Eigenfrequency2} are respectively equal to the dimensionalities of the 3 projectors $P$, $R_{\pm}$. These are $d_P = 2j+1$, $d_+ = 2j+3$, $d_- = 2j-1$, while the degeneracy of the eigenvalues of the decoupled $\theta$-oscillators is $d_{\theta} = 6\left(2j+1\right)$. Adding these degeneracies (by also multiplying with 2 to account for the square roots), we find that the total number of eigenvalues is $d_P + d_+ + d_- + d_{\theta} = 18\left(2j+1\right)$ as it should. For the extremal points $\left(\textbf{u}_0, \textbf{u}_{1/6}, \textbf{u}_{1/3}\right)$ in \eqref{So3Extrema1} we find:
\begin{IEEEeqnarray}{ll}
\textbf{u}_0: \ &\lambda_P^2 = \lambda_{\pm}^2 = \frac{1}{9}, \qquad \lambda_{\theta}^2 = \frac{1}{36} \qquad \label{MultipoleEigenvalues1} \\[6pt]
\textbf{u}_{1/6}: \ &\lambda_P^2 = 0, \qquad \lambda_+^2 =\frac{1}{36}\left(j + 1\right)\left(j + 4\right), \qquad \lambda_-^2 = \frac{j\left(j-3\right)}{36}, \qquad \lambda_{\theta}^2 = \frac{1}{36}\left(j^2 + j + 1\right) \qquad \label{MultipoleEigenvalues2} \\[6pt]
\textbf{u}_{1/3}: \ &\lambda_P^2 = 0, \qquad \lambda_+^2 =\frac{1}{9}\left(j + 1\right)^2, \qquad \lambda_-^2 = \frac{j^2}{9}, \qquad \lambda_{\theta}^2 = \frac{1}{36}\left(2j + 1\right)^2. \qquad \label{MultipoleEigenvalues3}
\end{IEEEeqnarray}
\indent We conclude that the fixed point $\textbf{u}_0$ (point-like membrane) is stable. The critical point $\textbf{u}_{1/3}$ has one vanishing eigenvalue ($2d_P$ degenerate) in the $P$-sector for $j = 1,2,\ldots$, and stable modes in the $R_{\pm}$ sectors. The fixed point $\textbf{u}_{1/6}$ has one zero mode (degeneracy $2d_P$) for all $j$'s, and a vanishing eigenvalue for $j=3$ (degeneracy 10). This point is unstable for $j=1$ (degeneracy 2) and $j=2$ (degeneracy 6) in the $R_-$ sector. The instabilities occur right on the local maximum of the double-well potential which becomes a saddle point in the radial direction. Notice also that, for $j=1$, the angular spectrum in \eqref{MultipoleEigenvalues1}--\eqref{MultipoleEigenvalues3} becomes identical to the radial spectrum (see table \ref{Table:RadialPerturbations}). \\[6pt]
\indent Our results nicely reproduce the ones which were found in the case of the BMN matrix model in \cite{DasguptaJabbariRaamsdonk02}. Let us also note in passing that, because the tetrahedral symmetry of the $SO(3)$ extremal points $\textbf{u}_{1/6}$ and $\textbf{u}_{1/3}$ in \eqref{So3Extrema1} encompasses the full membrane Hamiltonian \eqref{StaticHamiltonian} (see section \ref{SubSection:StaticDielectricMembrane}), their images will have the same spectra under the set of multipole perturbations \eqref{AngularPerturbations2a}--\eqref{AngularPerturbations2b}. \\[6pt]
\indent Based on what has been said above, the general solution of the fluctuation equations \eqref{AngularPerturbations3} takes the following form:
\begin{IEEEeqnarray}{l}
\eta_{i} = e^{i\lambda_{P}\tau}\xi_{i}^P + e^{-i\lambda_{P}\tau}\tilde{\xi}_{i}^P + e^{i\lambda_{+}\tau}\xi_{i}^+ + e^{-i\lambda_{+}\tau}\tilde{\xi}_{i}^+ + e^{i\lambda_{-}\tau}\xi_{i}^- + e^{-i\lambda_{-}\tau}\tilde{\xi}_{i}^- \label{GeneralSolution1a} \\[6pt]
\theta_{i} = e^{i\lambda_{\theta}\tau}\xi_{2i} + e^{-i\lambda_{\theta}\tau}\tilde{\xi}_{2i}, \label{GeneralSolution1b}
\end{IEEEeqnarray}
where $\xi_{i}^{P,\pm}$ and $\tilde{\xi}_{i}^{P,\pm}$ are the projections of two linearly independent eigenvectors $\xi_{1i}$, $\tilde{\xi}_{1i}$ of the first equation in \eqref{AngularPerturbations5},
\begin{IEEEeqnarray}{ll}
\xi_{i}^P \equiv P_{ik}\xi_{1k}, \qquad \xi_{i}^{\pm} \equiv R^{\pm}_{ik}\xi_{1k}, \qquad \tilde{\xi}_{i}^P \equiv P_{ik}\tilde{\xi}_{1k}, \qquad \tilde{\xi}_{i}^{\pm} \equiv R^{\pm}_{ik}\tilde{\xi}_{1k}, \label{Eigenvectors}
\end{IEEEeqnarray}
and $\xi_{2i}$, $\tilde{\xi}_{2i}$ are two linearly independent eigenvectors of the second equation in \eqref{AngularPerturbations5}.
\paragraph{Eigenvectors} Let us now specify all the eigenvectors of the projection operators $P$ and $R_{\pm}$. It turns out that these are directly related to the eigenvectors of the matrix $Q$ in \eqref{Definitions1}. To see this, take the square of $Q$ and then use the identities \eqref{AngularEigenvalueProblem3} to express $P$ and $R_{\pm}$ in terms of $Q$ as follows:
\begin{IEEEeqnarray}{l}
P = I + \frac{Q - Q^2}{j(j+1)}, \qquad
R_{\pm} = \frac{1}{2j +1} \bigg[\frac{1}{2}\left(2j+1\mp1\right) \cdot \frac{Q^2 - Q}{j(j+1)} \pm \left(I - Q\right)\bigg]. \qquad \label{AngularEigenvalueProblem4}
\end{IEEEeqnarray}
Obviously the eigenvectors of the projectors $P$ and $R_{\pm}$ are completely specified by those of $Q$. \\[6pt]
\indent Now notice that the operator $Q$ in \eqref{Definitions1} is the obrit-spin coupling operator for the orbital angular momentum $L = 1$ and the spin angular momentum $J = j$. This is more easily seen as follows. In the adjoint representation of $SU(2)$, the components of the orbital angular momentum $L = 1$ are:
\begin{IEEEeqnarray}{l}
\left(L_i\right)_{kl} = i\epsilon_{ilk} \quad \Rightarrow \quad
L_x = \begin{pmatrix}
0 & 0 & 0\\
0 & 0 & -i\\
0 & i &0
\end{pmatrix}, \quad
L_y = \begin{pmatrix}
0 & 0 & i\\
0 & 0 & 0\\
-i & 0 & 0
\end{pmatrix}, \quad
L_z = \begin{pmatrix}
0 & -i & 0\\
i & 0 & 0\\
0 & 0 & 0
\end{pmatrix}, \qquad \label{OrbitalAngularMomentum}
\end{IEEEeqnarray}
so that $Q$ is indeed the orbit-spin coupling operator (for $L = 1$ and $J = j$),
\begin{IEEEeqnarray}{c}
Q_{ik} = \left(L_l\right)_{ki} J_l \quad \Leftrightarrow \quad Q = -L_i \otimes J_i. \label{SpinOrbitCoupling1}
\end{IEEEeqnarray}
In terms of the $L = 1$ and $J = j$ raising and lowering operators $L_{\pm} \equiv L_x \pm i L_y$ and $J_{\pm} \equiv J_x \pm i J_y$, $Q$ can be written as
\begin{IEEEeqnarray}{c}
Q = -\frac{1}{2}\left(L_ + \otimes J_-\right) - \frac{1}{2}\left(L_-\otimes J_+\right) - L_z\otimes J_z. \qquad \label{SpinOrbitCoupling2}
\end{IEEEeqnarray}
The components of the total angular momentum $\textbf{J}_{\scaleto{T}{5pt}} = \textbf{L} + \textbf{J}$, for $J = j$, $L = 1$ read:
\begin{IEEEeqnarray}{c}
J_{\scaleto{T}{5pt}}^{i} = J_{i}\otimes I_{3} + I_{2j+1}\otimes L_{i},
\end{IEEEeqnarray}
or, written out in full,
\begin{IEEEeqnarray}{c}
J_{\scaleto{T}{5pt}}^x = \begin{pmatrix}
J_x & 0 & 0\\
0 & J_x & -i I\\
0 & i I & J_x
\end{pmatrix}, \qquad
J_{\scaleto{T}{5pt}}^y = \begin{pmatrix}
J_y & 0 & i I\\
0 & J_y & 0\\
-i I & 0 & J_y
\end{pmatrix}, \qquad
J_{\scaleto{T}{5pt}}^z = \begin{pmatrix}
J_z & -i I & 0\\
i I & J_z & 0\\
0 & 0 & J_z
\end{pmatrix}. \qquad
\end{IEEEeqnarray}
On the other hand, the square of the total angular momentum $\textbf{J}$ becomes
\begin{IEEEeqnarray}{c}
\textbf{J}_{\scaleto{T}{5pt}}^2 = \left(j(j+1)+2\right)I_{3(2j+1)}-2Q, \label{TotalAngularMomentum1}
\end{IEEEeqnarray}
while its eigenstates,
\begin{IEEEeqnarray}{l}
\vert j+1,m;j,1\rangle, \quad \vert j,m;j,1\rangle, \quad \vert j-1,m;j,1\rangle, \qquad
\end{IEEEeqnarray}
are shared by the spin-orbit coupling operator $Q$ and therefore diagonalize the operator $Q$ as well. \\[6pt]
\indent Now we can write down the full solution to the eigenvalue/eigenvector problem of the spin-orbit coupling operator $Q$. We proceed by directly diagonalizing $Q$ in \eqref{SpinOrbitCoupling2}. Equivalently, we may carry out the standard Clebsch-Gordan analysis (see e.g.\ \cite{Rose57} for more) for the (orbit/spin) angular momenta $L = 1$ and $J = j$:

\footnotesize\vspace{-.3cm}\begin{IEEEeqnarray}{l}
\vert j+1,m;j,1\rangle = \sqrt{\frac{\left(j+m\right)\left(j+m+1\right)}{2\left(j+1\right)\left(2j+1\right)}}\cdot\vert 1,1\rangle\vert j,m-1\rangle + \sqrt{\frac{\left(j+1\right)^2-m^2}{(j+1)\left(2j+1\right)}}\cdot\vert 1,0\rangle\vert j,m\rangle + \nonumber \\
\hspace{8.8cm} + \sqrt{\frac{\left(j-m\right)\left(j-m+1\right)}{2(j+1)\left(2j+1\right)}}\cdot\vert 1,-1\rangle\vert j,m+1\rangle, \qquad \label{TotalAngularMomentum2a} \\
\vert j,m;j,1\rangle = -\sqrt{\frac{\left(j+m\right)\left(j-m+1\right)}{2j\left(j+1\right)}}\cdot\vert1,1\rangle\vert j,m-1\rangle + \frac{m}{\sqrt{j(j+1)}}\cdot\vert1,0\rangle\vert j,m\rangle + \nonumber \\
\hspace{8.8cm} + \sqrt{\frac{\left(j-m\right)\left(j+m+1\right)}{2j\left(j+1\right)}}\cdot\vert1,-1\rangle\vert j,m+1\rangle, \qquad \label{TotalAngularMomentum2b} \\
\vert j-1,m;j,1\rangle = \sqrt{\frac{\left(j-m\right)\left(j-m+1\right)}{2j\left(2j+1\right)}}\cdot\vert 1,1\rangle\vert j,m-1\rangle - \sqrt{\frac{j^2-m^2}{j\left(2j+1\right)}}\cdot\vert 1,0\rangle\vert j,m\rangle + \nonumber \\
\hspace{8.8cm} + \sqrt{\frac{\left(j+m\right)\left(j+m+1\right)}{2j\left(2j+1\right)}}\cdot\vert 1,-1\rangle\vert j,m+1\rangle. \qquad\label{TotalAngularMomentum2c}
\end{IEEEeqnarray}\normalsize
Applying the spin-orbit coupling operator $Q$ on both sides of the Clebsch-Gordan system \eqref{TotalAngularMomentum2a}--\eqref{TotalAngularMomentum2c}, we come up with the following set of eigenvalues:
\begin{IEEEeqnarray}{l}
Q\cdot\vert j + 1,m;j,1\rangle = -j \,\vert j + 1,m;j,1\rangle, \qquad \\
Q\cdot\vert j,m;j,1\rangle = \vert j,m;j,1\rangle, \qquad \\
Q\cdot\vert j - 1,m;j,1\rangle = \left(j+1\right)\vert j - 1,m;j,1\rangle. \qquad
\end{IEEEeqnarray}
By further inserting the eigenstates \eqref{TotalAngularMomentum2a}--\eqref{TotalAngularMomentum2c} into the formulae \eqref{AngularEigenvalueProblem4} for the projectors $P$, $R_{\pm}$, we find that the eigenstates $\vert j,m;j,1\rangle$ span the subspace of the projection operator $P$, while the eigenstates $\vert j \pm 1,m;j,1\rangle$ span the subspaces of the projection operators $R_{\pm}$:
\begin{IEEEeqnarray}{c}
\vert P\rangle = \vert j,m;j,1\rangle, \qquad \vert \pm \rangle = \vert j \pm 1,m;j,1\rangle. \qquad
\end{IEEEeqnarray}
\paragraph{Gauss law constraint (LO)} To prove the statement made above that, if the Gauss law \eqref{GaussLaw2} can be shown to hold at one particular instant in time (say at $\tau = 0$), then it will hold at all times $\tau$, we insert the ansatz $x_i^0 = \mu u_0 e_i$ and the perturbation $\delta x_i(0)$ in \eqref{AngularPerturbations2a}, into the Gauss law constraint equation \eqref{GaussLaw2}. Then, by applying the property of spherical harmonics \eqref{SphericalHarmonics1} we get, at time $t=0$:
\begin{IEEEeqnarray}{c}
\sum_{j,m}\dot{\eta}_i^{jm}\left\{Y_{jm}, e_i\right\} = \sum_{j,m,m'}\dot{\eta}_i^{jm} \left(J_i\right)_{m'm}^{(j)} Y_{jm'} = 0 \Rightarrow \sum_{m}\dot{\eta}_i^{jm} \left(J_i\right)_{m'm}^{(j)} = 0, \label{CoplanarityConstraint1}
\end{IEEEeqnarray}
since the $Y_{jm}$'s form an orthonormal basis. Multiplying both sides of \eqref{CoplanarityConstraint1} with $J_k^{(j)}$ and using the definition of the projection operator $P$ in \eqref{AngularEigenvalueProblem1} we are led to,
\begin{IEEEeqnarray}{c}
\sum_{m, m'}\dot{\eta}_i^{jm} \left(J_k\right)_{m''m'}^{(j)}\left(J_i\right)_{m'm}^{(j)} = 0 \Rightarrow \sum_{m'} \left(P_{ik}\right)_{mm'}^{(j)} \dot{\eta}_k^{jm'} = 0. \label{ConstraintFirstOrder}
\end{IEEEeqnarray}
This formula constrains the general form of leading order perturbations $\eta_i^{jm}$ in \eqref{GeneralSolution1a}. Although the functions $\eta_i^{jm}$ generally span all three orthogonal subspaces $P, \ R_{\pm}$, the Gauss law constraint \eqref{ConstraintFirstOrder} forces them to live exclusively in the sectors $R_{\pm}$ at any one of the three fixed points $\left(\textbf{u}_0, \textbf{u}_{1/6}, \textbf{u}_{1/3}\right)$ in \eqref{So3Extrema1}. On the other hand, the $\eta_i^{jm}$ receive extra contributions at the critical points $\textbf{u}_{1/6}$ and $\textbf{u}_{1/3}$ from the vanishing eigenstate $\xi_P$ of the $P$-sector. \\[6pt]
\indent For all values of the angular momentum $j$, the linear system in $2j+1$ equations and unknowns $\dot{\eta}_{i}^{jm}\left(0\right)$, which is specified by \eqref{CoplanarityConstraint1}--\eqref{ConstraintFirstOrder}, always has a solution which can be expressed as a superposition of the eigenvectors $\boldsymbol{\xi}_1$ which solve the eigenvalue/eigenvector problem \eqref{AngularPerturbations6}. It then follows that the Gauss-law constraint \eqref{GaussLaw2} is always satisfied at time $t = 0$. Then, the perturbation equations \eqref{PerturbationEquation1} imply the validity of the constraint at any time $t$.
\subsubsection[\texorpdfstring{$SO(3)\times SO(6)$ sector}{$SO(3)xSO(6)$ sector}]{$SO(3)\times SO(6)$ sector \label{SubSubSection:AngularPerturbations2}}
\noindent We are now ready to examine the angular stability of a specific axially symmetric configuration in $SO\left(3\right)\times SO\left(6\right)$ that has the general form \eqref{Ansatz3}. The response of this system to angular perturbations will shed light on the generic behavior of axially symmetric configurations in plane-wave backgrounds of the form \eqref{MaximallySupersymmetricBackground1}--\eqref{MaximallySupersymmetricBackground2}. Let us first outline the setup. Inserting the axially symmetric ansatz \eqref{Ansatz3} into the full (type III) system of equations \eqref{EquationsOfMotion1}--\eqref{EquationsOfMotion3}, we are led to the following equations of motion:
\begin{IEEEeqnarray}{lll}
\ddot{u}_1 + \left(2u^2 + 2v^2 + \frac{1}{9}\right) u_1 = u_2 u_3, \qquad &\ddot{v}_i + \left(2u^2 + 2v^2 + \frac{1}{36}\right) v_i = 0, \qquad & i = 1,2 \qquad \label{AxiallySymmetricMembrane2a} \\[6pt]
\ddot{u}_2 + \left(2u^2 + 2v^2 + \frac{1}{9}\right) u_2 = u_3 u_1, \qquad &\ddot{v}_j + \left(2u^2 + 2v^2 + \frac{1}{36}\right) v_j = 0, \qquad & j = 3,4 \qquad \label{AxiallySymmetricMembrane2b} \\[6pt]
\ddot{u}_3 + \left(2u^2 + 2v^2 + \frac{1}{9}\right) u_3 = u_1 u_2, \qquad &\ddot{v}_k + \left(2u^2 + 2v^2 + \frac{1}{36}\right) v_k = 0, \qquad & k = 5,6, \qquad \label{AxiallySymmetricMembrane2c}
\end{IEEEeqnarray}
where we have used the dimensionless variables \eqref{Dimensionless} and $t \equiv \mu\tau$. As we have already remarked in section \ref{SubSection:AxiallySymmetricMembrane}, the non-static ($\ell \neq 0$) axially symmetric membrane \eqref{Ansatz3} can only be consistent with an $s_1 = s_2 = s_3 = 2$ split of the six $SO\left(6\right)$ coordinates $v_j$ in \eqref{SphericalAnsatz1y}--\eqref{SphericalAnsatz3y}. Having said that, we assume:
\begin{IEEEeqnarray}{ll}
\ell \neq 0, \qquad v = \sqrt{v_1^2 + v_2^2} = \sqrt{v_3^2 + v_4^2} = \sqrt{v_5^2 + v_6^2}. \qquad \label{Ansatz4}
\end{IEEEeqnarray}
Evidently, all the other $s_1 = s_2 = s_3 = 2$ possibilities are related to \eqref{Ansatz4} by a simple renaming of the $v$-coordinates.\footnote{The other six partitions of $s_1 + s_2 + s_3 = 6$ (i.e.\ $\left(3,3,0\right),\ \left(3,2,1\right),\ \left(4,2,0\right),\ \left(4,1,1\right),\ \left(5,1,0\right)$ and $\left(6,0,0\right)$) are only legitimate if $\ell_y = 0$, since one needs at least two nonzero coordinates in every direction $e_i$ in order to have non-vanishing angular momentum. As we have seen in section \ref{SubSection:AxiallySymmetricMembrane}, the $\ell_y = 0$ case reduces to the static and spherically symmetric dielectric membrane in $SO\left(3\right)$ ($u_1 = u_2 = u_3 = u_0$) that we studied in sections \ref{SubSection:StaticDielectricMembrane}, \ref{SubSubSection:RadialPerturbations1} and \ref{SubSubSection:AngularPerturbations1}, above.} \\[6pt]
\indent As in the previous subsection, we can only work out the angular spectrum of a specific solution (that is of type III, with an $s_1 = s_2 = s_3 = 2$ split of coordinates $v$) of the membrane equations of motion \eqref{AxiallySymmetricMembrane2a}--\eqref{AxiallySymmetricMembrane2c}. The axially symmetric solution that we are going to consider is \eqref{AnsatzSO3xSO6}, evaluated at the critical points $(u_0, v_0)$ of the axially symmetric potential \eqref{AxiallySymmetricPotential2}:
\begin{IEEEeqnarray}{lll}
u_i^0 = u_0, \quad i = 1,2,3, \qquad & v_j^0\left(t\right) = v_0 \cos\left(\omega t + \varphi_j\right), \quad & j = 1,3,5 \label{DielectricTop1a} \\[6pt]
& v_k^0\left(t\right) = v_0 \sin\left(\omega t + \varphi_k\right), \quad & k = 2,4,6, \label{DielectricTop1b}
\end{IEEEeqnarray}
albeit with a somewhat different split of coordinates.\footnote{That is $j = 1, 3, 5$ and $k = 2, 4, 6$, instead of $j = 1, 2, 3$ and $k = 4, 5, 6$, which was the case in \eqref{AnsatzSO3xSO6}.} As we have already shown in section \ref{SubSection:AxiallySymmetricMembrane}, the extrema $(u_0, v_0)$ satisfy \eqref{ExtremalBounds1},
\begin{IEEEeqnarray}{ll}
\frac{1}{6} \leq u_0 \leq \frac{1}{3} \quad \& \quad 0 \leq v_0 \leq \frac{1}{12}, \qquad \label{ExtremalBounds2}
\end{IEEEeqnarray}
as well as \eqref{AngularVelocity},
\begin{IEEEeqnarray}{l}
\omega^2 = 2u_0^2 + 2v_0^2 + \frac{1}{36} = u_0 - \frac{1}{12} \qquad \& \qquad \ell = \omega v_0^2. \qquad
\end{IEEEeqnarray}
Now suppose that we perform the general perturbation
\begin{IEEEeqnarray}{lll}
x_i = x_i^0 + \delta x_i, \quad i = 1, 2, 3 \qquad \& \qquad &y_i = y_i^0 + \delta y_i, \quad &i = 1,\ldots,6, \label{AngularPerturbations7} \qquad
\end{IEEEeqnarray}
to the equations of motion \eqref{xEquation}--\eqref{yEquation} and the Gauss law constraint \eqref{GaussLaw1}. Here are the resulting fluctuation equations:
\begin{IEEEeqnarray}{ll}
\delta\ddot{x}_i = &\left\{\left\{\delta x_i,x_j^0\right\},x_j^0\right\} + \left\{\left\{x_i^0,\delta x_j\right\},x_j^0\right\} + \left\{\left\{x_i^0,x_j^0\right\},\delta x_j\right\} + \left\{\left\{\delta x_i,y_j^0\right\},y_j^0\right\} + \left\{\left\{x_i^0,\delta y_j\right\},y_j^0\right\} + \nonumber \\[6pt]
& + \left\{\left\{x_i^0,y_j^0\right\},\delta y_j\right\} - \frac{\mu^2}{9}\,\delta x_i^0 + \mu\epsilon_{ijk}\left\{\delta x_j,x_k^0\right\} \\[12pt]
\delta\ddot{y}_i = &\left\{\left\{\delta y_i,y_j^0\right\},y_j^0\right\} + \left\{\left\{y_i^0,\delta y_j\right\},y_j^0\right\} + \left\{\left\{y_i^0,y_j^0\right\},\delta y_j\right\} + \left\{\left\{\delta y_i,x_j^0\right\},x_j^0\right\} + \left\{\left\{y_i^0,\delta x_j\right\},x_j^0\right\} + \nonumber \\[6pt]
& + \left\{\left\{y_i^0,x_j^0\right\},\delta x_j\right\} - \frac{\mu^2}{36}\,\delta y_i^0,
\end{IEEEeqnarray}
while the Gauss law becomes:
\begin{IEEEeqnarray}{c}
\left\{\delta\dot{x}_i, x_i^0\right\} + \left\{\dot{x}_i, \delta x_i^0\right\} + \left\{\delta\dot{y}_i, y_i^0\right\} + \left\{\dot{y}_i, \delta y_i^0\right\} = 0. \label{GaussLaw3}
\end{IEEEeqnarray}
Moreover, the form of the axially symmetric solution \eqref{DielectricTop1a}--\eqref{DielectricTop1b}, implies the following identifications:
\begin{IEEEeqnarray}{lll}
x_i^0 = \mu u_0 e_i, \quad i = 1, 2, 3 \qquad \& \qquad &y_i^0 = \mu v_i^0\left(t\right) e_1, \quad &i = 1,2 \qquad \\[6pt]
&y_j^0 = \mu v_j^0\left(t\right) e_2, \quad &j = 3,4 \qquad \\[6pt]
&y_k^0 = \mu v_k^0\left(t\right) e_3, \quad &k = 5,6. \qquad
\end{IEEEeqnarray}
To obtain the angular spectrum of the membrane \eqref{DielectricTop1a}--\eqref{DielectricTop1b}, we expand the perturbations $\delta x$ and $\delta y$ in spherical harmonics:
\begin{IEEEeqnarray}{lll}
\delta x_i = \mu\cdot\sum_{j,m}\eta_i^{jm}\left(\tau\right) Y_{jm}\left(\theta,\phi\right), \qquad &\eta_i^{jm}\left(0\right) = 0, \qquad &i = 1,2,3 \label{AngularPerturbations8a} \\[6pt]
\delta y_k = \mu\cdot\sum_{j,m}\epsilon_k^{jm}\left(\tau\right) Y_{jm}\left(\theta,\phi\right), \qquad &\epsilon_i^{jm}\left(0\right) = 0, \qquad &k = 1,3,5 \label{AngularPerturbations8b} \\[6pt]
\delta y_l = \mu\cdot\sum_{j,m}\zeta_l^{jm}\left(\tau\right) Y_{jm}\left(\theta,\phi\right), \qquad &\zeta_i^{jm}\left(0\right) = 0, \qquad &l = 2,4,6. \label{AngularPerturbations8c}
\end{IEEEeqnarray}
\indent As before, it can be shown that the Gauss-law constraint \eqref{GaussLaw2} will be satisfied by the perturbation \eqref{AngularPerturbations8a}--\eqref{AngularPerturbations8c} at all times. The fluctuation modes $\eta_i$, $\epsilon_i$ and $\zeta_i$ (we have omitted the indices $j,m$) satisfy the following equations of motion:
\begin{IEEEeqnarray}{l}
\ddot\eta_i + \omega_3^2 \eta_i = u_0 T_{ik}\left(u_0\eta_k + v_0\cos\left(\omega t + \varphi_k\right)\epsilon_k + v_0\sin\left(\omega t + \varphi_k\right)\zeta_k\right) + u_0 Q_{ik} \eta_k \label{AngularPerturbations9a} \\[6pt]
\ddot\epsilon_i + \omega_6^2 \epsilon_i = v_0\cos\left(\omega t + \varphi_i\right) T_{ik}\left(u_0\eta_k + v_0\cos\left(\omega t + \varphi_k\right)\epsilon_k + v_0\sin\left(\omega t + \varphi_k\right)\zeta_k\right) \label{AngularPerturbations9b} \\[6pt]
\ddot\zeta_i + \omega_6^2 \zeta_i = v_0\sin\left(\omega t + \varphi_i\right) T_{ik}\left(u_0\eta_k + v_0\cos\left(\omega t + \varphi_k\right)\epsilon_k + v_0\sin\left(\omega t + \varphi_k\right)\zeta_k\right), \label{AngularPerturbations9c}
\end{IEEEeqnarray}
where the definitions of the matrices $T_{ik}$ and $Q_{ik}$ can be found in \eqref{Definitions1}. Note also that all the repeated indices (with the exception of $i$) are summed. In addition,
\begin{IEEEeqnarray}{c}
\omega_3^2 \equiv \left(u_0^2 + v_0^2\right) j\left(j+1\right) + \frac{1}{9}, \qquad \omega_6^2 \equiv \left(u_0^2 + v_0^2\right) j\left(j+1\right) + \frac{1}{36}. \qquad \label{Definitions3}
\end{IEEEeqnarray}
The system of equations \eqref{AngularPerturbations9a}--\eqref{AngularPerturbations9c} can be transformed into a system with constant coefficients by means of the following change of variables:
\begin{IEEEeqnarray}{l}
\theta_i = \epsilon_i \cdot \cos\left(\omega t + \varphi_i\right) + \zeta_i \cdot \sin\left(\omega t + \varphi_i\right) \label{Rotation1} \\[6pt]
\chi_i = -\epsilon_i \cdot \sin\left(\omega t + \varphi_i\right) + \zeta_i \cdot \cos\left(\omega t + \varphi_i\right). \label{Rotation2}
\end{IEEEeqnarray}
The rotation \eqref{Rotation1}--\eqref{Rotation2} leads to the following set of fluctuation equations,
\begin{IEEEeqnarray}{l}
\ddot{H} + \left(\omega_3^2 - u_0^2 \, T - u_0 \, Q\right) \cdot H - u_0 \, v_0 \, T\cdot\Theta = 0 \\[6pt]
\ddot{\Theta} - 2\Omega\cdot\dot{X} + \left(\omega_6^2 - \Omega^2 - v_0^2 \, T\right)\Theta - u_0 \, v_0 \, T\cdot H = 0 \\[6pt]
\ddot{X} + 2\Omega\cdot\dot{\Theta} + \left(\omega_6^2 - \Omega^2\right)\cdot X = 0,
\end{IEEEeqnarray}
which we have expressed in a compact form by making the identifications $H \equiv \left(\eta_i\right)$, $\Theta \equiv \left(\theta_i\right)$, $X \equiv \left(\chi_i\right)$ and $\Omega \equiv \left(\omega\delta_{ik}\right)$, $Q \equiv \left(Q_{ik}\right)$, $T \equiv \left(T_{ik}\right)$. Further setting,
\begin{IEEEeqnarray}{c}
\left[\begin{array}{c} H \\ \Theta \\ X \end{array}\right] = e^{i\lambda t} \left[\begin{array}{c} \boldsymbol{\xi}_1 \\ \boldsymbol{\xi}_2 \\ \boldsymbol{\xi}_3 \end{array}\right], \label{Eigenvalues2}
\end{IEEEeqnarray}
we obtain the following eigenvalue/eigenvector problem:
\begin{IEEEeqnarray}{l}
\left(u_0^2\,T + u_0 Q\right)\boldsymbol{\xi}_1 + u_0 v_0\,T\cdot\boldsymbol{\xi}_2 = \left(\omega_3^2 - \lambda^2\right)\boldsymbol{\xi}_1 \label{AngularEigenvalueProblem5a} \\[6pt]
v_0^2\,T\cdot\boldsymbol{\xi}_2 = \left(\omega_6^2 - \Omega^2 - \lambda^2\right)\boldsymbol{\xi}_2 - 2i\lambda\,\Omega\cdot\boldsymbol{\xi}_3 - u_0 v_0\,T\cdot\boldsymbol{\xi}_1 \label{AngularEigenvalueProblem5b} \\[6pt]
2i\lambda\,\Omega\cdot\boldsymbol{\xi}_2 = \left(\lambda^2 - \omega_6^2 + \Omega^2\right)\boldsymbol{\xi}_3. \label{AngularEigenvalueProblem5c}
\end{IEEEeqnarray}
\indent Following the same procedure that we followed in section \ref{SubSection:AngularPerturbations} above for the $SO(3)$ symmetric membrane, we write out the matrices $T$, $Q$ and $I$ as a function of the 3 projectors $P$, $R_{\pm}$ in \eqref{AngularEigenvalueProblem1}. Then, the eigenvalue/eigenvector problem \eqref{AngularEigenvalueProblem5a}--\eqref{AngularEigenvalueProblem5c} becomes:
\begin{IEEEeqnarray}{c}
\left(\mathbb{A}_P \otimes P + \mathbb{A}_+ \otimes R_+ + \mathbb{A}_- \otimes R_-\right) \cdot \left[\begin{array}{c} \boldsymbol{\xi}_1 \\ \boldsymbol{\xi}_2 \\ \boldsymbol{\xi}_3 \end{array}\right] = 0, \qquad \label{AngularEigenvalueProblem6}
\end{IEEEeqnarray}
where we have defined,
\footnotesize\begin{IEEEeqnarray}{l}
\mathbb{A}_P = \left(\begin{array}{ccc}
\lambda^2 + s\left(u_0^2 - \frac{u_0}{2} + \frac{1}{18}\right) & s \, u_0 v_0 & 0 \\
s \, u_0 v_0 & \lambda^2 - s \, u_0^2 & 2 i \lambda \omega \\
0 & -2 i \lambda \omega & \lambda^2 - \frac{s}{2}\left(u_0 - \frac{1}{9}\right)
\end{array}\right), \qquad s \equiv j(j+1) - 2 \\[12pt]
\mathbb{A}_+ = \left(\begin{array}{ccc}
\lambda^2 + 2 j u_0^2 - j(j + 3) \frac{u_0}{2} + \frac{s}{18} & 2 j u_0 v_0 & 0 \\
2 j u_0 v_0 & \lambda^2 - 2 j u_0^2 - \frac{1}{2}\left(j^2 - j -2\right)\left(u_0 - \frac{1}{9}\right) & 2 i \lambda \omega \\
0 & -2 i \lambda \omega & \lambda^2 - \frac{s}{2}\left(u_0 - \frac{1}{9}\right)\end{array}\right) \\[12pt]
\mathbb{A}_- = \left(\begin{array}{ccc}
\lambda^2 - 2(j + 1)u_0^2 - [j(j-1) - 2]\frac{u_0}{2} + \frac{s}{18} & -2(j+ 1) u_0 v_0 & 0 \\
-2(j+ 1)u_0 v_0 & \lambda^2 + 2(j + 1)u_0^2 - \frac{1}{2}j(j+ 3)\left(u_0 - \frac{1}{9}\right) & 2 i \lambda \omega \\
0 & -2 i \lambda \omega & \lambda^2 - \frac{s}{2}\left(u_0 - \frac{1}{9}\right)\end{array}\right). \qquad
\end{IEEEeqnarray}\normalsize
\indent We observe that the matrices $\mathbb{A}_+$ and $\mathbb{A}_-$ transform into one another under the involution $j \mapsto - j - 1$, while $\mathbb{A}_P$ remains invariant under the same transformation. \eqref{AngularEigenvalueProblem6} then implies the following three eigenvalue/eigenvector problems, one for each subspace $P$ and $R_{\pm}$:
\begin{IEEEeqnarray}{c}
(\mathbb{A}_P \otimes P) \left[\begin{array}{c} \boldsymbol{\xi}_1 \\ \boldsymbol{\xi}_2 \\ \boldsymbol{\xi}_3 \end{array}\right]_P = (\mathbb{A}_{\pm} \otimes R_{\pm}) \left[\begin{array}{c} \boldsymbol{\xi}_1 \\ \boldsymbol{\xi}_2 \\ \boldsymbol{\xi}_3 \end{array}\right]_{\pm} = 0. \qquad \quad \label{AngularEigenvalueProblem7}
\end{IEEEeqnarray}
Taking into account the fact that the determinants of the projection operators $P$ and $R_{\pm}$ are equal to unity, and by using the following property for the determinant of Kronecker products,
\begin{IEEEeqnarray}{c}
\det\left(A \otimes B\right) = \left(\det A\right)^{\dim B}\left(\det B\right)^{\dim A}, \qquad
\end{IEEEeqnarray}
we convert the three eigenvalue/eigenvector problems in \eqref{AngularEigenvalueProblem7} into eigenvalue/eigenvector problems for the matrices $\mathbb{A}_P$, $\mathbb{A}_{\pm}$. The degeneracies of the corresponding eigenvalues are the same as in the $SO(3)$ case, i.e.\ $d_P = 2j+1$, $d_+ = 2j+3$, $d_- = 2j-1$. This way, the initial eigenvalue/eigenvector problem \eqref{AngularEigenvalueProblem5a}--\eqref{AngularEigenvalueProblem5c} gets reduced to three new (uncommon) eigenproblems for the matrices $\mathbb{A}_P$, $\mathbb{A}_{\pm}$. Similar eigenproblems, where the eigenvalues $\lambda$ (total $6(2j+1)$ in our case) show up in the non-diagonal parts of matrices ($\mathbb{A}_P$ and $\mathbb{A}_{\pm}$ in our case) have been found in relation to characteristic polynomials with matrix coefficients, see e.g.\ \cite{Schwarz00a, CerchiaiZumino00}. \\[-6pt]

\indent Here are the results for the spectrum of the matrices $\mathbb{A}_P$ and $\mathbb{A}_{\pm}$. For $\mathbb{A}_P$ we find that exactly one eigenvalue is always zero while the remaining two eigenvalues are given by a closed formula:
\small\begin{IEEEeqnarray}{lll}
\lambda_{P}^2 = \frac{1}{2}\left(j^2 + j + 2\right)u_0 - \frac{1}{18}\left(1 + j\left(j + 1\right) \pm 3\sqrt{144\left(j^2 + j - 2\right)u_0^3 - 12\left(j^2 + j - 14\right)u_0^2 - 24u_0 + 1}\right). \qquad \quad \label{LyapunovExponents1}
\end{IEEEeqnarray}\normalsize
Setting $j = 1$ in \eqref{LyapunovExponents1}, we find that the minus-signed eigenvalue $\lambda_{P(-)}$ vanishes. All the remaining eigenvalues $\lambda_{P(\pm)}^2$ in \eqref{LyapunovExponents1} are greater than zero inside the interval \eqref{ExtremalBounds2} ($1/6 < u_0 < 1/3$), for all values of the angular momentum $j$. Therefore these eigenvalues correspond to stable directions in the membrane spectrum. \\[6pt]
\indent Let us now discuss the spectra of $\mathbb{A}_{\pm}$. Because $\mathbb{A}_{\pm}$ map into one another with the involution $j \leftrightarrow -j-1$, the corresponding spectra will also be related by the same transformation. As it turns out however, the eigenvalues of $\mathbb{A}_{\pm}$ are given by rather involved expressions of $j$ and $u_0$. We can still study them by examining their characteristic polynomials. The characteristic equation of $\mathbb{A}_{\pm}$ is
\begin{IEEEeqnarray}{lll}
x^3 + \aaa_{\pm} x^2 + \bbb_{\pm} x + \ccc_{\pm} = 0, \qquad x \equiv \lambda^2, \qquad \label{CharacteristicEquation}
\end{IEEEeqnarray}
where $\aaa_{\pm}, \ \bbb_{\pm}, \ \ccc_{\pm}$ are some real polynomial functions of $u_0$ and $j$ (too complicated to be included here) which satisfy
\begin{IEEEeqnarray}{lll}
\aaa_{\pm} < 0, \qquad \bbb_{\pm} > 0, \qquad \ccc_{\pm} < 0,
\end{IEEEeqnarray}
for all $j \geq 3$ inside the interval \eqref{ExtremalBounds2}. Then, by the Descartes rule of signs for polynomial functions, the roots of \eqref{CharacteristicEquation} will either all be real and greater than zero, or one root will be real and greater than zero and the other two roots will be complex for any $j \geq 3$ in \eqref{ExtremalBounds2}. If we further take into account the sign of the discriminant of the characteristic polynomial \eqref{CharacteristicEquation},
\begin{IEEEeqnarray}{lll}
\Delta = 18\aaa \bbb \ccc - 4\aaa^3 \ccc + \aaa^2 \bbb^2 - 4\bbb^3 - 27\ccc^2, \qquad
\end{IEEEeqnarray}
which is greater than zero for all $j \geq 1$ inside the interval \eqref{ExtremalBounds2}, we conclude that the equation \eqref{CharacteristicEquation} can only possess three real positive roots, for $j \geq 3$ in \eqref{ExtremalBounds2}. Therefore, for $j \geq 3$, the critical points $\left(u_0,v_0\right)$ of the potential \eqref{AxiallySymmetricPotential2} are stable under angular perturbations. We would obtain the same result if we had instead examined the monotonicity of each term of the characteristic polynomial \eqref{CharacteristicEquation}, by also taking into account the relative positions of its extremal points (maxima/minima) on the real axis. \\[6pt]
\indent Let us now examine the multipole stability of the axially symmetric solution \eqref{DielectricTop1a}--\eqref{DielectricTop1b} for $j = 1, 2$. The $j = 1$ case turns out to be related to the radial spectrum of the same configuration which was studied in section \ref{SubSubSection:AngularPerturbations2} (more precisely, in appendix \ref{Appendix:RadialPerturbations}). For $j = 1$ we find that one of the eigenvalues of $\mathbb{A}_{\pm}$ is always zero, while the other two are identical to the ones we found in the spectrum of radial perturbations, cf.\ \eqref{AxiallySymmetricEigenvalues1}:\footnote{Even better, see \eqref{AxiallySymmetricEigenvalues2a}--\eqref{AxiallySymmetricEigenvalues2b} in appendix \ref{Appendix:RadialPerturbations}. The overall sign difference is due to the fact that in \eqref{Eigenvalues2} we are using $e^{i\lambda t}$, while in \eqref{Eigenvalues1} and \eqref{Eigenvalues3} we used $e^{\lambda t}$.}\\[-20pt]

\small\begin{IEEEeqnarray}{l}
\lambda_{+}^2 = \frac{5u_0}{2} - \frac{1}{9} \pm \sqrt{\frac{1}{9^2} - \frac{u_0}{9} - \frac{5u_0^2}{12} + 4 u_0^3}, \qquad
\lambda_{-}^2 = \frac{5u_0}{2} - \frac{5}{18} \pm \sqrt{\frac{5^2}{18^2} - \frac{35u_0}{18} + \frac{163u_0^2}{12} - 20u_0^3}. \qquad \label{LyapunovExponents2}
\end{IEEEeqnarray}\normalsize
As we have already discussed in section \ref{SubSubSection:AngularPerturbations2} (as well as in appendix \ref{Appendix:RadialPerturbations}), the eigenvalues \eqref{LyapunovExponents2} are all greater than zero inside the interval \eqref{ExtremalBounds2} and so they correspond to stable directions of the membrane. An exception is the minus-signed eigenvalue $\lambda_{-(-)}^2$ which is only greater than zero for $u_{\text{crit}} \approx 0.25971 < u_0 < 1/3$ and so it is stable inside this interval and unstable outside it ($1/6 < u_0 < u_{\text{crit}} \approx 0.25971$). See also the graphs in figures \ref{Graph:Eigenvalues1} and \ref{Graph:Eigenvalues3}.
\begin{figure}[H]
\begin{center}
\includegraphics[scale=0.4]{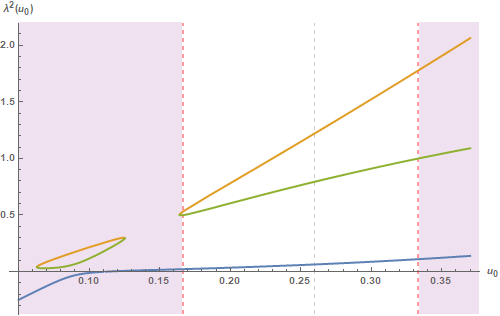} \qquad \includegraphics[scale=0.4]{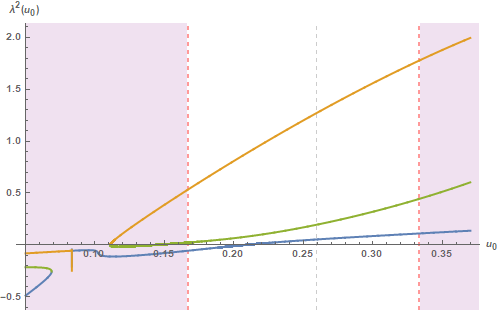}
\caption{Plots of the squares of the $j = 2$ eigenvalues of $\mathbb{A}_{\pm}$ in terms of the coordinate $u_0$.} \label{Graph:Eigenvalues2}
\end{center}
\end{figure}
\vspace{-0.4cm}\indent For $j = 2$ we can write down analytic (albeit too involved to include here) formulas for the roots of the cubic equation \eqref{CharacteristicEquation}. By plotting the squares of the eigenvalues of $\mathbb{A}_{\pm}$ as functions of $u_0$ (see figure \ref{Graph:Eigenvalues2} above), we see that all the eigenvalues of $\mathbb{A}_+$ are greater than zero inside the interval \eqref{ExtremalBounds2}. Therefore, the corresponding modes of the membrane are stable. Conversely, there is one eigenvalue of $\mathbb{A}_-$ which becomes less than zero inside the interval \eqref{ExtremalBounds2},
\begin{IEEEeqnarray}{lll}
\frac{1}{6} \leq u_0 \leq 0.207245 < u_{\text{crit}} \approx 0.25971.
\end{IEEEeqnarray}
This eigenvalue naturally corresponds to an unstable mode of the axially symmetric configuration \eqref{DielectricTop1a}--\eqref{DielectricTop1b}. We observe that the domain of instability shrinks as the angular momentum is increased from $j=1$ to $j=2$. There are no instabilities for larger values of the angular momentum (i.e.\ for $j \geq 3$), as we have already described. Note however that this can only by true at the linear level, as we will see in the following section. \\[6pt]
\indent As in the $SO(3)$ case above (cf.\ section \ref{SubSubSection:AngularPerturbations1}), the linearized Gauss law constraint \eqref{GaussLaw3} will be valid at all times if it holds at any given moment (e.g.\ at $t=0$). In the rotated coordinate frame of $\theta$ and $\chi$ (as defined in \eqref{Rotation1}--\eqref{Rotation2}), we proceed to eliminate time from \eqref{GaussLaw3}. We obtain a linear system,
\begin{IEEEeqnarray}{ll}
\sum_{m}\sum_{i=1}^3 \left(J_i\right)_{m'm}\left(u_0\dot{\eta}_{i}^{jm}\left(0\right) + v_0\dot{\theta}_{i}^{jm}\left(0\right)\right) = 0, \qquad \label{CoplanarityConstraint2}
\end{IEEEeqnarray}
which is very similar to the system \eqref{CoplanarityConstraint1} which we obtained in the $SO(3)$ case (see section \ref{SubSubSection:AngularPerturbations1}). Therefore, and by the same token as in section \ref{SubSubSection:AngularPerturbations1}, the linear system of equations \eqref{CoplanarityConstraint2} will always have a unique solution which will be expressible as a superposition of the three eigenvectors $\boldsymbol{\xi}_1$, $\boldsymbol{\xi}_2$ and $\boldsymbol{\xi}_3$ in \eqref{AngularEigenvalueProblem7}. As a result, the Gauss law constraint \eqref{GaussLaw3} will always be satisfied at $t = 0$ and so it will always be satisfied at any time $t$. \\[6pt]
\indent To sum up, the angular spectrum of the axially symmetric configuration \eqref{DielectricTop1a}--\eqref{DielectricTop1b} always includes one zero eigenvalue (that is a zero mode of the matrix $\mathbb{A}_P$). For $j = 1$, there is a second zero eigenvalue in the spectrum of $\mathbb{A}_P$, namely $\lambda_{P(-)}$ in \eqref{LyapunovExponents1}. For $j =1,2$ the membrane \eqref{DielectricTop1a}--\eqref{DielectricTop1b} has exactly one unstable mode inside the intervals $1/6 < u_0 < u_{\text{crit}} \approx 0.25971$ (for $j = 1$) and $1/6 < u_0 < 0.207245 < u_{\text{crit}}$ (for $j=2$). These instabilities show up as purely imaginary eigenvalues in the spectrum of $\mathbb{A}_-$ (more precisely, for $j = 1$, the unstable eigenvalue is given by $\lambda_{-(-)}$ in \eqref{LyapunovExponents2}). For $j \geq 3$ (and except one zero eigenvalue in the spectrum of $\mathbb{A}_P$) every eigenvalue of the matrices $\mathbb{A}_P$ and $\mathbb{A}_{\pm}$ is purely real; therefore the system is stable for $j \geq 3$.
\vspace{-.4cm}\begin{table}[H]
\begin{center}
\begin{eqnarray}
\begin{array}{|c|c|c|c|c|}
\hline &&&& \\
\text{eigenvalues} & j=1 & j=2 & j \geq 3 & \text{stability} \\[6pt]
\hline &&&& \\
\lambda_P^2 & 0,0,+ & 0,+,+ & 0,+,+ & \text{stable} \\[12pt]
\lambda_+^2 & 0,+,+ & +,+,+ & +,+,+ & \text{stable} \\[12pt]
\lambda_-^2 & 0,+,+ & +,+,+ & +,+,+ & \text{stable} \\
& \text{\footnotesize{$\left(u_{\text{crit}} \approx 0.25971 < u_0 \leq 1/3\right)$}} & \text{\footnotesize{$\left(0.207245 < u_{\text{crit}} < u_0 \leq 1/3\right)$}} && \\[6pt]
& 0,+,- & +,+,- && \text{saddle} \\
& \text{\footnotesize{$\left(1/6 \leq u_0 < 0.25971 \approx u_{\text{crit}} \right)$}} & \text{\footnotesize{$\left(1/6 \leq u_0 < 0.207245 < u_{\text{crit}}\right)$}} && \\[3pt]
\hline
\end{array} \nonumber
\end{eqnarray}
\end{center}
\caption{Angular spectrum and stability of the axially symmetric membrane in $SO(3)\times SO(6)$. \label{Table:AngularPerturbations}}
\end{table}
\indent Our findings are summarized in table \ref{Table:AngularPerturbations} above. This table contains the sign of the square of each eigenvalue $\lambda_{P,\pm}^2$ inside the allowed interval \eqref{ExtremalBounds2} $1/6 < u_0 < 1/3$, for all values of the angular momentum $j = 1,2,3, \ldots$.
\section[NLO stability analysis]{NLO stability analysis \label{Section:NLOperturbations}}
\noindent In the present section we study the angular stability of static dielectric membranes in $SO(3)$ (as described by the ansatz \eqref{StaticMembrane1}), at the next-to-leading order (NLO) in perturbation theory. Already, in section \ref{SubSubSection:RadialPerturbations1}, we examined the radial stability of these configurations to leading order (LO) in perturbation theory. We found that, among the 9 critical points \eqref{So3Extrema1} of the effective potential energy \eqref{StaticPotential}, $\textbf{u}_0$ (point-like membranes) and $\textbf{u}_{1/3}$ (Myers dielectric spheres) are stable centers, while the critical points $\textbf{u}_{1/6}$ are saddle points. This way we confirmed our earlier conclusions (section \ref{SubSection:StaticDielectricMembrane}) which were based on the inspection of the Hessian (second derivative) matrix. We also determined the values of the corresponding eigenvalues, along with their degeneracies. Refer to table \ref{Table:RadialPerturbations} for more. \\[6pt]
\indent In section \ref{SubSubSection:AngularPerturbations1} we studied leading order (LO) angular/multipole perturbations for the same configuration. Our conclusions confirmed and extended the ones which were obtained from the LO radial stability analysis of section \ref{SubSubSection:RadialPerturbations1}. Specifically, the $j = 1$ angular spectrum turned out to be identical to the radial spectrum. For the other values of the angular momentum ($j = 1,2,\ldots$) we found that the critical point $\textbf{u}_0$ is always stable, while the critical point $\textbf{u}_{1/3}$ has one zero mode in the $P$-sector (degeneracy $2d_P$), and stable eigenvalues in all the other sectors. The saddle point $\textbf{u}_{1/6}$ was found to contain one zero mode (degeneracy $2d_P$) for every $j = 1,2,\ldots$, and one zero mode for $j=3$ (10-fold degenerate). In the $R_-$ sector, $\textbf{u}_{1/6}$ was found to be unstable for $j=1$ (2-fold degenerate) and $j=2$ (6-fold degenerate).
\subsection[Higher-order perturbations]{Higher-order perturbations}
\noindent Let us now examine the angular/multipole spectrum of the configuration \eqref{StaticMembrane1} at the next-to-leading order (NLO). To proceed, we must generalize the LO series \eqref{AngularPerturbations1} as follows:
\begin{IEEEeqnarray}{l}
x_i = \sum_{n = 0}^{\infty}\varepsilon^n\delta x_i^n = x_i^0 + \sum_{n = 1}^{\infty}\varepsilon^n\delta x_i^n, \quad i = 1, 2, 3 \label{HigherOrderPerturbations1a} \\[6pt]
y_i = \sum_{n = 0}^{\infty}\varepsilon^n\delta y_i^n = y_i^0 + \sum_{n = 1}^{\infty}\varepsilon^n\delta y_i^n, \quad i = 1,\ldots,6. \label{HigherOrderPerturbations1b}
\end{IEEEeqnarray}
We then expand around the spherically symmetric solution \eqref{AngularPerturbations1},
\begin{IEEEeqnarray}{l}
x_i^0 = \mu u_0 e_i, \quad i = 1,2,3 \qquad \& \qquad y_i^0 = 0, \quad i = 1,\ldots,6, \label{SO3solution}
\end{IEEEeqnarray}
where $u_0$ can be any of the critical points \eqref{So3Extrema1}. The $t = 0$ initial conditions for the perturbations $\delta x_i^{(n)}$ and their derivatives are determined from the initial conditions of the full solution \eqref{AngularPerturbations1}. We find,
\begin{IEEEeqnarray}{ll}
\delta x_i^{(1)}(0) \neq 0, \quad \delta \dot{x}_i^{(1)}(0) \neq 0, \qquad \delta x_i^{(n)}(0) = \delta \dot{x}_i^{(n)}(0) = 0, \quad n = 2,3,\ldots, \label{InitialConditionPerturbations}
\end{IEEEeqnarray}
i.e.\ all the initial conditions vanish unless $n = 0,1$. Plugging \eqref{HigherOrderPerturbations1a}--\eqref{HigherOrderPerturbations1b} into the equations of motion \eqref{xEquation}--\eqref{yEquation}, we are led to the following system of equations:
\begin{IEEEeqnarray}{ll}
\delta\ddot{x}_i^n = &\{\{\delta x_i^n,x_k\},x_k\} + \{\{x_i,\delta x_k^n\},x_k\} + \{\{x_i,x_k\},\delta x_k^n\} - \frac{\mu^2}{9}\delta x_i^n + \epsilon_{ikl}\{x_k,\delta x_l^n\} + \nonumber \\[6pt]
& + \sum_{p = 1}^{n-1}\Bigg[\{\{x_i,\delta x_k^{n-p}\},\delta x_k^p\} + \{\{\delta x_i^{n-p}, x_k\},\delta x_k^p\} + \{\{\delta x_i^{n-p},\delta x_k^p\},x_k\} + \frac{\mu}{2}\epsilon_{ikl}\{\delta x_k^{n-p},\delta x_l^p\} + \nonumber \\[6pt]
& \hspace{1.2cm} + \{\{x_i,\delta y_k^{n-p}\},\delta y_k^p\}\Bigg] + \sum_{p = 1}^{n-1} \sum_{q = 1}^{p-1}\Bigg[\{\{\delta x_i^{n-p},\delta x_k^{p-q}\},\delta x_k^q\} + \{\{\delta x_i^{n-p},\delta y_k^{p-q}\},\delta y_k^q\}\Bigg] \qquad \label{HigherOrderPerturbations2a} \\[6pt]
\delta\ddot{y}_i^n = &\{\{\delta y_i^n,x_k\},x_k\} - \frac{\mu^2}{36}\delta y_i^n + \sum_{p = 1}^{n-1}\Bigg[\{\{\delta y_i^{n-p}, x_k\},\delta x_k^p\} + \{\{\delta y_i^{n-p},\delta x_k^p\},x_k\}\Bigg] + \nonumber \\[6pt]
& + \sum_{p = 1}^{n-1} \sum_{q = 1}^{p-1}\Bigg[\{\{\delta y_i^{n-p},\delta y_k^{p-q}\},\delta y_k^q\} + \{\{\delta y_i^{n-p},\delta x_k^{p-q}\},\delta x_k^q\}\Bigg], \qquad \label{HigherOrderPerturbations2b}
\end{IEEEeqnarray}
where we have omitted the superscript from $x_i^0$ in order to keep our expressions simple. To access the multipole spectrum, we expand the perturbations \eqref{HigherOrderPerturbations2a}--\eqref{HigherOrderPerturbations2b} in spherical harmonics as
\begin{IEEEeqnarray}{lll}
\delta x_i^n = \mu\cdot\sum_{j,m}\eta_i^{njm}\left(\tau\right) Y_{jm}\left(\theta,\phi\right), \qquad &\eta_i^{njm}\left(0\right) = 0, \qquad &i = 1,2,3 \label{AngularPerturbations10a} \\[6pt]
\delta y_i^n = \mu\cdot\sum_{j,m}\theta_i^{njm}\left(\tau\right) Y_{jm}\left(\theta,\phi\right), \qquad &\theta_i^{njm}\left(0\right) = 0, \qquad &i = 1,\ldots,6. \label{AngularPerturbations10b}
\end{IEEEeqnarray}
\indent Now, besides the basic property \eqref{SphericalHarmonics1}, spherical harmonics $Y_{jm}\left(\theta,\phi\right)$ obey
\begin{IEEEeqnarray}{c}
\left\{Y_{jm}\left(\theta,\phi\right),Y_{j'm'}\left(\theta',\phi'\right)\right\} = f_{jm,j'm'}^{j''m''} \cdot Y_{j''m''}\left(\theta'',\phi''\right), \label{SphericalHarmonics2}
\end{IEEEeqnarray}
where $f_{\alpha\beta}^{\gamma}$ are the structure constants of area-preserving transformations which leave the Hamiltonian \eqref{ppWaveHamiltonian1}--\eqref{ppWaveHamiltonian2} of the $SO(3)$ membrane \eqref{SO3solution} invariant. Small Greek indices $(\alpha,\beta,\gamma)$ will henceforth be used to group the (total) angular momentum quantum numbers as $\alpha \equiv jm$, $\beta \equiv j'm'$, $\gamma \equiv j''m''$. The structure constants $f_{\alpha\beta}^{\gamma}$ can be computed by means of a closed formula which holds for all values of $\alpha$, $\beta$, $\gamma$. Precise expressions for the $j = 1,2$ structure constants $f_{1m,\beta}^{\gamma}$ and $f_{2m,\beta}^{\gamma}$ can be found in appendix \ref{Appendix:StructureConstants}.
\subsection[Next-to-leading order perturbations]{Next-to-leading order perturbations}
\begin{figure}[H]
\begin{center}
\includegraphics[scale=0.4]{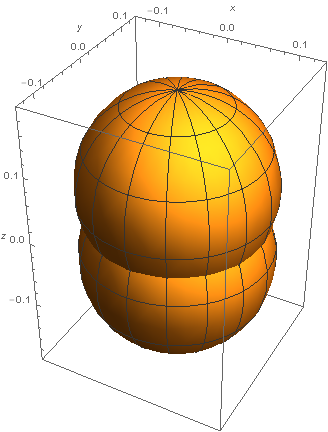}
\end{center}
\caption{Parametric plot of the forcing \eqref{Forcing3}.\label{Figure:ForcingPlot3}}
\end{figure}

\noindent By first setting $n = 2$ in the system \eqref{HigherOrderPerturbations2a}--\eqref{HigherOrderPerturbations2b} of higher-order perturbations and then plugging the $SO(3)$ solution \eqref{SO3solution} and the spherical harmonic expansion \eqref{AngularPerturbations10a}--\eqref{AngularPerturbations10b}, we are led to:\footnote{For $n = 2$, the system \eqref{HigherOrderPerturbations2a}-\eqref{HigherOrderPerturbations2b} of higher-order perturbations becomes:
\begin{IEEEeqnarray}{ll}
\delta\ddot{x}_i^{(2)} = &\{\{\delta x_i^{(2)},x^{(0)}_k\},x^{(0)}_k\} + \{\{x^{(0)}_i,\delta x_k^{(2)}\},x^{(0)}_k\} + \{\{x^{(0)}_i,x^{(0)}_k\},\delta x_k^{(2)}\} - \frac{\mu^2}{9}\delta x_i^{(2)} + \epsilon_{ikl}\{x^{(0)}_k,\delta x_l^{(2)}\} + \nonumber \\
&+ \Bigg[\{\{x^{(0)}_i,\delta x_k^{(1)}\},\delta x_k^{(1)}\} + \{\{\delta x_i^{(1)}, x^{(0)}_k\},\delta x_k^{(1)}\} + \{\{\delta x_i^{(1)},\delta x_k^{(1)}\},x^{(0)}_k\} + \frac{\mu}{2}\epsilon_{ikl}\{\delta x_k^{(1)},\delta x_l^{(1)}\}\Bigg] \qquad \label{HigherOrderPerturbations3a} \\
\delta\ddot{y}_i^{(2)} = &\{\{\delta y_i^{(2)},x^{(0)}_k\},x^{(0)}_k\} - \frac{\mu^2}{36}\delta y_i^{(2)} + \Bigg[\{\{\delta y_i^{(1)}, x^{(0)}_k\},\delta x_k^{(1)}\} + \{\{\delta y_i^{(1)},\delta x_k^{(1)}\},x^{(0)}_k\}\Bigg]. \qquad \label{HigherOrderPerturbations3b}
\end{IEEEeqnarray}}
\begin{IEEEeqnarray}{c}
\ddot\eta_i^{(2)} + \omega_3^2 \eta_i^{(2)} = u_0^2 T_{ik} \eta_k^{(2)} + u_0 Q_{ik} \eta_k^{(2)} + F_i^{(2)} \qquad \& \qquad \ddot\theta_i^{(2)} + \omega_6^2 \theta_i^{(2)} = G_i^{(2)}, \label{AngularPerturbations11}
\end{IEEEeqnarray}
where we have suppressed the indices $j,m$, as usual. Notice that the system \eqref{AngularPerturbations11} is just the first order ($n = 1$) system of equations \eqref{AngularPerturbations3}, driven by the forcing terms $F_i, \ G_i$. The latter are given by:
\begin{IEEEeqnarray}{ll}
F_i^{2\gamma} = &-i u\left[f_{j\dot{m},\beta}^{\gamma}\left(\eta_k^{1\alpha}\eta_k^{1\beta} + \theta_k^{1\alpha}\theta_k^{1\beta}\right)\left(J_i\right)^{(j)}_{\dot{m}m} - \eta_i^{1\alpha}\eta_k^{1\beta}\left(f_{j\dot{m},\beta}^{\gamma}\left(J_k\right)^{(j)}_{\dot{m}m} + f_{\alpha\beta}^{j''\dot{m}}\left(J_k\right)^{(j'')}_{m''\dot{m}}\right)\right] + \nonumber \\
& + \frac{1}{2}\epsilon_{ikl}f_{\alpha\beta}^{\gamma}\eta_k^{1\alpha}\eta_l^{1\beta} \label{Forcing1a} \\[6pt]
G_i^{2c} = &i u\left[f_{j\dot{m},\beta}^{\gamma}\left(J_k\right)^{(j)}_{\dot{m}m} + f_{\alpha\beta}^{j''\dot{m}}\left(J_k\right)^{(j'')}_{m''\dot{m}}\right]\theta_i^{1\alpha}\eta_k^{1\beta}, \label{Forcing1b}
\end{IEEEeqnarray}
where again $\alpha \equiv jm$, $\beta \equiv j'm'$, $\gamma \equiv j''m''$. As before, we have switched to dimensionless time $t \equiv \mu\tau$ and have employed the Einstein convention for the repeated indices. In addition, all time dependencies have been made implicit in \eqref{AngularPerturbations11}--\eqref{Forcing1b}, instead of explicit. \\[6pt]
\indent For what follows, we will just focus on the $SO(3)$ modes $\eta_i$, by neglecting the dynamics that is due to the $SO(6)$ modes $\theta_i$. In other words, we set $\theta_i^{(n)} = 0$ in \eqref{AngularPerturbations11}--\eqref{Forcing1b}. The forcing term $F_i^{2\gamma}$ in \eqref{Forcing1a} is given by the following bilinear form,
\begin{IEEEeqnarray}{l}
F_i^{2\gamma}\left(t\right) = \eta_{k}^{1\alpha} K_{ikl;\alpha\beta}^{\gamma} \eta_{l}^{1\beta}, \label{Forcing2}
\end{IEEEeqnarray}
where we have left out the sums over the spatial indices $k,l$ and the sums over the spin indices $\alpha,\beta$. For simplicity, we are also writing $\eta_k^{1\alpha}$ instead of $\eta_k^{1\alpha}\left(t\right)$. The matrix which corresponds to the bilinear form $K$ in \eqref{Forcing2} reads:
\begin{IEEEeqnarray}{l}
K_{ikl;\alpha\beta}^{\gamma} = \left(\mathcal{J}_a\right)^{(j)}_{\dot{m}m} f_{j\dot{m},\beta}^{\gamma}\left(\epsilon_{bak}\epsilon_{bil} + \epsilon_{bai}\epsilon_{bkl}\right) + \frac{1}{2}\epsilon_{ikl}f_{\alpha\beta}^{\gamma}, \hspace{.7cm} \label{ForcingOrder2}
\end{IEEEeqnarray}
where $\mathcal{J}_a \equiv - i u_0 J_a$ and we omitted the sums over the spatial indices $a,b$ and the spin index $\dot{m}$. In compact form, the $\eta_i^{(2)}$ equation in \eqref{AngularPerturbations11} and the corresponding forcing term \eqref{Forcing2} can be written as
\begin{IEEEeqnarray}{c}
\ddot{H}^{(2)} + \mathcal{K} \cdot H^{(2)} = F^{(2)}, \quad F^{(2)} \equiv H^{(1)}\,K \, H^{(1)}, \qquad \label{AngularPerturbations12}
\end{IEEEeqnarray}
where most of the conventions that are employed in writing \eqref{AngularPerturbations12} can be found in \eqref{AngularPerturbations4}--\eqref{Definitions2}. \\[6pt]
\indent Let us now see how unstable modes propagate from the LO towards the NLO (and all the higher perturbative orders). First recall that the only instabilities at LO are the $j = 1,2$ modes of the critical point $\textbf{u}_{1/6}$. These instabilities correspond to purely imaginary eigenvalues, for which $\lambda_{-}^2<0$ in the general solution \eqref{GeneralSolution1a}. All the unstable LO modes contribute to the NLO forcing term \eqref{Forcing2}, since there is an explicit summation over the angular momenta $j,j' = 1,2,\ldots$\ As a result, the NLO forcing term \eqref{Forcing2} is capable of injecting unstable modes to the NLO, for all values of the spin $j$. The form of the structure constants $f_{1m,\beta}^{\gamma}$ and $f_{2m,\beta}^{\gamma}$ (cf.\ \eqref{StructureConstants2}--\eqref{StructureConstants5} in appendix \ref{Appendix:StructureConstants}) then indicates that by coupling unstable ($j = 1$) LO modes to stable LO modes $j'$, we can make all $j''= j'$ modes at NLO unstable. In the same vein, by coupling unstable ($j = 2$) LO modes to stable LO modes $j'$, we can make all $j''= j'\pm 1$ modes at NLO unstable. It is then clear that this avalanche/cascade of perturbative instabilities carries over to higher perturbative orders as well. In other words, the fluctuation modes $\eta_i^{(n)}$ (at any given order of perturbation theory $n$) can be destabilized by the LO instabilities of the critical point $\textbf{u}_{1/6}$ at $j = 1,2$. \\[6pt]
\indent Yet another potential source of instabilities emerges whenever the frequency of the external forcing $F_i^{2\gamma}\left(t\right)$ in \eqref{Forcing2}--\eqref{ForcingOrder2} becomes equal to one of the system's natural eigenfrequencies in \eqref{MultipoleEigenvalues1}--\eqref{MultipoleEigenvalues3}. Unlike the cascade of instabilities that we have just described however, these resonances are not in any way related to the instabilities of the critical point $\textbf{u}_{1/6}$ for $j = 1,2$. Instead, these resonances can destabilize stable critical points, e.g.\ the Myers dielectric sphere $\textbf{u}_{1/3}$. Take for example a zero mode in the $P$ sector which couples to another mode in either of the $R_{\pm}$ sectors so that their combined forcing in \eqref{ForcingOrder2} includes at least one of natural eigenfrequencies in \eqref{MultipoleEigenvalues1}--\eqref{MultipoleEigenvalues3}. The system undergoes a frequency resonance which causes the fluctuation amplitude $\eta_i^{(n)}$ to increase. Then the whole system becomes unstable. Of course, a complete discussion of resonances would have to include a thorough analysis of the Gauss law constraint at both the LO and the NLO. The role of these constraints is essential, since they allow for the proper incorporation of symmetries into the solutions of the fluctuation equations (order by order in perturbation theory).
\paragraph{Gauss law constraint (NLO)} As we have already mentioned, the LO Gauss law constraint \eqref{ConstraintFirstOrder} points out that the initial conditions for the velocity of any LO mode $\eta_i^{(1)}$ are orthogonal to the vectors of the $P$ sector. Thus these modes can be written as a superposition of the eigenvectors in the other two sectors, namely $R_{\pm}$. \\[6pt]
\indent On the other hand, the NLO Gauss law constraint \eqref{GaussLaw2} is given by
\begin{equation}
\left\{\dot{x}^{(2)}_i,x_i^{(0)} \right\} + \left\{\dot{x}^{(1)}_i,x_i^{(1)} \right\} = 0,
\end{equation}
because $\dot{x}^{(0)}_i = 0$ in \eqref{SO3solution}. Substituting the values of $x_i^{(0,1)}$, $\dot{x}^{(1,2)}_i$ from \eqref{SO3solution}, \eqref{AngularPerturbations10a} we are led to
\begin{IEEEeqnarray}{l}
u_0 \dot{\eta}_i^{2jm} \left\{Y_{jm},e_i \right\} + \dot{\eta}_i^{1jm}\eta_i^{1j^\prime m^\prime} \left\{Y_{jm},Y_{j^\prime m^\prime} \right\} = 0, \qquad \label{GaussLaw4}
\end{IEEEeqnarray}
where we omit the sums over the repeated indices for simplicity. Computing the Poisson brackets by using the properties of spherical harmonics \eqref{SphericalHarmonics1}, \eqref{SphericalHarmonics2}, we can write the NLO constraint \eqref{GaussLaw4} as
\begin{equation}
i \, u_0 \,\dot{\eta}_i^{2jm} \left(J_i\right)^{(j)}_{m^\prime m} Y_{jm^\prime} + \dot{\eta}_i^{1jm} \eta_i^{1j^\prime m^\prime} f_{jm,j^\prime m^\prime}^{j^{\prime\prime}m^{\prime\prime}} Y_{j^{\prime\prime}m^{\prime\prime}} = 0,
\end{equation}
which obviously implies the following identity
\begin{IEEEeqnarray}{c}
i \, u_0 \, \dot{\eta}_i^{2j^{\prime\prime}m} \left(J_i\right)^{(j^{\prime\prime})}_{m^{\prime\prime}m} + \dot{\eta}_i^{1jm} \eta_i^{1j^\prime m^\prime} f_{jm,j^\prime m^\prime}^{j^{\prime\prime}m^{\prime\prime}} = 0, \qquad \label{ConstraintSecondOrder}
\end{IEEEeqnarray}
after factoring out the spherical harmonics. Similarly to what we did in the case of the LO Gauss constraint in \eqref{CoplanarityConstraint1}--\eqref{ConstraintFirstOrder}, we multiply both sides of \eqref{ConstraintSecondOrder} with $J_k^{(j)}$ and then use the definition of the projection operator $P$ in \eqref{AngularEigenvalueProblem1} to obtain
\begin{IEEEeqnarray}{c}
\left(P_{ik}\right)^{(j^{\prime\prime})}_{m^{\prime\prime\prime}m} \dot{\eta}_i^{2j^{\prime\prime}m} = i \, j^{\prime\prime}\left(j^{\prime\prime} + 1\right) u_0^{-1} \,\dot{\eta}_i^{1jm} \eta_i^{1j^\prime m^\prime} f_{jm,j^\prime m^\prime}^{j^{\prime\prime}m^{\prime\prime}}\left(J_k\right)^{(j^{\prime\prime})}_{m^{\prime\prime\prime}m^{\prime\prime}}. \qquad \label{GaussPcomponent}
\end{IEEEeqnarray}
\indent We conclude that the initial velocity $\dot{\eta}_i^{(2)}$ can be written as a superposition of the eigenvectors in the $P$ and $R_{\pm}$ sectors. The component of $\dot{\eta}_i^{(2)}$ in the $P$ sector is specified by \eqref{GaussPcomponent}. This component depends on the coupling of the initial velocity $\dot{\eta}_i^{(1)}$ to the corresponding position $\eta_i^{(1)}$, for different values of the spin quantum number $j$.
\paragraph{General solution} Let us now write out the generic form of the solution to the system of NLO perturbation equations \eqref{AngularPerturbations12}. Typically, the solution is the sum of the general solution of the homogeneous system \eqref{AngularPerturbations4} at LO and a special solution of the forced NLO system \eqref{AngularPerturbations12}:
\begin{IEEEeqnarray}{c}
H^{(2)}\left(t\right) = H^{(2)}_{\text{h}}\left(t\right) + H^{(2)}_{\text{s}}\left(t\right). \label{GeneralSolution2}
\end{IEEEeqnarray}
The general solution $H^{(2)}_{\text{h}}\left(t\right)$ of the LO homogeneous system \eqref{AngularPerturbations4} is given by \eqref{GeneralSolution1a}. Further setting $\mathcal{K} \equiv \omega_3^2 - u_0^2 T - u_0 Q = \Omega_0^2$, we may write the homogeneous LO solution $H^{(2)}_{\text{h}}\left(t\right)$ in the following form:
\begin{IEEEeqnarray}{c}
H^{(2)}_{\text{h}}\left(t\right) = H^{(2)}_{\text{h}}\left(0\right)\cos\Omega_0 t + \dot{H}^{(2)}_{\text{h}}\left(0\right)\Omega_0^{-1}\sin\Omega_0 t. \qquad \label{HomogeneousSolution}
\end{IEEEeqnarray}
On the other hand, the special NLO solution $H^{(2)}_{\text{s}}\left(t\right)$ can be written as follows:
\begin{IEEEeqnarray}{l}
H^{(2)}_{\text{s}}\left(t\right) = \Omega_0^{-1}\sin\left(\Omega_0 t\right)\int_{0}^{t}ds\cos\left(\Omega_0 s\right)F^{(2)}\left(s\right) - \Omega_0^{-1}\cos\left(\Omega_0 t\right)\int_{0}^{t}ds\sin\left(\Omega_0 s\right)F^{(2)}\left(s\right). \qquad \label{SpecialSolution}
\end{IEEEeqnarray}
We observe that the general form of the full NLO solution \eqref{GeneralSolution2}--\eqref{SpecialSolution} will practically remain the same at all higher orders of perturbation theory $n = 2,3,\ldots$\ The corresponding (higher-order) forcing terms $F^{(n)}(s)$ will generally depend on the solutions of the fluctuation equations in all the previous perturbative orders (i.e.\ $1,2,\ldots,n-1$).
\paragraph{Example} The instability cascade phenomenon which we described above is best exemplified by a simple solution which shows how LO instabilities propagate to NLO. To get things going, let us compute the forcing term $F_i^{2j''m''}$, right at the critical point $u_0 = 1/6$. For simplicity, we turn on only the $(n = 1,\ j = 2, \ m = 0)$ LO mode $\xi^{2,0}_{-} \equiv \xi$. This LO instability will propagate to $j'' = 1$ and $j'' = 3$ at the NLO, as we will see shortly. The LO general solution \eqref{GeneralSolution1a} becomes:
\begin{IEEEeqnarray}{c}
H\left(t\right) = e^{i\lambda_{-}t}\xi_{-}, \qquad \xi_{-} = \xi\cdot\vert - \rangle\Big|_{j=2, m=0}. \qquad
\end{IEEEeqnarray}
We obviously need only the value of the eigenvector $\vert - \rangle$ in \eqref{TotalAngularMomentum2c}, for $j=2$, $m=0$. Then it is just the following components of $\eta_i^{1jm}$ with $j=2$ which are nonzero:
\begin{IEEEeqnarray}{l}
\eta_x^{1,2,\pm1} = \pm\frac{\xi}{2} \, \sqrt{\frac{3}{5}} \cdot e^{t/(3\sqrt{2})}, \qquad \eta_y^{1,2,\pm1} = -\frac{i\xi}{2}\,\sqrt{\frac{3}{5}} \cdot e^{t/(3\sqrt{2})}, \qquad \eta_z^{1,2,0} = -\xi\,\sqrt{\frac{2}{5}} \cdot e^{t/(3\sqrt{2})}. \qquad \label{SimpleSolution}
\end{IEEEeqnarray}
\indent The solution \eqref{SimpleSolution} satisfies, not only the reality condition \eqref{RealityCondition}, but also the leading order Gauss constraint \eqref{ConstraintFirstOrder}. If we insert \eqref{SimpleSolution} into the formulae \eqref{Forcing2}--\eqref{ForcingOrder2} which give the forcing term at NLO, we obtain the following non-vanishing components of $F_i^{2j''m''}\left(t\right)$, for $j'' = 3$:
\begin{IEEEeqnarray}{l}
F_x^{2,3,\pm 1} = \pm\frac{2\xi^2}{5} \sqrt{\frac{3}{7\pi}}\cdot e^{\sqrt{2}t/3}, \quad F_y^{2,3,\pm 1} = -\frac{2i\xi^2}{5} \sqrt{\frac{3}{7\pi}}\cdot e^{\sqrt{2}t/3}, \quad F_z^{2,3,0} = -\frac{6\xi^2}{5\sqrt{7\pi}} \cdot e^{\sqrt{2}t/3}, \qquad \label{Forcing3}
\end{IEEEeqnarray}
where we made use of the spin-2 representation of the angular momentum matrices $J_i$. For the structure constants $f_{\alpha\beta}^{\gamma}$ ($\alpha \equiv jm$, $\beta \equiv j'm'$, $\gamma \equiv j''m''$), we actually didn't have to compute every possible term from their analytic expressions in \eqref{StructureConstants3}--\eqref{StructureConstants5}. On the contrary, in order to specify the forcing terms in \eqref{Forcing3}, we only needed (even though the value of $f_{2,2;2,1}^{3,3}$ was not really used),
\begin{IEEEeqnarray}{l}
f_{2,0;2,1}^{3,1} = -f_{2,1;2,0}^{3,1} = -f_{2,0;2,-1}^{3,-1} = f_{2,-1;2,0}^{3,-1} = -3i\sqrt{\frac{2}{7\pi}}, \qquad f_{2,1;2,-1}^{3,0} = -f_{2,-1;2,1}^{3,0} = \frac{6i}{\sqrt{7\pi}} \qquad \\[6pt]
f_{2,2;2,-1}^{3,1} = -f_{2,-2;2,1}^{3,-1} = 3i\sqrt{\frac{3}{7\pi}}, \qquad f_{2,2;2,1}^{3,3} = f_{2,2;2,0}^{3,2} = -f_{2,-1;2,-1}^{3,-3} = -f_{2,-2;2,0}^{3,-2} = 3i\sqrt{\frac{5}{7\pi}}, \qquad
\end{IEEEeqnarray}
because the only nonzero components of the LO modes $\eta_i^{1jm}$ are those in \eqref{SimpleSolution}. The three components of forcing in \eqref{Forcing3} have been plotted (in a parametric plot) in figure \ref{Figure:ForcingPlot3}.
\indent Once we have obtained all the required forcing terms \eqref{Forcing3}, we may proceed to insert them into the NLO perturbation equation for $\eta_i^{(2)}$ in \eqref{AngularPerturbations11}. We want to solve this equation for the $n = 2$ (NLO), $j = 3$ mode $\eta_i^{2,3,m}$ which was metastable for $n = 1$ (LO), in the $P_-$ sector (recall that $\lambda_-^2 < 0$, for $j=3$ in \eqref{MultipoleEigenvalues2}). The general solution of the NLO equation for $\eta_i^{(2)}$ in \eqref{AngularPerturbations11} is given by:
\begin{IEEEeqnarray}{l}
\eta_k^{2\gamma}\left(t\right) = \tilde{\eta}_{k}^{\gamma}\left(t\right) + \zeta_{k}^{\gamma} \, e^{\sqrt{2}t/3}, \quad k = 1,2,3, \qquad \label{GeneralSolution3}
\end{IEEEeqnarray}
where $\tilde{\eta}_{k}^{\gamma}\left(t\right)$ is the general solution of the homogeneous part of \eqref{AngularPerturbations11}. Because the homogeneous part of the NLO equation for $\eta_i^{(2)}$ in \eqref{AngularPerturbations11} (cf.\ \eqref{AngularPerturbations12}) coincides with the LO equation for $\eta_i^{(1)}$ in \eqref{AngularPerturbations3} (cf.\ \eqref{AngularPerturbations4}), its solution $\tilde{\eta}_{k}^{\gamma}\left(t\right)$ will also be given by an expression of the form \eqref{GeneralSolution1a}. On the other hand, $\zeta_{k}^{\gamma}$ in \eqref{GeneralSolution3} is a special solution of the NLO equation \eqref{AngularPerturbations11}. We can specify it by inserting the general solution \eqref{GeneralSolution3} into the $\eta_i^{(2)}$ equation in \eqref{AngularPerturbations11}, by also using the fact that $\tilde{\eta}_{k}^{\gamma}\left(t\right)$ solves the homogeneous equation \eqref{AngularPerturbations4}. We are led to,
\begin{IEEEeqnarray}{l}
\left(\frac{2}{9} + \omega_3^2\right) \zeta_{i} - u_0\left(u_0 T_{ik} + Q_{ik}\right) \zeta_{k} = \tilde{f}_i, \qquad \label{Forcing4}
\end{IEEEeqnarray}
where $F_k^{2\gamma} \equiv \tilde{f}_k^{\gamma} e^{\sqrt{2}t/3}$ in \eqref{Forcing3}. Obviously, the coefficients $\tilde{f}_k^{\gamma}$ in \eqref{Forcing4} can be immediately read-off from the values of the forcing in \eqref{Forcing3}. Then, writing down the solution of \eqref{Forcing4} is rather straightforward:
\begin{IEEEeqnarray}{l}
\zeta_x^{3,\pm 1} = \pm \frac{9\xi^2}{5}\sqrt{\frac{3}{7\pi}}, \qquad \zeta_y^{3,\pm 1} = -\frac{9i\xi^2}{5}\sqrt{\frac{3}{7\pi}}, \qquad \zeta_z^{3,0} = -\frac{27\xi^2}{5\sqrt{7\pi}}. \label{Solutionj3}
\end{IEEEeqnarray}
We should also take into account the NLO initial conditions \eqref{InitialConditionPerturbations} which read:
\begin{IEEEeqnarray}{l}
\eta_k^{2\gamma}\left(0\right) = \dot{\eta}_k^{2\gamma}\left(0\right) = 0.
\end{IEEEeqnarray}
Plugging these initial conditions into the general solution \eqref{GeneralSolution3} of the NLO fluctuation equations \eqref{AngularPerturbations11}, we obtain a set of constraints for the solution of the homogeneous equation $\tilde{\eta}_{k}^{\gamma}\left(t\right)$:
\begin{IEEEeqnarray}{l}
\tilde{\eta}_{x,y}^{3,\pm 1}\left(0\right) = -\zeta_{x,y}^{3,\pm 1}, \quad \tilde{\eta}_z^{3,0}\left(0\right) = -\zeta_z^{3,0}, \quad \dot{\tilde{\eta}}_{x,y}^{3,\pm 1}\left(0\right) = -\frac{\sqrt{2}}{3}\zeta_{x,y}^{3,\pm 1}, \quad \dot{\tilde{\eta}}_z^{3,0}\left(0\right) = -\frac{\sqrt{2}}{3}\zeta_z^{3,0}. \qquad \label{NLOconstraints}
\end{IEEEeqnarray}
By using the constraints \eqref{NLOconstraints}, it can be demonstrated that the general NLO solution \eqref{GeneralSolution3} obeys the NLO Gauss law constraint \eqref{ConstraintSecondOrder} at the initial time $t = 0$. \\[6pt]
\indent Let us now consider the case $j'' = 1$. Inserting the LO solution \eqref{SimpleSolution} into the formula \eqref{ForcingOrder2} for the NLO forcing term, we find the following non vanishing forcing components for $F_i^{2j''m''}\left(t\right)$:
\begin{IEEEeqnarray}{l}
F_x^{2,1,\pm 1} = \pm\frac{3\xi^2}{40} \sqrt{\frac{3}{2\pi}}\cdot e^{\sqrt{2}t/3}, \quad F_y^{2,1,\pm 1} = -\frac{3i\xi^2}{40} \sqrt{\frac{3}{2\pi}}\cdot e^{\sqrt{2}t/3}, \quad F_z^{2,1,0} = -\frac{\xi^2}{10} \sqrt{\frac{3}{\pi}} \cdot e^{\sqrt{2}t/3}, \qquad \label{Forcing5}
\end{IEEEeqnarray}
where, to obtain the forcing terms \eqref{Forcing5}, we made use of the following (nonzero) values of the structure constants:
\begin{IEEEeqnarray}{l}
f_{2,-0;2,-1}^{1,-1} = f_{2,1;2,0}^{1,1} = -f_{2,-1;2,0}^{1,-1} = -f_{2,0;2,1}^{1,1} = \frac{3i}{2\sqrt{\pi}}, \qquad f_{2,1;2,-1}^{1,0} = -f_{2,-1;2,1}^{1,0} = \frac{i}{2}\sqrt{\frac{3}{\pi}} \qquad \\[6pt]
f_{2,-2;2,1}^{1,-1} = -f_{2,2;2,-1}^{1,1} = i\sqrt{\frac{3}{2\pi}}. \qquad
\end{IEEEeqnarray}
The parametric plot of the forcing \eqref{Forcing5} has been drawn in figure \ref{Figure:ForcingPlot1} below. The special solution of \eqref{Forcing4} reads, in the case of $j'' = 1$:
\begin{IEEEeqnarray}{l}
\zeta_x^{1,\pm 1} = \pm \frac{29\xi^2}{20\sqrt{6\pi}}, \qquad \zeta_y^{1,\pm 1} = -\frac{29i\xi^2}{20\sqrt{6\pi}}, \qquad \zeta_z^{1,0} = -\frac{8\xi^2}{5\sqrt{3\pi}}, \label{Solutionj1}
\end{IEEEeqnarray}
where we can still enforce the NLO initial conditions \eqref{InitialConditionPerturbations} in the same way we did for $j'' = 3$ above. \\[6pt]
\begin{figure}[H]
\begin{center}
\includegraphics[scale=0.4]{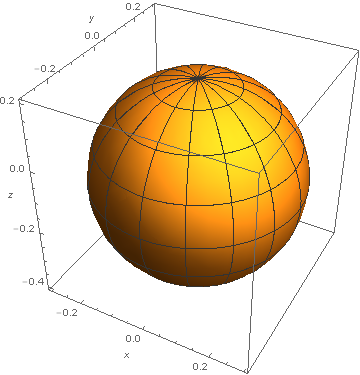}
\end{center}
\caption{Parametric plot of the forcing \eqref{Forcing5}.\label{Figure:ForcingPlot1}}
\end{figure}
\indent Let us now obtain the components of the $j'' = 1$ and $j'' = 3$ solutions that we have just found along each of the three subspaces $P$, $R_{\pm}$. We first recall that the $(n = 1,\ j = 2,\ m = 0)$ LO mode $\xi^{2,0}_{-} \equiv \xi$ (the one we turned on initially) lives exclusively in the $R_-$ sector. Conversely, the decomposition of the solutions \eqref{Solutionj3} and \eqref{Solutionj1} inside the $3\times(2j+1)$-dimensional space reads:
\begin{equation}
Z_{3}=\frac{9\xi^2}{5}\sqrt{\frac{3}{\pi}}\cdot \vert - \rangle\Big|_{j=3, m=0}, \qquad Z_{1}=-\frac{\xi^2}{10\sqrt{2\pi}}\cdot \vert + \rangle\Big|_{j=1, m=0} + \frac{3\xi^2}{2\sqrt{\pi}}\cdot \vert - \rangle\Big|_{j=1, m=0},
\end{equation}
where we have defined the $3\times(2j+1)$ vectors
\begin{IEEEeqnarray}{l}
Z_3 = \begin{pmatrix} \zeta_x^{3m} \\ \zeta_y^{3m} \\ \zeta_z^{3m} \end{pmatrix}, \qquad Z_1 = \begin{pmatrix} \zeta_x^{1m} \\ \zeta_y^{1m} \\ \zeta_z^{1m} \end{pmatrix}.
\end{IEEEeqnarray}
\indent We therefore conclude that LO instabilities are not only capable of propagating towards NLO modes of higher angular momenta $j$ (compared to the angular momenta of the LO instabilities), but they can also spread to different sectors as well. Note however that the magnitude of the $R_+$ instability of the $j'' = 1$ solution is somewhat suppressed compared to the ones in the $R_-$ sector, since $\vert\zeta^3_-\vert\simeq 44\vert\zeta^1_+\vert$ and $\vert\zeta^1_-\vert\simeq 21\vert\zeta^1_+\vert$. \\[6pt]
\indent Our analysis supports a claim, which was first put forward in \cite{AxenidesFloratosLinardopoulos17b} and later supported by further evidence in \cite{AxenidesFloratosKatsinisLinardopoulos21}, that a cascade of instabilities that originates from the unstable $j = 1,2$ sectors at LO, can propagate towards higher multipoles of the unstable sectors. The various (constant $j$) multipoles at a given order in perturbation theory couple to all the $j$’s of the previous orders through the forcing terms, so that the lowest order instabilities (at $j = 1,2$) will couple to all the modes (i.e.\ with different $j$'s) of the first order. The way that these instabilities propagate towards higher modes depends on the algebra of area-preserving diffeomorphisms of the perturbed configuration. Eventually, chaos is expected to emerge in all orders of perturbation theory. At the quantum level, these instabilities could be manifested as a spontaneous emission of higher spin states.
\section[Conclusions and discussion]{Conclusions and discussion \label{Section:ConclusionsDiscussion}}
\noindent Let us conclude with a discussion of our results and a list of interesting research projects. We have studied spherical M2-branes in the large-$N$ limit of the BMN matrix model. These are described by classical bosonic membranes which spin inside the 11-dimensional plane-wave background \eqref{MaximallySupersymmetricBackground1}--\eqref{MaximallySupersymmetricBackground2}. We have classified all possible membrane configurations (three types, I, II, III), based on the relative distribution of their components inside the $SO(3)\times SO(6)$ symmetric background \eqref{MaximallySupersymmetricBackground1}--\eqref{MaximallySupersymmetricBackground2}. For two very representative ansätze, namely the static dielectric membrane in $SO(3)$ and the axially symmetric membrane in $SO(3)\times SO(6)$, we have performed leading order (LO) perturbations in the radial and the angular direction, obtaining the corresponding spectra. By studying perturbations at the next-to-leading order (NLO), we have also demonstrated the instability cascade phenomenon by which dipole ($j=1$) and quadrupole ($j=2$) instabilities propagate from leading order ($n=1$) towards all higher multipoles ($j = 1,2,\ldots$) of higher-order perturbation theory ($n = 2,3,\ldots$). \\[6pt]
\indent Another characteristic feature of the instability cascade phenomenon is the coupling between different multipoles in higher orders of perturbation theory (i.e.\ modes with $j = 1$ couple with $j=2, 3, \ldots$, and modes with $j=2$ couple with $j=3, 4, \ldots$ and so on, for every $n = 2,3,\ldots$). In this sense, the cascade phenomenon can be used as a model for controlling the onset of weak chaos which is similar to the onset of weak turbulence in hydrodynamics. Here, the role of breaking long-wavelength vortices into shorter-wavelength ones is played by unstable multipoles of small $j$’s (i.e.\ long wavelengths $\sim1/j$) which couple to multipoles of larger $j$'s (i.e.\ short wavelengths $\sim1/j$). This analogy can play a crucial role in the physical understanding of the cascade mechanism and the possible existence of energy/wavelength scaling laws, analogous to Kolmogorov scaling in turbulence. \\[6pt]
\indent The results from the study of the cascade/avalanche mechanism can be used to construct a concrete model for the quantum chaotic dynamics of the BH degrees of freedom. These reside on the BH horizon and are governed by the BMN matrix model, as we have argued in the introduction. In this context, we can quantify the rate of fast scrambling and fast information processing by measuring the speed at which an initial perturbation can be transferred to the BH horizon. This way, we can verify the validity of the Hayden-Preskill-Sekino-Susskind limit \cite{HaydenPreskill07, SekinoSusskind08}, according to which BH horizons are the fastest scramblers in nature with a scrambling rate proportional to the logarithm of the BH entropy. All other known chaotic physical systems with local interactions have lower scrambling rates, proportional to fractional powers of the entropy. \\[6pt]
\indent Our main guide in constructing new solutions inside the plane-wave background \eqref{MaximallySupersymmetricBackground1}--\eqref{MaximallySupersymmetricBackground2}, was the form of external and internal symmetries of the membrane. For the static dielectric membrane in $SO(3)$ (see \eqref{StaticMembrane1}), the external symmetries are given by the group of space rotations, time translations, reflections and Galilean transformations. The internal symmetries are somewhat more important and are specified by the parameters of the (infinite-dimensional) group of area-preserving transformations SDiff$\left(\text{S}^2\right)$. It is obvious from the analysis of section \ref{Section:NLOperturbations} that in order to carry out a full-scale investigation of NLO perturbations, we need to know all the structure constants $f_{\alpha\beta}^{\gamma}$ of the Lie-Poisson algebra of spherical harmonics on S$^2$. These in turn provide the structure constants of the Lie algebra of the infinite-dimensional group of area-preserving transformations SDiff$\left(\text{S}^2\right)$. \\[6pt]
\indent An explicit, closed-form (albeit rather lengthy) expression for all the structure constants $f_{\alpha\beta}^{\gamma}$ for spherical harmonics on S$^2$ has been known for a long time (see \cite{ArakelianSavvidy89}). In this paper we only needed the $j = 1,2$ structure constants, the closed-form expressions of which are significantly simpler (cf.\ appendix \ref{Appendix:StructureConstants}). The S$^2$ structure constants give rise to a number of selection rules which constrain the interactions between the various NLO multipoles. Our detailed study of SDiff$\left(\text{S}^2\right)$ however is certainly not the end of the story. We believe that a deeper understanding of the area-preserving symmetry is the key to elucidating the mechanisms of chaos on the surface of BHs. As such, new results in this direction are needed, e.g.\ with regard to the hierarchy of structure constants in the general case, or possible applications of SDiff$\left(\text{S}^2\right)$ to incompressible fluid flow in 2 dimensions. \\[6pt]
\indent As we have already mentioned in the introduction, we view the BMN matrix model (and its large-$N$ limit realization which is described by membranes on the maximally supersymmetric 11-dimensional plane-wave background) as a probe to the chaotic and nonlocal dynamics of the degrees of freedom which reside on the horizons of BHs. By this token, it would be especially interesting to attempt to make an estimate for the scrambling time. Based on what we have said above, the scrambling time should be related to the time evolution of high-order high-multipole instabilities which are generated by dipole and quadrupole instabilities at LO perturbation theory. In particular, we need to compute the rate by which energy diffuses from dipole and quadrupole perturbations towards higher multipole perturbations (energy-multipole relation). Given that the chaotic phenomena which occur on BH horizons are mainly quantum in nature (our present paper only discusses classical aspects of BMN membranes), the above steps presuppose a purely quantum mechanical treatment of BMN membranes and their instabilities. \\[6pt]
\indent We believe that the instability cascade/avalanche phenomenon is related to the onset of weak chaos in membrane dynamics which is in many ways analogous to the onset of weak turbulence in hydrodynamics. Having computed the time evolution of instabilities and the rate by which energy diffuses towards higher multipoles (shorter wavelengths), one could envisage formulating scaling laws à la Kolmogorov. A computer could be used to visualize the time evolution of disturbances and simulate the cascade/avalanche phenomenon. For a single multipole excitation with predefined energy, it would be interesting to model the mechanism for the diffusion of energy to neighboring multipoles and provide a numerical estimate for the time needed to de-excite a multipole by simultaneously diffusing its energy to neighboring multipoles. \\[6pt]
\indent Although our study of the radial and angular/multipole membrane spectra was very detailed, it is still rather incomplete in the sense that only local Lyapunov exponents around the various critical points have been computed. It would be interesting to obtain the full (global) Lyapunov spectrum of our system in the infinite-dimensional phase space of multipole modes (i.e.\ $\eta^{jm}\left(\tau\right)$, $\epsilon^{jm}\left(\tau\right)$, $\zeta^{jm}\left(\tau\right)$). Moreover, the application of a wide set of diagnostic tools from dynamical systems, classical chaos, KAM and Melnikov theory (e.g.\ phase space diagrams, Poincaré maps, separatrix theory etc.), could lead to a deeper understanding of the stability/instability properties, as well as the mechanism by which classical chaos emerges in the case of membranes in plane-wave spacetimes. For the BFSS/BMN matrix models, such studies have already been carried out, see for instance \cite{AsanoKawaiYoshida15, Gur-AriHanadaShenker15, CotlerCotlerGur-AriHanadaPolchinskiSaadShenkerStanfordStreicherTezuka16, BuividovichHanadaSchafer18}. \\[6pt]
\indent Another closely related problem to the calculations we have performed in this paper, is that of post-instability, post-chaotic state of membranes. The various instabilities and strongly chaotic behavior of relativistic membranes are due to their self-interactions. These induce changes in their shapes and may eventually modify their topologies. More generally, explaining the processes by which membranes (self-)interact is an important open problem which reflects our present level of understanding of membrane field theory (or its lack thereof). So far, only a few (Euclidean-time) membrane solutions are known and have been shown to undergo topology changes \cite{KovacsSatoShimada15}. Studying membrane interactions and the topology changing phenomenon for the solutions we have analyzed in this paper would be an interesting followup project. Multimembrane configurations (not considered in this paper) is one of the key ingredients of matrix models, which is expected to have a crucial role in describing the chaotic dynamics of black hole horizons. The instabilities which are associated with the single-membrane systems we treated in this paper, as well as multi-membrane systems (both at the level of matrix models and their classical membrane limits) are responsible for the onset of chaos and are inevitably related to the study of topology changes in the shapes of membranes. \\[6pt]
\indent Yet another interesting followup project, would be to study NLO perturbations (and instability cascade) by considering the full $SO(3)\times SO(6)$ geometry of plane-waves. Similarly, since our present study focused only on the tree-level, leading and next-to-leading order properties of bosonic membranes (i.e.\ by omitting the fermionic components), it would be interesting to repeat our analyses by including fermions in the membrane action. The membrane ansätze we have considered break the supersymmetry of the background \eqref{MaximallySupersymmetricBackground1}--\eqref{MaximallySupersymmetricBackground2}, so that it would also be interesting to determine what fraction of the initial supersymmetry is preserved. Apart from the 3-dimensional extended objects (M2-branes) that we studied in the present paper, M-theory is also known to include higher ($p+1$) dimensional extended objects, Mp-branes.\footnote{In fact supersymmetry only allows specific Mp-brane dimensionalities to exist in various spacetime dimensions. The corresponding table (mapping $p$, the allowed Mp-brane dimensionality, to the corresponding spacetime dimension $D$) is known as the "brane scan" \cite{AchucarroEvansTownsendWiltshire87}. See also \cite{Duff96}.} Since 11 spacetime dimensions can, apart from M2-branes also host M5-branes, it would be interesting to extend our present considerations to the case of M5-branes as well. It would also be interesting to explore the Carroll ultra-relativistic limit of our configurations, much along the lines of \cite{Roychowdhury19d} who considered it for M2-branes in 11-dimensional supergravity backgrounds. \\[6pt]
\indent Of course there are plenty other interesting future directions, such as for example to consider the Penrose limit of various classes of relativistic membranes, for example the solutions \cite{Hoppe22a, Hoppe23a, Hoppe23b, Hoppe23c, Hoppe23d}, as well as certain membranes which are toroidally compactified on M$_9\times \text{T}^2$ \cite{AlvarezGarciaMoralPenaPrado21, AlvarezMoralPenaPrado21}. The study of minimal surfaces \cite{HoppeLinardopoulosTurgut2016, Hoppe20, Hoppe21e, Hoppe21g}\footnote{See also \cite{Hoppe13b, Hoppe21f}.} and their quantization \cite{ArnlindHoppeKontsevich19, Hoppe21h} is also highly relevant to the purposes of the present work. \\[6pt]
\indent Finally let us explore some applications of our work to cosmology. M2 and M5 branes show up in M-theory and 11 dimensional supergravity as 2d and 5d (electric and magnetic) sources of the 3-form antisymmetric gauge field $A_{IJK}$. Closed finite-energy membrane configurations (such as those in our paper and in \cite{Myers99b}) have long been known to behave as bubbles of false vacuum \cite{ColemanDeLuccia80} with a positive cosmological constant $\Lambda$, whenever the gravitational field is coupled to a 3-index antisymmetric gauge potential \cite{AuriliaNicolaiTownsend80, DuffNieuwenhuizen80, FreundRubin80}. The cosmological constant $\Lambda$ emerges as a constant of integration to the classical equations of motion and it can subsequently be treated as a dynamical variable which relaxes to a small positive value. \\[6pt]
\indent Several relaxation mechanisms of $\Lambda$ have been proposed in the literature, with bubble nucleation \cite{BrownTeitelboim87} and multiple compactification fluxes \cite{BoussoPolchinski00, FengMarchRussellSethiWilczek00} being more closely related to the scopes of the present work. Bubble nucleation behaves as pair creation when an external electric field is present, causing the initially large and positive cosmological constant to decrease in value. A similar mechanism which leaves behind BHs in the aftermath of vacuum bubble thermal decay has also been worked out \cite{Teitelboim85b, GomberoffHenneauxTeitelboim01, GomberoffHenneauxTeitelboimWilczek03}. \\[6pt]
\indent M2 branes are classical counterparts of matrix models \cite{BMN02, BFSS97} which are conjectured to describe M-theory. In early universe cosmology, matrix models capture many subtle features of the quantum gravitational origins of inflation. Their dynamical role as launching pads of our emergent universe also constitutes a very active current area of research \cite{Ketov19b, BrahmaBrandenbergerLaliberte21}. As we have seen, certain classical counterparts of the BMN matrix model (that is membranes in the plane wave background \eqref{MaximallySupersymmetricBackground1}--\eqref{MaximallySupersymmetricBackground2}) exhibit an interesting interplay between attractive and repulsive (e.g.\ Myers flux) terms in their potential energies. Their precise role as springboards of the inflationary universe, as well as the origin of solitonic configurations (Q-balls, see \cite{Coleman85c, SafianColemanAxenides86, Axenides86}) deserves a closer scrutiny in our future work. 
\section[Acknowledgements]{Acknowledgements}
\noindent We thank C.\ Bachas, T.\ Bountis and D.\ Katsinis for discussions. Special thanks to M.\ Roberts for comments on the final form of the manuscript. G.L.\ would like to thank M.\ Bordemann, D.\ O'Connor, J.\ Hoppe, B.-H.\ Lee, H.\ Steinacker and T.\ Turgut for illuminating discussions. G.L.\ also thanks the organizers of the conference \textit{Space Time Matrices} which was held in IHES Paris in 2019, for support. The work of G.L.\ was supported in part by the National Research Foundation of Korea (NRF) grant funded by the Korea government (MSIT) (No.\ 2023R1A2C1006975), as well as by an appointment to the JRG Program at the APCTP through the Science and Technology Promotion Fund and Lottery Fund of the Korean Government.
\appendix
\section[Membranes in plane-wave backgrounds as classical tops]{Membranes in plane-wave backgrounds as classical tops \label{Appendix:ClassicalTopProperty}}
\noindent The purpose of the present appendix is to express the energy of membranes in the maximally supersymmetric background \eqref{MaximallySupersymmetricBackground1}--\eqref{MaximallySupersymmetricBackground2} in such a way that their classical top property is made manifest. We follow closely \cite{AxenidesFloratos07}, where the corresponding exercise was carried out in flat Minkowski space. Let us start from the generic ansatz
\begin{IEEEeqnarray}{clclclcl}
x^i &= R_x^{ij}\left(\tau\right) & x_0^j\left(\boldsymbol{\sigma}\right) & \qquad \& \qquad & y^i & = R_y^{ij}\left(\tau\right) & y_0^j\left(\boldsymbol{\sigma}\right), \label{Ansatz5} \\
\downarrow && \downarrow && \downarrow && \downarrow \nonumber \\
\begin{array}{c} \text{space} \\ \text{frame} \end{array} && \begin{array}{c} \text{brane} \\ \text{frame} \end{array} && \begin{array}{c} \text{space} \\ \text{frame} \end{array} && \begin{array}{c} \text{brane} \\ \text{frame} \end{array} \nonumber
\end{IEEEeqnarray}
where the coordinates $x$ and $y$ are in the so-called "space" frame, while the zero subscripts denote the "brane" frame of the membrane. $R_x$ and $R_y$ are the rotation matrices
\begin{IEEEeqnarray}{c}
R_x\left(\tau\right) \equiv \exp\left(\Omega_x \tau\right) \qquad \& \qquad R_y\left(\tau\right) \equiv \exp\left(\Omega_y \tau\right),
\end{IEEEeqnarray}
while $\Omega_x$, $\Omega_y$ are any antisymmetric matrices (the same in the space and the brane frame)
\begin{IEEEeqnarray}{c}
\Omega_x^T = -\Omega_x^{-1} \qquad \& \qquad \Omega_y^T = -\Omega_y^{-1},
\end{IEEEeqnarray}
such that $R_x$ and $R_y$ (which are again the same in the space/brane frame) are orthogonal as they should:
\begin{IEEEeqnarray}{c}
R_x^T = R_x^{-1} \qquad \& \qquad R_y^T = R_y^{-1}.
\end{IEEEeqnarray}
\indent A special instance of the ansatz \eqref{Ansatz5} is the spherical configuration \eqref{SphericalAnsatz1x}--\eqref{SphericalAnsatz3y}, \eqref{Ansatz1a}--\eqref{Ansatz1b} that we introduced in section \ref{Section:DielectricMembranes}. We also define:
\begin{IEEEeqnarray}{c}
v_x \equiv R_x^{-1} \ddot{R}_x = \Omega_x^2 \qquad \& \qquad v_y \equiv R_y^{-1} \ddot{R}_y = \Omega_y^2.
\end{IEEEeqnarray}
The moments of inertia in the brane frame of the membrane (subscript "B") are defined as follows:
\begin{IEEEeqnarray}{c}
\textrm{I}_{x(\text{B})}^{ij} = T\int d^2\sigma \ x_0^i x_0^j, \qquad \textrm{I}_{y(\text{B})}^{ij} = T\int d^2\sigma \ y_0^i y_0^j,
\end{IEEEeqnarray}
while the corresponding (conserved) angular momenta are given by
\begin{IEEEeqnarray}{c}
L_{x(\text{B})}^{ij} = T\int d^2\sigma \left(\dot{x}_0^i x_0^j - \dot{x}_0^j x_0^i\right), \qquad L_{y(\text{B})}^{ij} = T\int d^2\sigma \left(\dot{y}_0^i y_0^j - \dot{y}_0^j y_0^i\right).
\end{IEEEeqnarray}
\indent Both quantities can be transformed between the brane and the space frame (subscript "S") by the following transformations:
\begin{IEEEeqnarray}{c}
\textrm{I}_{\text{S}} \equiv R\cdot \textrm{I}_{\text{B}} \cdot R^{-1} \qquad \& \qquad L_{\text{S}} \equiv R\cdot L_{\text{B}} \cdot R^{-1}.
\end{IEEEeqnarray}
Plugging the ansatz \eqref{Ansatz5} into the expression \eqref{ppWaveHamiltonian1}--\eqref{ppWaveHamiltonian2} for the energy of the membrane we find:
\begin{IEEEeqnarray}{c}
E = -\frac{3}{4}\left(\frac{1}{2}\cdot\frac{\text{Tr}\left[\Omega_x L_x\right]^2}{\text{Tr}\left[\Omega_x^2 \textrm{I}_x\right]} + \frac{1}{2}\cdot\frac{\text{Tr}\left[\Omega_y L_y\right]^2}{\text{Tr}\left[\Omega_y^2 \textrm{I}_y\right]}\right) + V_{\text{ext}}. \label{ClassicalTopEnergy}
\end{IEEEeqnarray}
The angular momenta and the moments of inertia can be in either frame ("space" or "brane") and
\begin{IEEEeqnarray}{c}
V_{\text{ext}} = \frac{T}{4} \int d^2\sigma \left[\frac{\mu^2}{9} \, x_0^i x_0^i + \frac{\mu^2}{36} \, y_0^i y_0^i + \mu\left(\frac{1}{2}R^{nk} - \frac{2}{3}R^{kn}\right)\epsilon_{ijk}R_x^{il} R_x^{jm}\{x_0^l,x_0^m\}x_0^n \right].
\end{IEEEeqnarray}
In the case of flat Minkowski space, $\mu = 0 \Rightarrow V_{\text{ext}} = 0$ and the energy \eqref{ClassicalTopEnergy} of the membrane is very similar to that of an Euler top:
\begin{IEEEeqnarray}{ll}
E = \frac{\ell_x^2}{2\textrm{I}_x} + \frac{\ell_y^2}{2\textrm{I}_y} + \frac{\ell_z^2}{2\textrm{I}_z}.
\end{IEEEeqnarray}
In plane-wave spacetimes $\mu \neq 0$ and the membrane is a classical (non-Eulerian) top that moves under the influence of the external torques that are contained in $V_{\text{ext}}$.
\section[Radial perturbations revisited]{Radial perturbations revisited \label{Appendix:RadialPerturbations}}
\noindent In section \ref{SubSubSection:RadialPerturbations2} we studied the stability of the two allowed extremal points of the $SO(3)\times SO(6)$ axially symmetric potential \eqref{AxiallySymmetricPotential2}. Our analysis was rather generic, in the sense that it made no reference to a particular solution of the membrane equations of motion. With this approach however, we were essentially obliged to treat the angular momentum $\ell$ as a constant, and we ignored the fact that it must also fluctuate when the radii are perturbed. In the present appendix (a condensed form of which was included in the publication \cite{AxenidesFloratosLinardopoulos17a}), we take the other route by considering a particular solution of the membrane equations of motion \eqref{AxiallySymmetricMembrane2a}--\eqref{AxiallySymmetricMembrane2c} and allowing the angular momentum $\ell$ to vary along with the variations of the target space coordinates. In this way, we complement the analysis of radial perturbations which was carried out in section \ref{SubSubSection:RadialPerturbations2}, as well as section \ref{SubSection:AxiallySymmetricMembrane}, where the stability of the membrane was inferred by the eigenvalues of the matrix of second derivatives of the potential (Hessian). \\[6pt]
\indent To proceed, we consider the following axially symmetric (type III) solution of the membrane equations of motion \eqref{AxiallySymmetricMembrane2a}--\eqref{AxiallySymmetricMembrane2c} (with $s_1 = s_2 = s_3 = 2$):
\begin{IEEEeqnarray}{c}
u_i^0 = u_0, \quad v_j^0\left(t\right) = v_0 \cos\left(\omega t + \varphi_j\right), \quad w_j^0\left(t\right) \equiv v_{j+3}^0\left(t\right) = v_0 \sin\left(\omega t + \varphi_j\right), \quad i,j = 1,2,3, \quad \label{DielectricTop2}
\end{IEEEeqnarray}
where $u_0$ and $v_0$ are the extrema of the axially symmetric potential \eqref{AxiallySymmetricPotential2} that obey the left equation in \eqref{AxiallySymmetricExtrema1}, as well as \eqref{AngularVelocity}. More about the dielectric-top solution \eqref{DielectricTop2} can be found in section \ref{SubSection:AxiallySymmetricMembrane}, as well as in section \ref{SubSubSection:AngularPerturbations2} where its angular stability was studied. Let us now set out to examine the radial stability of this solution. Setting, for the $SO(3) \times SO(6)$ variables,
\begin{IEEEeqnarray}{l}
u_i = u_i^0 + \delta u_i\left(t\right), \qquad v_i = v_i^0\left(t\right) + \delta v_i\left(t\right), \qquad w_i = w_i^0\left(t\right) + \delta w_i\left(t\right), \qquad i = 1,2,3, \label{AxialPerturbations2}
\end{IEEEeqnarray}
and subsequently inserting these relations into the equations of motion \eqref{EquationsOfMotion1}--\eqref{EquationsOfMotion3}, by also using the minimization condition \eqref{AxiallySymmetricExtrema1} (for $u_0 \neq 0$), we obtain the following system of linearized equations:
\begin{IEEEeqnarray}{l}
\left[\begin{array}{c} \delta\ddot{\textbf{u}} \\ \delta\ddot{\textbf{v}} \\ \delta\ddot{\textbf{w}} \end{array}\right] + \left[\begin{array}{ccc} A_1 & A_2' & A_3 \\ A_2' & B_1' & B_2 \\ A_3 & B_2 & B_3 \end{array}\right]\cdot\left[\begin{array}{c} \delta\textbf{u} \\ \delta\textbf{v} \\ \delta\textbf{w} \end{array}\right] = 0, \label{LinearizedSystem3}
\end{IEEEeqnarray}
where we have defined (for simplicity, we omit the constant phase factors $\varphi_i$),
\begin{IEEEeqnarray}{ll}
A_1 = u_0 \, I_3 + u_0\left(2u_0-1\right) \cdot \mathfrak{g}_3, \qquad & B_1' = \left(u_0 - \frac{1}{12}\right) I_3 + 2v_0^2\cos\omega t \cdot \mathfrak{g}_3 \\[6pt]
A_2' = 2u_0v_0\cos\omega t \cdot \mathfrak{g}_3, \qquad & B_2 = v_0^2\sin\omega t \cdot \mathfrak{g}_3 \\[6pt]
A_3 = 2u_0v_0\sin\omega t \cdot \mathfrak{g}_3, \qquad & B_3 = \left(u_0 - \frac{1}{12}\right) I_3 + 2v_0^2\sin\omega t \cdot \mathfrak{g}_3, \label{LinearizedSystem4}
\end{IEEEeqnarray}
while $I_n$ stands for the $n$-dimensional identity matrix and the $3\times 3$ matrix $\mathfrak{g}_3$ is given by
\begin{IEEEeqnarray}{lll}
\mathfrak{g}_3 \equiv
\left(\begin{array}{ccc}
0 & 1 & 1 \\
1 & 0 & 1 \\
1 & 1 & 0
\end{array}\right).
\end{IEEEeqnarray}
\indent Following \cite{AxenidesFloratosPerivolaropoulos00, AxenidesFloratosPerivolaropoulos01}, we may transform the linear inhomogeneous system \eqref{LinearizedSystem3}--\eqref{LinearizedSystem4} into a constant-coefficient system by performing the following rotation:
\begin{IEEEeqnarray}{l}
\delta v'_{i} = \cos\omega t \cdot \delta v_i + \sin\omega t \cdot \delta w_i, \qquad \delta w'_{i} = -\sin\omega t \cdot \delta v_i + \cos\omega t \cdot \delta w_i, \qquad i = 1,2,3, 
\end{IEEEeqnarray}
which leads to:
\begin{IEEEeqnarray}{l}
\left[\begin{array}{c} \delta\ddot{\textbf{u}} \\ \delta\ddot{\textbf{v}}' \\ \delta\ddot{\textbf{w}}' \end{array}\right] + 2 \, \omega \left[\begin{array}{ccc} 0 & 0 & 0 \\ 0 & 0 & -I_3 \\ 0 & I_3 & 0 \end{array}\right] \cdot \left[\begin{array}{c} \delta\dot{\textbf{u}} \\ \delta\dot{\textbf{v}}' \\ \delta\dot{\textbf{w}}' \end{array}\right] + \left[\begin{array}{ccc} A_1 & A_2 & 0 \\ A_2 & B_1 & 0 \\ 0 & 0 & 0 \end{array}\right] \cdot \left[\begin{array}{c} \delta\textbf{u} \\ \delta\textbf{v}' \\ \delta\textbf{w}' \end{array}\right] = 0, \label{LinearizedSystem5}
\end{IEEEeqnarray}
and we have defined
\begin{IEEEeqnarray}{c}
A_2 = 2u_0v_0 \cdot \mathfrak{g}_3, \qquad \& \qquad B_1 = 2v_0^2 \cdot \mathfrak{g}_3. \label{LinearizedSystem6} \qquad
\end{IEEEeqnarray}
In order to solve the reduced system of equations \eqref{LinearizedSystem5}--\eqref{LinearizedSystem6}, we plug the following general solution into \eqref{LinearizedSystem5}:
\begin{IEEEeqnarray}{l}
\left[\begin{array}{c} \delta\textbf{u} \\ \delta\textbf{v}' \\ \delta\textbf{w}' \end{array}\right] = \sum_{i=1}^{18} c_i \, e^{\lambda_i t} \, \boldsymbol{\xi}_i, \label{Eigenvalues3}
\end{IEEEeqnarray}
where the $c_i$ are constants (which are specified by the initial conditions) and $\lambda_i$, $\boldsymbol{\xi}_i$ are the eigenvalues/eigenvectors of the resulting eigenproblem, for all $i = 1,\ldots,18$. The calculation shows that there are 6 zero modes (which are related to the global symmetry of the $SO(6)$ potential) and the following 4 modes which are generically nonzero:
\begin{IEEEeqnarray}{l}
\lambda_{1\pm}^2 = \frac{1}{9} - \frac{5u_0}{2} \pm \sqrt{\frac{1}{9^2} - \frac{u_0}{9} - \frac{5u_0^2}{12} + 4 u_0^3}, \label{AxiallySymmetricEigenvalues2a} \\[6pt]
\lambda_{2\pm}^2 = \frac{5}{18} - \frac{5u_0}{2} \pm \sqrt{\frac{5^2}{18^2} - \frac{35u_0}{18} + \frac{163u_0^2}{12} - 20u_0^3}. \label{AxiallySymmetricEigenvalues2b}
\end{IEEEeqnarray}
These eigenvalues have degeneracies 4 and 2 respectively (so that $6 + 2\cdot4 + 2\cdot2 = 18$). Note that the eigenvalues $\lambda_{2\pm}$ in \eqref{AxiallySymmetricEigenvalues2b} are identical to the ones in \eqref{AxiallySymmetricEigenvalues1}, a clear indication that the present analysis is in complete agreement with the conclusions of section \ref{SubSubSection:RadialPerturbations2}. A plot of the squares of the eigenvalues \eqref{AxiallySymmetricEigenvalues2a}--\eqref{AxiallySymmetricEigenvalues2b} appears in the following figure \ref{Graph:Eigenvalues3}. Notice the striking resemblance with the corresponding plot of radial eigenvalues which appeared in figure \ref{Graph:Eigenvalues1}.
\begin{figure}[H]
\begin{center}
\includegraphics[scale=0.4]{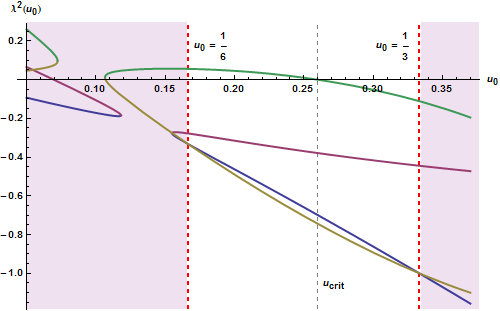}
\caption{Plot of the eigenvalues \eqref{AxiallySymmetricEigenvalues2a}--\eqref{AxiallySymmetricEigenvalues2b} as a function of the coordinate $u_0$.} \label{Graph:Eigenvalues3}
\end{center}
\end{figure}
\vspace{-0.4cm}\indent As we anticipated, our conclusions are generally identical to those that we found by means of the analysis in section \ref{SubSubSection:RadialPerturbations2}. In the domain of allowed $u_0$'s \eqref{ExtremalBounds1}, the spectrum of the (type III) axially symmetric configuration \eqref{DielectricTop2} always includes 3 purely imaginary modes (for which $\lambda^2<0$). These correspond to stable directions. By contrast, the square of the non-degenerate mode $\lambda_{2+}$ can be either greater or less than zero depending on whether $u_0$ is respectively smaller or greater than $u_{\text{crit}} = \left(11+\sqrt{21}\right) / 60 \approx 0.25971$. For $u_0 = u_{\text{crit}}$, $\lambda_{2+}^2$ flips sign, causing the corresponding direction to change from being stable ($\lambda_{2+}^2<0$) to being stable/unstable, depending on whether the sign is negative/positive (while $\lambda_{2+}^2>0$). Therefore the rightmost extremum ($u_0 > u_{\text{crit}}$) is always stable, whereas the leftmost extremum ($u_0 < u_{\text{crit}}$) is stable/unstable depending on the sign of the real eigenvalue $\lambda_+$.
\section[Structure constants]{Structure constants \label{Appendix:StructureConstants}}
\noindent The definition of the spherical harmonic structure constants $f_{\alpha\beta}^{\gamma}$ (showing up in the forcing terms \eqref{Forcing1a}--\eqref{Forcing1b}) was provided in \eqref{SphericalHarmonics2}. By inverting \eqref{SphericalHarmonics2} we are led to
\begin{equation}
f_{\alpha\beta}^{\gamma} = \int_{\text{S}^2} Y^{*}_{\gamma}\left(\theta,\phi\right)\left\{Y_{\alpha}\left(\theta,\phi\right),Y_{\beta}\left(\theta,\phi\right)\right\}d\Omega, \label{StructureConstants1}
\end{equation}
which allows us to work out a closed formula for the structure constants $f_{\alpha\beta}^{\gamma}$, for all values of the quantum numbers $\alpha \equiv jm$, $\beta \equiv j'm'$ and $\gamma \equiv j''m''$. See \cite{ArakelianSavvidy89}. The closed expressions for the $j = 1,2$ structure constants are much simpler. They read:

\footnotesize\begin{IEEEeqnarray}{l}
f^{j'm'}_{1,\pm 1;jm} = \pm i \, \sqrt{\frac{3}{8\pi}} \cdot \sqrt{\left(j\mp m\right) \left(j \pm m + 1\right)}\,\delta_{j'j}\,\delta_{m',m\pm 1}, \qquad
f^{j'm'}_{1,0;jm} = -i m \cdot\sqrt{\frac{3}{4\pi}} \, \delta_{j'j}\,\delta_{m'm} \qquad \label{StructureConstants2} \\[12pt]
f^{j'm'}_{2,0;jm} = -3i m \sqrt{\frac{5}{4\pi}} \cdot \left[\sqrt{\frac{\left(j + 1\right)^2 - m^2}{\left(2j + 1\right) \left(2j + 3\right)}} \, \delta_{j',j+1} + \sqrt{\frac{j^2 - m^2}{\left(2j + 1\right) \left(2j - 1\right)}} \, \delta_{j',j-1}\right]\cdot\delta_{m'm} \qquad \label{StructureConstants3} \\[12pt]
f^{j'm'}_{2,\pm 1;jm} = \pm i \sqrt{\frac{15}{8\pi}} \cdot \Bigg[\left(j \mp 2m\right)\cdot \sqrt{\frac{\left(j \pm m + 1\right) \left(j \pm m + 2\right)}{\left(2j + 1\right) \left(2j + 3\right)}} \, \delta_{j',j+1} + \nonumber \\
\hspace{6cm} + \left(j \pm 2m + 1\right) \cdot \sqrt{\frac{\left(j \mp m - 1\right) \left(j \mp m\right)}{\left(2j + 1\right) \left(2j - 1\right)}} \, \delta_{j',j-1}\Bigg] \cdot \delta_{m',m \pm 1} \qquad \label{StructureConstants4} \\[12pt]
f^{j'm'}_{2,\pm2;jm} = \pm i \sqrt{\frac{15}{8\pi}}\cdot \Bigg[\sqrt{\frac{\left(j \mp m\right) \left(j \pm m + 1\right) \left(j \pm m + 2\right) \left(j \pm m + 3\right)}{\left(2j + 1\right) \left(2j + 3\right)}}\,\delta_{j',j+1} - \nonumber \\
\hspace{5cm} - \sqrt{\frac{\left(j \mp m\right) \left(j \mp m - 1\right) \left(j \mp m - 2\right) \left(j \pm m + 1\right)}{\left(2j + 1\right) \left(2j - 1\right)}}\,\delta_{j',j-1}\Bigg]\cdot\delta_{m',m \pm 2}. \qquad \label{StructureConstants5}
\end{IEEEeqnarray}\normalsize
From \eqref{StructureConstants1} we infer that the structure constants $f_{\alpha\beta}^{\gamma}$ obey the following set of sum rules:
\begin{equation}
m + m' = m'', \qquad j + j' + j'' = \text{odd},
\end{equation}
where the rightmost equation can be derived from \eqref{StructureConstants1} by sending $(\theta,\phi)\rightarrow(\pi-\theta,\phi+\pi)$. Moreover, the quantum numbers $j,\thinspace j'$ and $j''$ can be shown to obey a set of triangle inequalities. For instance, it can be shown that the following inequality holds,
\begin{equation}
\left|j - j'\right| + 1 \leq j'' \leq \left|j + j'\right| - 1, \label{TriangleInequality}
\end{equation}
along with all its cyclic permutations in $\left[j,j',j''\right]$. In relation to the present work, \eqref{TriangleInequality} then implies that, when the fluctuation modes $\eta^{1jm}$ are switched on up to a maximum angular momentum $j_{\textrm{max}}$ (i.e.\ $j,j' \leq j_{\textrm{max}}$), the forcing term $F^{2j''m''}$ becomes zero for all $j''\geq 2j_{\textrm{max}}$.

\bibliographystyle{style}
\bibliography{Phys_Bibliography,Math_Bibliography}
\end{document}